\documentclass[twocolumn]{aastex62}

\usepackage{natbib}

\usepackage{amsmath}

\usepackage[T1]{fontenc}

\graphicspath{{./}{figures/}}

\shorttitle{NGC 1977 VLA}
\shortauthors{Boyden \& Eisner}

\begin{document}

\title{{\bf {\large Constraining free-free emission and photoevaporative mass loss rates for known proplyds and new VLA-identified candidate proplyds in NGC 1977}}}

\correspondingauthor{Ryan Boyden}
\email{rboyden@virginia.edu}

\author[0000-0001-9857-1853]{Ryan D. Boyden}
\altaffiliation{Virginia Initiative on Cosmic Origins Fellow.}
\affiliation{Department of Astronomy, University of Virginia, Charlottesville, VA 22904, USA}
\affiliation{Space Science Institute, Boulder, CO 80301, USA}
\affil{Steward Observatory, University of Arizona, 933 N. Cherry Ave, Tucson, AZ, 85719, USA}

\author[0000-0002-1031-4199]{Josh Eisner}
\affil{Steward Observatory, University of Arizona, 933 N. Cherry Ave, Tucson, AZ, 85719, USA}

\begin{abstract}

{
We present Karl G. Jansky Very Large Array observations covering the NGC 1977 region at 3.0, 6.4, and 15.0 GHz. We search for compact radio sources and detect continuum emission from {34} NGC 1977 cluster members and {37} background objects.  Of the {34} radio-detected cluster members, 3 are associated with known proplyds in NGC 1977, 22 are associated with additional young stellar objects in NGC 1977, and {9} are newly-identified cluster members. {We examine the radio spectral energy distributions, circular polarization, and variability of the detected NGC 1977 sources, and identify 
10 new candidate proplyds whose radio fluxes are dominated by optically thin free-free emission. 
{We use measurements of free-free emission to calculate the mass-loss rates of known proplyds and new candidate proplyds in NGC 1977},} and find values $\sim10^{-9}-10^{-8}$ M$_{\odot}$ yr$^{-1}$, which are lower than the mass-loss rates measured towards proplyds in the Orion Nebula Cluster, but consistent with the mass-loss rates predicted by external photoevaporation models for spatially-extended disks that are irradiated by the typical external UV fields encountered in NGC 1977. 
Finally, we show that photoevaporative disk winds in NGC 1977 may be illuminated by internal or external sources of ionization, depending on their positions within the cluster. This study provides new constraints on disk properties in a clustered star-forming region with a weaker UV environment than the Orion Nebula Cluster, but a stronger UV environment than low-mass star-forming regions like Taurus. Such intermediate UV environments represent the typical conditions of Galactic star and planet formation.
}

\end{abstract}


\section{\bf{Introduction}} \label{sec:intro}

Most star formation occurs 
in dense, massive clusters \citep[e.g.,][]{Lada93, Lada03, Krumholz19}, and the formation and evolution of planets is likely influenced by the stellar cluster environment. In particular, clusters host massive OB stars that irradiate their surroundings with ultraviolet (UV) photons, and the intense UV radiation from massive stars is capable of heating and dispersing protoplanetary disks through a process known as ``external photevaporation'' \citep[][and references therein]{Winter22}. 
Theoretical work suggests that external photoevaporation can severely reduce the masses, sizes, and lifetimes of disks over a range of realistic disk-cluster conditions \citep[][]{Johnstone98, Storzer99, Scally01, Adams04, Clarke07, Facchini16, Haworth18, Haworth18b, Winter18,  Haworth19, Parker21b, Coleman22, Qiao22}, which can in turn influence the timescale and building blocks of planet formation  \cite[e.g., ][]{Throop05, Walsh13, Ndugu18, Nicholson19, Sellek20, Haworth21a, Winter22b, Boyden23, Qiao23}.
Indirect evidence for external photoevaporation is also routinely observed in nearby clusters, with surveys finding lower disk fractions near massive OB stars \citep[e.g.,][]{Balog07, Guarcello07, Fang12, Guarcello16, vanTerwisga23}, lower disk masses in clustered vs. lower-mass star-forming regions \citep[e.g.,][]{Ansdell17, Eisner18, Terwisga19, vanTerwisga20, Mauco23}, and a lack of spatially extended disks in nearby O-star-hosting clusters \citep[e.g.,][]{Eisner18, Boyden20, Otter21}.

The Orion Nebula Cluster (ONC) has long been regarded as the prototypical region for studies of disk evolution in clustered star-formation environments. At a distance of the 400 pc \cite[][]{Hirota07, Kraus07, Menten07, Sandstrom07, Kounkel17, Grob18, Kounkel18}, the ONC hosts several-thousand young ($1-2$ Myr), low-mass (${<}2$ M$_{\odot}$) stars \citep{Hillenbrand97, Fang21} that are irradiated by the massive Trapezium stars, most notably the O-star $\theta^1$ Ori C. Compared with other nearby clusters, the ONC contains the largest known population of {\it proplyds}\textemdash disks that are surrounded by cocoons of ionized gas with a cometary morphology. 
The presence and morphologies of proplyds are a direct result of the external photoevaporation process, in which strong far-ultraviolet (FUV) and extreme-ultraviolet (EUV) radiation from massive stars drive material off the disks in the form of ionized photoevaporative winds \citep[e.g.,][]{Johnstone98, Storzer99}.
{The high surface brightnesses of the ONC proplyds have enabled large samples of photoevaporating disks in the ONC to be detected and characterized with a range of facilities, including the Very Large Array \citep[e.g.,][]{Churchwell87, Garay87, Zapata04a, Forbrich16, Sheehan16}, the Hubble Space Telescope \cite[HST; e.g.,][]{Odell94, Bally98, Bally00, Ricci08} the {Atacama Large Millimeter/submillimeter Array \citep[ALMA; e.g.,][]{Eisner18, Ballering23}}, {VLT-MUSE \citep[e.g.,][]{Haworth23, Aru24}}, and more recently, JWST \citep[e.g.,][]{Berne22, Habart23, McCaughrean23, Berne24}.}

While studies of the ONC proplyds have provided crucial information on how strong UV fields from massive stars launch photoevaporative winds from the surfaces of circumstellar disks, most stars are born in localized regions of clusters that harbor less extreme irradiation conditions than the proplyd-hosting regions of the ONC.
The external UV field strength of a star-forming region is characterized in units of the Habing Field, $G_0$, where $G_0 = 1.6 \times 10^{-3}$ erg cm$^{-2}$ s$^{-1}$ and is defined as the local ISM FUV field strength over the wavelength range $930-2000$ \r{A} \citep{Habing68}. 
While disks and proplyds in the ONC are irradiated by intense external FUV fields with typical values $> 10^4 - 10^5$ $G_0$ \citep[e.g.,][]{Johnstone98, Storzer99, Odell17}, 
{most stars in the Galaxy, including ones that form in high-mass clusters,}
are likely irradiated by ``intermediate'' FUV  fields of  $\sim 10-10^4$ $G_0$  \citep[e.g.,][]{Fatuzzo08, Winter20b,  Parker21a, Winter22}, which are weaker than than 
the FUV fields found in the proplyd-hosting regions of the ONC, 
but stronger than those found in nearby low-mass star-forming regions like Taurus \citep[$\lesssim 10$ $G_0$; e.g.,][]{Mooley13}.

The expected prevalence of 
intermediately-irradiated {star-forming environments}  
makes NGC 1977 an ideal region to study Galactic star and planet formation. 
Located ${\sim}0.5^{\circ}$ north of the ONC, {NGC 1977} hosts several hundred low-mass stars with masses and ages similar to those found in the ONC \cite[][]{Peterson08, Megeath12, DaRio16}. Unlike other nearby star-forming clusters in Orion, the most massive stars in NGC 1977 are B-type stars rather than O-type stars. 
This means that as disk-bearing stars evolve in NGC 1977, they are never exposed to the extreme $10^4 - 10^5$ $G_0$ 
UV fields found in the intensely-irradiated regions of the ONC, and are instead irradiated by $\sim 10 - 10^4$ $G_0$ UV fields that mirror the UV environments of typical Galactic star-forming regions. The absence of $>10^4$ $G_0$ UV fields in NGC 1977 also marks an important distinction from disks currently located in the intermediately-irradiated outskirts of the ONC, {as dynamical evolution implies that these disks were likely closer to the massive Trapezium stars and, thus, exposed to stronger UV fields in the past than they are now \citep[e.g.,][]{Scally01, Winter19}.}

{Recently, \cite{Bally12} and \cite{Kim16}  discovered $\sim 10$ candidate proplyds in NGC 1977 with similar morphologies as the ONC proplyds. } This discovery confirms that intermediate UV fields are sufficient to trigger external photoevaporation in planet-forming disks, as predicted by external photoevaporation theory \citep[e.g.,][]{Adams04, Clarke07, Facchini16, Haworth18, Haworth18b, Haworth19}. However, the HST observations used to discover the NGC 1977 proplyds 
only mapped a small area of NGC 1977, 
so it remains unclear whether external photoevaporation is truly a widespread phenomenon in the intermediately-irradiatedly environment of NGC 1977.

\begin{figure*}[ht!]
\epsscale{1.1}
\hspace{-0.3in}

    \vspace{-0.5in}
    \plotone{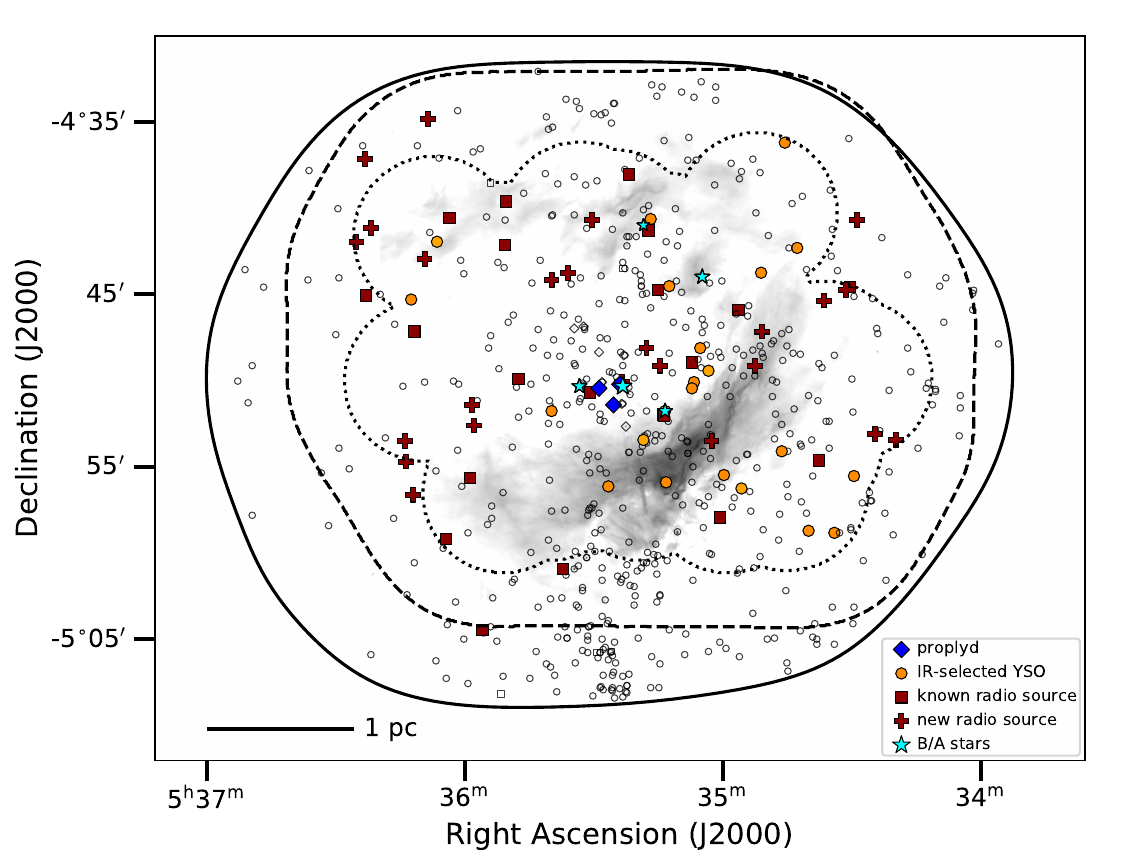}
\caption{{The NGC 1977 region, as seen at 8 $\mu$m with {\it Spitzer}-IRAC (grayscale). Solid, dashed, and dotted contours depict the field of views of our 3.0, 6.4, and 15.0 GHz VLA observations, respectively (see Section \ref{sec:data}). 
The positions of 42 Ori ($\sim$B1V star), HD 37058 ($\sim$B3V star), HD 294264 ($\sim$B3V star), HD 369658 ($\sim$B3V star), and HD 294262 ($\sim$A0 star) are marked with cyan stars, with larger star sizes corresponding to earlier spectral types. 
{Our VLA maps cover ${\sim} 600$ sources previously identified at optical, infrared, and/or radio wavelengths (see Section \ref{sec:Sample}), and {we detect} a total of 71 sources. We use diamond markers to indicate the positions of known proplyds in NGC 1977 \citep{Bally12, Kim16}, circle markers to indicate the positions of infrared-selected YSOs \citep{Peterson08, Megeath12, DaRio16}, and square markers to denote the positions of known radio sources \citep{Kounkel14}. Filled diamonds, circles, and squares correspond to sources that are detected in our maps, while open diamonds, circles, and squares correspond to nondetected sources. Finally, we use plus symbols to denote the positions of sources that are detected in our maps but not associated with the proplyd, YSO, and known radio source catalogs.  \label{fig:NGC1977_region}}   
}}
\end{figure*}

Here we present new {NSF's} Karl G. Jansky Very Large Array (henceforth denoted as VLA) observations that have mapped the entire NGC 1977 region at 3.0 GHz (13 cm), 6.4 GHz (4.6 cm), and 15.0 GHz (2.0 cm). At long centimeter wavelengths, dust emission from protoplanetary disks declines significantly due to its steep spectral index ($F_{\nu} \propto \nu^{2 + \beta}$, $\beta \approx 0-2$), and when there is substantial ionized gas, free-free emission dominates the continuum. Centimeter wavelength observations can therefore be used to identify which disk-bearing stars in a clustered star-forming region are launching ionized winds driven by external photoevaporaton. 
{With our deep, multi-band VLA observations, we can search for free-free-emitting proplyds in NGC 1977, and establish a sample of intermediately-irradiated young stellar objects (YSOs) that are influenced by external photoevaporation.}

\section{\bf Sample}\label{sec:Sample}

The NGC 1977 region comprises several-hundred pre-main-sequence stars ranging from brown dwarfs to B-type stars \citep[see review in][]{Peterson08}. Figure \ref{fig:NGC1977_region} shows the spatial distribution of YSOs in NGC 1977. The lower-mass cluster members are distributed throughout all of NGC 1977, while the higher-mass B- and A-type stars are concentrated towards the inner regions of the cluster. In particular, the B1V star 42 Ori is located at the cluster center along with the B3V stars HD 37058 and HD 294264; and, the B3V star HD 36958 and A0 double star HD 294262 lie to the northwest of 42 Ori.

\begin{figure}[ht!]
\epsscale{1.1}
\hspace{-0.15in}
    \plotone{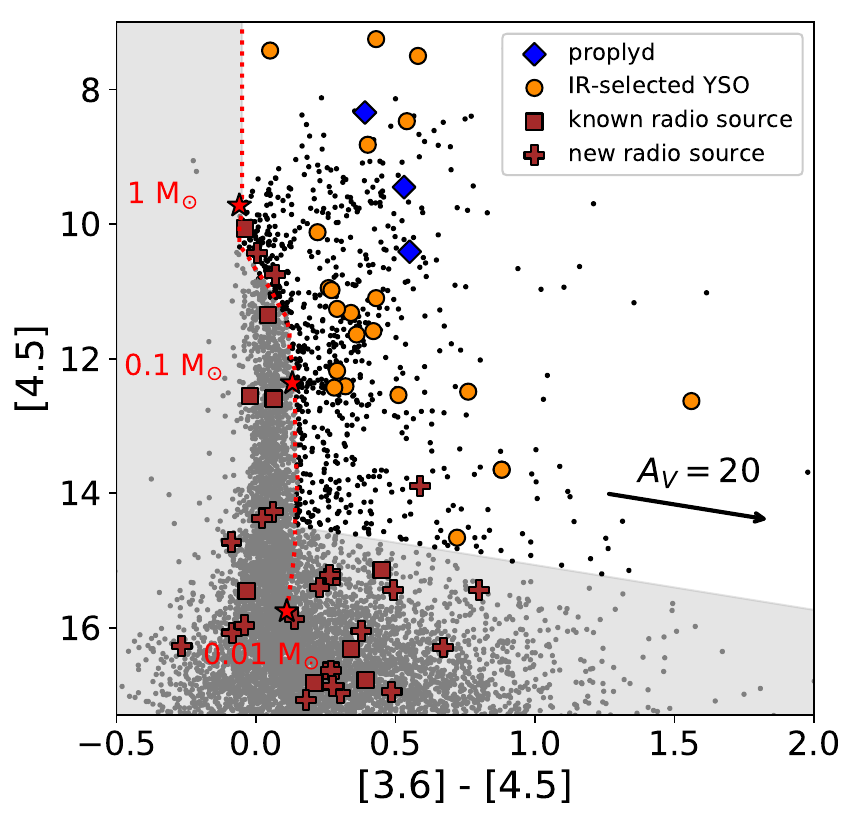}
\caption{Color-magnitude diagram for sources detected in our VLA maps and all {\it Spitzer} point sources within our VLA maps. 
Photometric data for VLA detections previously classified as YSOs are taken from \cite{Megeath12}. Photometric data for {unclassified radio detections} and additional {\it Spitzer} point sources are taken from the online {\it Spitzer} point source catalog. The dotted red line shows the 1 Myr isochrone for the pre-main-sequence evolutionary models from \cite{Baraffe15}, while the red star symbols indicate the individual model points for a 0.01 $M_{\odot}$ star, a 0.1 $M_{\odot}$ star, and a 1 $M_{\odot}$ star. 
The solid black arrow shows the extinction vector for $A_V = 20$. 
{The shaded gray region indicates the region of color-magnitude magnitude space where we classify VLA-detected sources as candidate background objects \citep[c.f.,][]{Gutermuth08, Harvey08, Gutermuth09, Megeath12}. If a {\it Spitzer} point source falls within the shaded grey region, we use a filled gray circle to plot its position. Otherwise, we plot its position using a filled black circle.} For the VLA-detected sources, we plot their positions using the same marker symbols as in  Figure \ref{fig:NGC1977_region}. \label{fig:CMD} 
}
\end{figure}

Here we assemble catalogs that can be used 
{to identify radio emission from known sources in NGC 1977.}
Our first catalog consists of all 10 NGC 1977 proplyds that were identified by \cite{Bally12} and \cite{Kim16} with HST and/or {\it Spitzer} imaging. The positions of these proplyds are indicated in Figure \ref{fig:NGC1977_region} with diamond markers. All 10 proplyds are located in the same inner region of NGC 1977, as the bulk of them were identified with HST observations that mapped a small area to the east of 42 Ori. One of the proplyds, however, lies outside of the region mapped with HST, but this source \citep[identified with {\it Spitzer} imaging, see][]{Kim16} is located directly north of 42 Ori, and is therefore positioned in the inner region of NGC 1977.

Our second catalog consists of all infrared-selected YSOs {within the central $\sim 30' \times 45'$ region of NGC 1977.}
This includes $85$ disk-bearing stars detected at 24 $\mu$m with {\it Spitzer}-MIPS \citep{Peterson08}, 291 YSOs detected at $H$-band with SDSS-APOGEE \citep{DaRio16}, and 386 {\it Spitzer}-selected YSOs from the photometric catalog of \cite{Megeath12}, for a total of 559 unique sources after accounting for overlap amongst the different infrared catalogs. In Figure \ref{fig:NGC1977_region}, we use circle markers to indicate the positions of all infrared-selected YSOs in our VLA maps.

To obtain updated source coordinates for the infrared-selected YSOs, we use Gaia DR3 \citep{GAIA2022}. 
Namely, we search the Gaia archive for objects within $1\rlap{.}''5$ of the published source coordinates, finding matches for $\sim 90\%$ of the considered sources. The updated coordinates from Gaia show a systematic offset of around $-1''$ to $-0\rlap{.}''5$ in right ascension and $-0\rlap{.}''25$ to $-0\rlap{.}''75$ in declination from the published {\it Spitzer} catalog coordinates \citep{Peterson08, Megeath12}. They also show a similar offset in right ascension but a smaller, $-0\rlap{.}''25$ to $0''$ offset in declination from the published {SDSS-APOGEE} catalog coordinates \citep{DaRio16}. We use these systematic offsets to extrapolate the coordinates of the $\sim 10 \%$ of sources without GAIA counterparts onto the GAIA reference frame. In particular, we apply an offset of $0\rlap{.}''75$ in right ascension and $0\rlap{.}''5$ in declination to the coordinates of {\it Spitzer}-selected objects, and an offset of $0\rlap{.}''5$ in right ascension to the coordinates of {SDSS-APOGEE} selected objects.

\begin{deluxetable*}{rlcclcccc}\tablenum{1}
\tablewidth{0pt}
\tablecaption{Summary of VLA observations \tablenotemark{ }}\label{tab:VLA_obs}
\tablehead{ 
    \colhead{Band} & 
    \colhead{Obs. Type}  &
    \colhead{Date}  &  
    \colhead{rms} &  
    \colhead{Peak rms}  & 
    \colhead{Beam size} & 
    \colhead{Beam P.A.}  & 
    \nocolhead{Total}  &
    \nocolhead{${>}6{\sigma}$}    \\[-0.2cm]
    \vspace{-0.15in} \\	
    \colhead{}        &  
    \colhead{} & 
    \colhead{}    & 
    \colhead{($\mu$Jy)}     & 
    \colhead{($\mu$Jy)}   &
    \colhead{}  &        
    \colhead{($^{\circ}$)}  & 
    \nocolhead{Detections\tablenotemark{a}}  & \nocolhead{Detections}                      
}
\startdata 
3.0 GHz & 
Mosaic & 
2021 September 29 &
27 &
${\sim}$150 &
$1\rlap{.}''5 \times 2\rlap{.}''3$ &
16 \\ 
6.4 GHz & 
Mosaic & 
2021 September 29 &
24 &
${\sim}$130 &
$1 \rlap{.}''0 \times 1\rlap{.}''5$ &
-27 \\
15.0 GHz & 
Individual Pointings & 
2021 September 29 &
30 &
${\sim}20 \times 10^3$  &
$0\rlap{.}''3 \times 0\rlap{.}''5$ &
-7 \\
\enddata
\end{deluxetable*}

The {\it Spitzer} point source catalog\footnote{
\dataset[https://irsa.ipac.caltech.edu/Missions/spitzer.html]{https://irsa.ipac.caltech.edu/Missions/spitzer.html}}
contains thousands of additional sources that fall within the central $\sim 30' \times 45'$ region of NGC 1977. 
However, most of these are background objects not associated with NGC 1977. 
Figure \ref{fig:CMD} shows a color-magnitude diagram for all {\it Spitzer} point sources in NGC 1977 with available photometry at 3.6 and 4.5 $\mu$m. The {\it Spitzer} point sources are color-coded by their location in color-magnitude space, with black points denoting sources that are consistent with being YSOs, and gray points denoting sources that are consistent with being background objects. We consider a {\it Spitzer} point source to be a contaminant background source if it has {\it Spitzer}-band colors that are bluer than the expected colors of a $\sim$1 Myr-old pre-main-sequence star (and thus, inconsistent with infrared excess), or if it has {\it Spitzer}-band magnitudes that are fainter than the expected magnitude of a $\sim$1 Myr-old YSO with a mass ${>}0.05$ M$_{\odot}$  \citep[consistent with previous YSO classification schemes; e.g.,][]{Gutermuth08, Harvey08, Gutermuth09, Megeath12}. 
To determine the expected colors and magnitudes for ${>}0.05$ M$_{\odot}$ YSOs, we use the 1 Myr isochrone from the pre-main-sequence evolutionary models of \cite{Baraffe15},  shown in Figure \ref{fig:CMD}, and we assume a distance of 400 pc. 

While our selection criteria for YSOs versus background objects may reject potential sub-stellar objects in NGC 1977, they are sufficient to identify YSOs with disks, which are the main objects of interest in this study. 
Under these criteria, we classify the majority of {\it Spitzer} point sources as background objects (see Figure \ref{fig:CMD}). Of the remaining sources that are not classified as background objects, most are objects that have been previously classified as YSOs by \cite{Megeath12}. Some, however, are consistent with YSOs in two or three photometric bands, but lack photometric coverage in additional {\it Spitzer} bands due to contaminating nebulosity and/or local extinction effects. These candidate YSOs are not associated with the \cite{Megeath12} catalog, which only includes YSOs with photometric coverage in four or more bands.

Finally, we consider the catalog of 40 known radio sources that fall within the NGC 1977 region and were detected previously at 4.5 and/or 7.5 GHz by \cite{Kounkel14}. Most of these radio sources have {\it Spitzer} point source counterparts whose photometry are consistent with background objects under our selection criteria. A small subset, however, has two-, three-, or four-band photometry that is consistent with a YSO, while another subset has counterparts in one of more of the previously-assembled YSO catalogs for NGC 1977. In Figures \ref{fig:NGC1977_region} and \ref{fig:CMD}, we use square markers to plot sources associated with the \cite{Kounkel14} radio source catalog; however, if a source is associated with both the \cite{Kounkel14} catalog and our YSO catalog, we plot its position with the same circle markers used to denote YSOs.

\section{\bf{Observations and Data Reduction}} \label{sec:data}

We imaged the NGC 1977 cluster at 3.0 GHz (13 cm), 6.4 GHz (4.6 cm), and 15.0 GHz (2.0 cm) with the 
VLA. Observations were taken on 2021 September 29 under project code 21A-015. At this date, the VLA was in its B-configuration, {which provided baselines ranging from 0.21 km to 11.1 km}.

Table \ref{tab:VLA_obs} provides an overview of our multi-wavelength VLA dataset, and in Figure \ref{fig:NGC1977_region}, we show the field of views of our observations at each wavelength. 
{The 15.0 GHz observations consisted of 46 individual pointings that were centered on the positions of the known proplyds and the 24 $\mu$m-selected YSOs in NGC 1977 (see Section \ref{sec:Sample}), which was sufficient to cover the majority of additional NGC 1977 sources in our search catalogs (see Figure \ref{fig:NGC1977_region}).}
At 3.0 and 6.4 GHz, the primary beams are wide enough such that large portions of NGC 1977 can be mosaicked efficiently. We {utilized} 10 pointings at 3.0 GHz and 66 pointings at 6.4 GHz to mosaic the central $\sim 30' \times 45'$ region of NGC 1977 over these frequencies, and we {employed} a hexagonal mosaic pattern with a pointing spacing of FWHM$/\sqrt{3}$ in order to achieve Nyquist sampling across these maps.

The spectral setups at each observing wavelength were designed to maximize sensitivity to continuum emission. Observations at 3.0 GHz were taken using the VLA's S-band receivers, with 8 subbands arranged continuously from 1.988 - 4.012 GHz for a total bandwidth of 2 GHz. The 6.4 GHz data were taken using the VLA's C-band receivers. We centered two 2 GHz basebands at 5.25 GHz and 7.5 GHz in order to avoid strong radio-frequency interference (RFI) in the range 4.0 - 4.2 GHz from satellites in the Clarke Belt, and the 16 128 MHz subbands in each baseband were arranged from 4.226 - 6.724 GHz and 6.7467-8.534 GHz to provide a bandwidth of 4 GHz. For the 15.0 GHz observations, we utilized the VLA's Ku-band receivers. The 64 128 MHz subbands available at Ku-band were arranged from 11.756 - 18.412 GHz. Most of the data between 12-12.8 GHz and 17.2-17.6 GHz were, however, {were} affected by significant RFI, so the 15.0 GHz observations achieved an effective bandwidth of around 5 GHz.

\begin{deluxetable}{lcc}
\tabletypesize{\small}
\tablewidth{200pt}
\tablenum{2}
\tabcolsep=0.05cm
\tablecaption{Catalog Associations of Detected Sources\tablenotemark{ }\label{tab:source_detections}}
\tablehead{ 
    \colhead{Catalog} &
    \colhead{No.} & 
    \colhead{No.}  \\[-0.7cm] \\ 
    \colhead{} &
    \colhead{Sources\tablenotemark{a}} & 
    \colhead{Detections\tablenotemark{b}}  
   }
\startdata 
Proplyd catalogs              & 10 & 3 \\
{Full YSO catalog}\tablenotemark{c}                   & 559 & 24 \\
\hspace{0.1in} {Spitzer} 24$\mu$m YSO catalog (P08)   & 85 & 5 \\
\hspace{0.1in} {SDSS-APOGEE} YSO catalog (D16)        & 291 & 17 \\
\hspace{0.1in}  Full {Spitzer} YSO catalog  (M12)       & 386 & 12 \\
Known radio sources (K14)          & 40 & 27 \\
New radio sources\tablenotemark{d}       & \nodata & 24 \\ All VLA-detected sources          &   & 71 \\
\hspace{0.1in} Confirmed/Candidate NGC 1977 sources    &  & {34} \\
\hspace{0.1in} Background sources        &  & {37} 
\enddata
\tablenotetext{ }{{\bf Notes:} 
References for proplyd catalogs: \cite{Bally12, Kim16}. 
Reference for {Spitzer} 24 $\mu$m catalog: \cite{Peterson08}, i.e., ``P08''. 
Reference for {Spitzer} SDSS-APOGEE catalog: \cite{DaRio16}, i.e., ``D16''. 
Reference for {Spitzer} YSO catalog: \cite{Megeath12}, i.e., ``M12''. 
Reference for known radio source catalog: \cite{Kounkel14}, i.e., ``K14''. }
\tablenotetext{a}{Indicates all sources covered within our VLA maps.}
\tablenotetext{b}{Indicates all detections in our VLA maps.}
\tablenotetext{c}{Indicates all unique YSO sources in our maps after accounting for overlap amongst the P08, M12, and D16 catalogs.}
\tablenotetext{d}{Indicates all radio detections not associated with a {proplyd, YSO, or known radio source} catalog.}
\end{deluxetable}

All observations were reduced using the VLA continuum data reduction pipeline, which {included} procedures for automatic and manual flagging of RFI as well as standard flux, bandpass, and gain calibrations. The bright quasar J0319+4130 was used to derive bandpass solutions for each antenna. To calculate antenna-based complex gains, we used periodic observations of J0503+0203 for observations at 3.0 GHz, and J0541-0541 for observations at 6.4 and 16.0 GHz. Moreover, 3C147 was used as the flux density calibrator for observations at 3.0 and 6.4 GHz, and 3C48 was used as the flux density calibrator for observations at 15.0 GHz.

We imaged the radio-continuum observations using the CASA {\tt tclean} task in MFS (Multi-Frequency Synthesis) mode.
The 3.0 and 6.4 GHz mosaics were imaged  
with a phase center of 05:35:23.16 -4:50:18.09, i.e., the coordinates of 42 Ori. For the 15.0 GHz observations, we imaged the individual fields separately out to a primary beam value of $0.001$, and then generated smaller sub-images towards the positions of individual sources using the closest field pointing. 
{Clean boxes were determined using an iterative process where we first placed clean boxes around all objects detected above 10$\sigma$, generated a cleaned image, searched the residuals for detections above 5$\sigma$, and then generated a new cleaned image with additional clean boxes placed around all additional detections.}
{For our final cleaned images, we used }
the “mtmfs” \citep[Multi-Term Multi-Frequency Synthesis;][]{Rau11} clean algorithm with nterms = 2, a Briggs weighting method with a robust parameter of 0.5, and a {\it uv} cut of $>$$25$ k$\lambda$. The {\it uv} cut was employed to spatially filter extended emission in NGC 1977 and improve the noise levels in the vicinity of compact radio sources. Our chosen {\it uv} cut of $>$$25$ k$\lambda$ eliminated spatial scales ${>}8{''}$. 

{
Our main 3.0, 6.4, and 15.0 GHz image products were generated in the Stokes I plane using a mosaic gridder for the 3.0 and 6.4 GHz observations, and a standard gridder for the 15.0 GHz observations. To search for signatures of circular polarization towards VLA-detected sources (see Section \ref{sec:pol}), we also generated an additional set of 3.0, 6.4, and 15.0 GHz  images in the Stokes V plane following the same cleaning procedure outlined above. We used the awproject gridder to generate the Stokes V images, as this gridder corrects for beam squinting and allows for a more reliable examination of Stokes V signals that are spatially offset from a pointing center (see EVLA memo 113). 
Finally, to improve the spectral characterization of a few VLA-detected sources, we also imaged 5.25 and 7.5 GHz C-band basebands separately (see Section \ref{sec:SED_modeling}). }

The synthesized beam sizes of our final cleaned images are $1\rlap{.}''5 \times 2\rlap{.}''3$ at 3.0 GHz, $1\rlap{.}''0 \times 1\rlap{.}''5$ at 6.4 GHz, and $0\rlap{.}''3 \times 0\rlap{.}''5$ at 15.0 GHz (see Table \ref{tab:VLA_obs}). 
At the {400 pc distance to Orion}, these beam sizes correspond to spatial resolutions of about $\sim$120 AU, $\sim$400 AU, and $\sim$600 AU, respectively.

\section{\bf{Results}}\label{sec:fitting:results}

\subsection{Source Detections}\label{sec:fitting:detections}

To identify compact radio sources in NGC 1977, we first perform a blind detection search over the full {Stokes I} VLA images. Our maps contain a large number of synthesized beams ($>10^7$), and so we must employ a conservative noise {threshold} to ensure that none of the blindly-detected sources are noise spikes.  However, if a source is blindly detected at multiple wavelengths, then the probability that the detection is a noise spike would decrease. We therefore use an $>$$8$$\sigma$ detection limit for sources that are blindly detected at a single wavelength, which ensures that $<<$1 of the detections are Gaussian noise spikes. For blind detections over multiple wavelengths, we use a lower detection threshold of $>$$6$$\sigma$, at which level $<<$1 Gaussian noise spikes are expected at the same position in multiple images.

We also perform a detection search towards the predetermined positions of all sources from the proplyd, YSO, and known radio source catalogs that we compiled in Section \ref{sec:Sample}. We limit our catalog search radius to $1''$, reflective of the typical positional uncertainties of sources without GAIA counterparts, as well as the expected sizes of ionized proplyd structures in NGC 1977 \citep[e.g.,][]{Kim16}. 
Due to the smaller number of synthesized beams being probed in our catalog search versus our blind detection search, we are also able to employ a lower detection threshold in our catalog search. 
{We adopt a ${>}3.5{\sigma}$ detection threshold for the YSO and known radio source catalog searches at 3.0 and 6.4 GHz, as we expect $<<$1 Gaussian noise spikes above this level. For the proplyd catalog search, we employ a lower detection threshold of ${>}2.5{\sigma}$ at 3.0 and 6.4 GHz, as $<<$1 Gaussian noises spikes are still expected at this threshold for a catalog search of $10$ objects. }
For the 15.0 GHz catalog searches, we use a detection threshold of ${>}3.75{\sigma}$ due to the larger number of beams probed at 15.0 GHz versus 3.0 GHz or 6.4 GHz. However, we relax the 15.0 GHz detection threshold down to ${>}3.5{\sigma}$ if the source is also detected in 3.0 GHz or 6.4 GHz, following the discussion above.

The rms noise is calculated from a $120 \times 120$ {pixel} box around each pixel in the residual map. The 3.0 and 6.4 GHz images typically have local rms levels of ${\sim}25-35$ $\mu$Jy, while regions closer to the image outskirts and/or the bright radio source J053558.88-045537.7 have rms levels of around ${\sim}100$ $\mu$Jy (see Table \ref{tab:VLA_obs}). 
For the 15.0 GHz images, the rms levels vary from ${\sim}30$ $\mu$Jy to ${\sim}30$ mJy depending of the position of a source relative to a 15.0 GHz pointing center. 
{All of the sources in our proplyd catalog have 15.0 GHz rms levels of ${\sim}30-40$ $\mu$Jy, 
and towards the full sample of cataloged sources 
that fall within the field of view of our 15.0 GHz maps (see Figure \ref{fig:NGC1977_region}),  ${\sim}50\%$ have 15.0 GHz rms levels ${<}50$ $\mu$Jy, ${\sim}66\%$ have 15.0 GHz rms levels ${<}70$ $\mu$Jy, and ${\sim}80\%$ have 15.0 GHz rms levels ${<}1$ mJy.}

{Figure \ref{fig:NGC1977_region} shows the positions of all detected sources in our maps. }
We detect a total of 71 distinct objects through our blind and catalog searches. Of these 71 objects, 56 are detected at 3.0 GHz, 65 are detected at 6.4 GHz, and 26 are detected at 15.0 GHz.

{Table \ref{tab:source_detections} summarizes the catalog associations of the detected sources. {25 of the 71 detections are associated with an infrared-selected YSO, and 3 of these radio-detected YSOs are also associated with known proplyds in NGC 1977.}
The remaining detections in our sample are either known radio sources in our VLA maps (27 of the 71 detections), or sources that are identified from our blind detection search but not associated with the proplyd, YSO, or known radio source catalogs (24 of the 71 detections).}  {Of the 27 known radio sources detected in our maps, 5 are associated with an infrared-selected YSO.  }

\begin{figure}[ht!]
\epsscale{1.1}
\hspace{-0.15in}
    \plotone{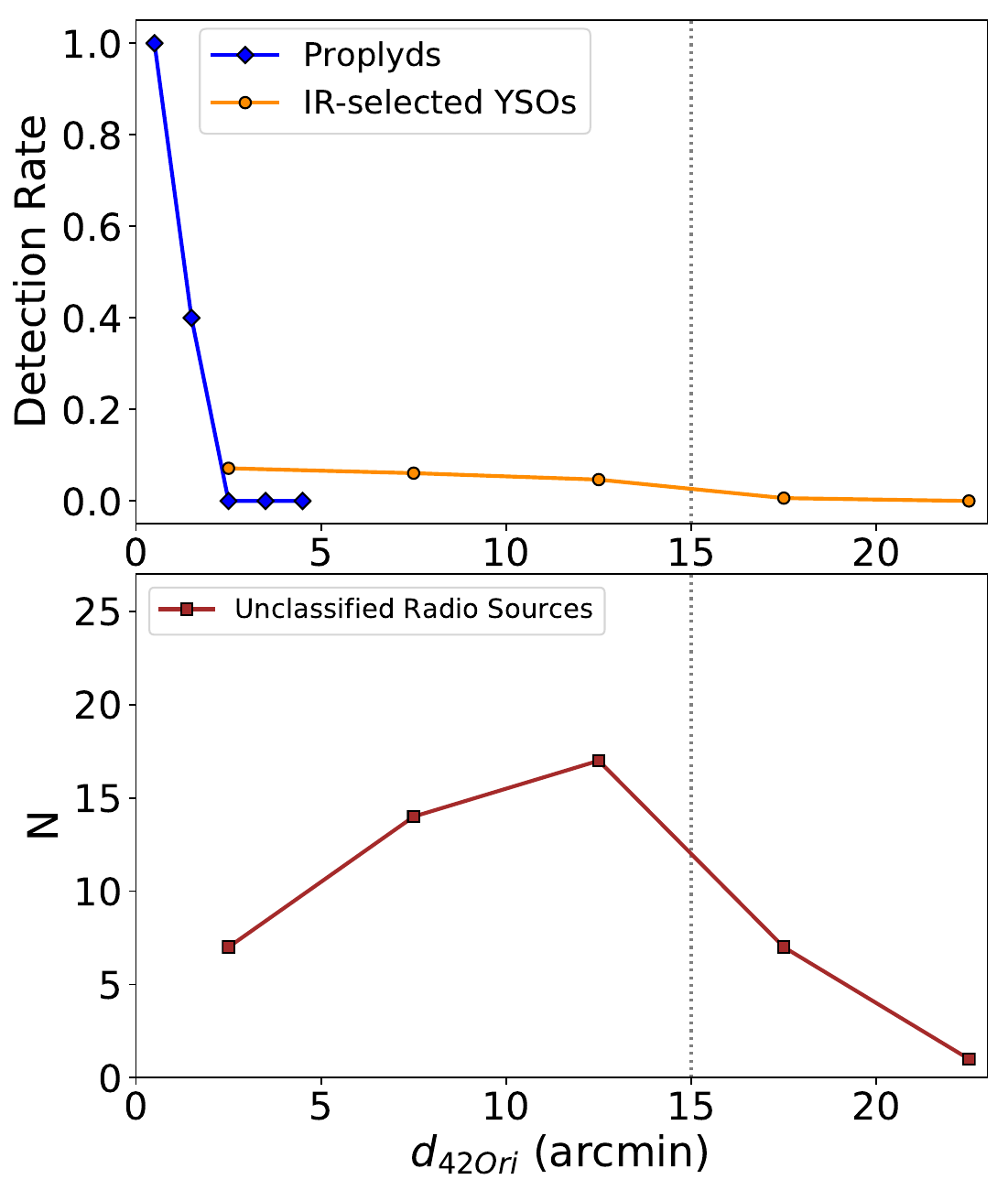}
\caption{Top: Radially-averaged detection fractions observed towards the HST- and Spitzer-identified proplyd catalog (blue) and the infrared-selected YSO catalog (orange, see Section \ref{sec:Sample}), plotted as a function of projected separation from 42 Ori. We use $1'$ bins to compute the detection rates for the proplyd catalog, and $5'$ bins to compute the detection rates for the YSO catalog. The gray dotted line depicts the approximate primary beam full-width-at-half maximum of our 6.4 GHz VLA mosaic. 
Bottom: Radially-averaged spatial distribution of radio sources  that are detected in our VLA maps but not associated with a known proplyd or YSO catalog 
(i.e., ``Unclassified Radio Sources''). 
\label{fig:detection_profile}  }
\end{figure}

In Figure \ref{fig:detection_profile}, we show radial profiles of the spatial distributions of sources detected in our VLA maps. 
{The detection rate among infrared-selected YSOs is low (${\sim}5-10\%$), but we see a slight increase in the overall detection rate towards the cluster center. The detection rate among known proplyds is also low (${\sim}30\%$), but here the dependence on projected separation is even steeper than what is seen towards the infrared-selected YSOs.} For the 46 radio-detected sources that are not associated with a proplyd or YSO catalog, we initially see an increase in the number of detections at larger projected separations. However, towards the outskirts of our maps, the rms noise is higher, and so we see a decline in the number of {these detections} at projected separations ${\gtrsim}15’$.

\begin{deluxetable*}{lllcrrrrr}\tablenum{3}
\tabletypesize{\footnotesize}
\tablewidth{0pt}
\tablecaption{Properties of VLA-detected sources in NGC 1977 \tablenotemark{ }}\label{tab:VLA_source_properties}
\tablehead{ 
    \colhead{ID} & 
    \colhead{R.A.}  &
    \colhead{Decl.}  & 
    \colhead{Catalog}  & 
    \colhead{$F_{\nu, \ \textrm{\footnotesize 3.0 GHz}}$} &  
    \colhead{$F_{\nu, \ \textrm{\footnotesize 6.4 GHz}}$}  & 
    \colhead{$F_{\nu, \ \textrm{\footnotesize 15.0 GHz}}$} & 
    \colhead{Radio Classification} & 
    \colhead{Notes} \\ 
    \colhead{}        &  
    \colhead{(J2000)} & 
    \colhead{(J2000)}    & 
    \colhead{}    & 
    \colhead{ (mJy)}   &    
    \colhead{(mJy)}  & 
    \colhead{(mJy)} & \colhead{} & \colhead{} }
\colnumbers
\startdata 
1  &  05:35:24.10 &  --4:50:09.60 &  KCFF1, P08, D16, M12 &   $0.023 \pm 0.102$ &  $-0.001 \pm 0.086$ &     $0.109 \pm 0.038$ &        Known Proplyd &               \\
2  &  05:35:25.52 &  --4:51:20.65 &  KCFF2, P08, D16, M12 &   $0.108 \pm 0.030$ &   $0.121 \pm 0.025$ &    $-0.029 \pm 0.129$ &        Known Proplyd &          E, P \\
3  &  05:35:28.82 &  --4:50:22.60 &            KCFF3, P08 &  $-0.001 \pm 0.104$ &   $0.061 \pm 0.024$ &     $0.043 \pm 0.131$ &        Known Proplyd &               \\
4  &  05:36:06.42 &  --4:41:53.83 &         P08, D16, M12 &   $0.050 \pm 0.114$ &   $0.106 \pm 0.023$ &     $0.062 \pm 0.124$ &     Possible Proplyd &             E \\
5  &  05:35:03.26 &  --4:49:21.00 &         P08, D16, M12 &   $0.021 \pm 0.095$ &   $0.073 \pm 0.022$ &     $0.128 \pm 0.037$ &        Candidate Jet &            PS \\
6  &  05:35:18.38 &  --4:53:23.55 &         P08, D16, M12 &   $0.004 \pm 0.111$ &   $0.051 \pm 0.025$ &    $-0.035 \pm 0.124$ &     Possible Proplyd &             O \\
7  &  05:34:55.67 &  --4:56:12.22 &    P08, D16, M12, K14 &   $0.007 \pm 0.125$ &   $0.028 \pm 0.088$ &     $0.243 \pm 0.043$ &         Possible Jet &               \\
8  &  05:34:45.55 &  --4:36:07.69 &                   D16 &   $0.241 \pm 0.072$ &   $0.471 \pm 0.049$ &   $68.052 \pm 85.217$ &        Candidate Jet &            PS \\
9  &  05:35:16.77 &  --4:40:32.53 &                   D16 &   $0.497 \pm 0.049$ &   $0.766 \pm 0.058$ &     $0.710 \pm 0.147$ &    Candidate Proplyd &        FS, NP \\
10 &  05:34:42.68 &  --4:42:14.69 &              D16, K14 &   $0.108 \pm 0.039$ &   $0.143 \pm 0.024$ &    $-0.670 \pm 3.296$ &     Possible Proplyd &          E, V \\
11 &  05:34:51.06 &  --4:43:41.46 &                   D16 &   $0.091 \pm 0.032$ &   $0.133 \pm 0.024$ &    $-0.176 \pm 1.677$ &     Possible Proplyd &               \\
12 &  05:35:12.46 &  --4:44:25.93 &              D16, K14 &   $0.232 \pm 0.033$ &   $0.372 \pm 0.028$ &     $0.513 \pm 0.113$ &    Candidate Jet &            PS \\  
13 &  05:35:05.25 &  --4:48:02.89 &                   D16 &   $0.097 \pm 0.028$ &   $0.218 \pm 0.025$ &     $0.244 \pm 0.043$ &    Candidate Proplyd &            FS \\ 
14 &  05:35:07.24 &  --4:50:25.51 &                   D16 &   $0.125 \pm 0.030$ &   $0.120 \pm 0.025$ &     $0.214 \pm 0.041$ &        Candidate Jet &            PS \\
15 &  05:35:39.74 &  --4:51:41.68 &                   D16 &   $0.533 \pm 0.050$ &   $1.912 \pm 0.138$ &     $1.846 \pm 0.142$ &    Candidate Proplyd &     E, FS, NP \\
16 &  05:34:46.41 &  --4:54:02.02 &              D16, K14 &   $0.216 \pm 0.035$ &   $0.553 \pm 0.046$ &     $0.405 \pm 0.056$ &    Candidate Proplyd &        FS, NP \\
17 &  05:35:13.10 &  --4:55:52.47 &              D16, M12 &   $0.042 \pm 0.125$ &   $0.138 \pm 0.027$ &     $0.166 \pm 0.277$ &     Possible Proplyd &             O \\
18 &  05:35:26.59 &  --4:56:06.78 &                   D16 &   $0.047 \pm 0.128$ &   $0.167 \pm 0.026$ &     $0.183 \pm 0.223$ &     Possible Proplyd &               \\
19 &  05:34:40.14 &  --4:58:39.91 &                   D16 &   $0.079 \pm 0.166$ &   $0.284 \pm 0.034$ &     $2.521 \pm 0.466$ &            Protostar &         E, SS \\
20 &  05:36:12.40 &  --4:45:15.88 &                   M12 &   $0.386 \pm 0.042$ &   $0.299 \pm 0.028$ &  $-13.030 \pm 20.212$ &    Candidate Proplyd &            FS \\ 
21 &  05:35:06.79 &  --4:50:01.98 &                   M12 &   $0.279 \pm 0.035$ &   $0.276 \pm 0.030$ &     $0.087 \pm 0.161$ &    Candidate Proplyd &         E, FS \\
22 &  05:35:13.34 &  --4:51:44.94 &              M12, K14 &   $0.038 \pm 0.117$ &   $0.118 \pm 0.030$ &     $0.261 \pm 0.346$ &           Radio Star &          P, V \\
23 &  05:34:59.89 &  --4:55:27.32 &              M12, K14 &   $0.220 \pm 0.037$ &   $0.064 \pm 0.083$ &     $0.553 \pm 0.060$ &        Candidate Jet &         E, PS \\
24 &  05:34:29.56 &  --4:55:30.20 &                   M12 &   $0.111 \pm 0.046$ &   $0.018 \pm 0.099$ &     $3.840 \pm 8.691$ &  Possible Radio Star &               \\
25 &  05:34:34.08 &  --4:58:49.89 &                   M12 &   $0.213 \pm 0.056$ &   $0.144 \pm 0.030$ &   $-1.732 \pm 11.525$ &     Possible Proplyd &               \\  
26 &  05:35:17.21 &  --4:41:13.50 &         K14 $\dagger$ &   $0.010 \pm 0.114$ &   $0.087 \pm 0.022$ &     $0.181 \pm 0.223$ &           Radio Star &             V \\
27 &  05:35:15.12 &  --4:44:42.90 &         K14 $\dagger$ &   $0.217 \pm 0.033$ &   $0.317 \pm 0.031$ &     $0.274 \pm 0.051$ &    Candidate Proplyd &      E, FS, V \\
28 &  05:34:56.33 &  --4:45:48.80 &         K14 $\dagger$ &   $0.162 \pm 0.031$ &   $0.135 \pm 0.024$ &     $0.112 \pm 0.124$ &         Possible Jet &   E, FS, P, V \\
29 &  05:34:37.64 &  --4:54:36.00 &         K14 $\dagger$ &   $0.622 \pm 0.057$ &   $0.716 \pm 0.056$ &     $0.878 \pm 0.084$ &        Candidate Jet &   E, PS, P, V \\
30 &  05:36:21.86 &  --4:41:06.34 &     \nodata $\dagger$ &   $0.259 \pm 0.054$ &   $0.280 \pm 0.043$ &   $-0.016 \pm 12.587$ &     Possible Proplyd &               \\
31 &  05:35:39.82 &  --4:44:04.48 &     \nodata $\dagger$ &   $0.035 \pm 0.102$ &   $0.309 \pm 0.031$ &     $0.265 \pm 0.052$ &    Candidate Proplyd &            FS \\
32 &  05:34:52.55 &  --4:49:06.45 &     \nodata $\dagger$ &   $0.273 \pm 0.034$ &   $0.265 \pm 0.029$ &     $0.195 \pm 0.044$ &    Candidate Proplyd &            FS \\
33 &  05:35:23.55 &  --4:50:01.25 &     \nodata $\dagger$ &   $0.223 \pm 0.034$ &   $0.160 \pm 0.027$ &     $0.137 \pm 0.151$ &    Candidate Proplyd &            FS \\
34 &  05:35:02.76 &  --4:53:27.73 &     \nodata $\dagger$ &   $0.209 \pm 0.033$ &   $0.273 \pm 0.030$ &    $-0.065 \pm 1.099$ &        Candidate Jet &         E, PS \\
\enddata
\tablenotetext{ }{{\bf Notes.} 
Column (1): source IDs used in this article. Columns (2) and (3): phase center coordinates. 
Column (3): catalog association of VLA-detected sources. Sources with HST- and/or Spitzer-identified proplyds are indicated using the full IDs assigned by \cite{Kim16}, which consists of a ``KCFF'' followed by an additional digit. Sources detected at 24 $\mu$m with {\it Spitzer}-MIPS \citep{Peterson08} are labeled with a ``P08’’. Sources detected at {\it H}-band with SDSS-APOGEE \citep{DaRio16} are labeled with a ``D16.'' An ``M12’’ is used to specify association with the full catalog of {\it Spitzer}-selected YSOs from \cite{Megeath12}. A ``K14’’ is used to specify associations with the radio catalog of \cite{Kounkel14}. Sources not associated with any of the above catalogs are indicated with an ellipses.
Columns (5), (6), and (7): measured fluxes at 3.0 GHz, 6.4 GHz, and 15.0 GHz.
{Column (8): Source classification, based on the observed properties of the detected radio emission (see Section \ref{sec:classification}).}
{Column (9): Notes on the morphologies, radio spectral energy distributions (SEDs), polarization, and/or variability characteristics of individual sources. 
`E' and `O' correspond to notes related to source morphology, with `E' denoting sources with spatially resolved and elongated radio emission in one or more bands, and `O' denoting sources with radio emission that is spatially offset from the central star (see Section \ref{sec:resolved_new}).
`FS', `PS', and `SS' correspond to notes related to the radio spectral indices, with `FS' denoting flat spectrum sources with SEDs consistent with optically thin free-free emission, `PS' denoting positive spectrum sources with SEDs consistent with optically thick free-free emission, and `SS' denoting steep spectrum sources with SEDs consistent with dust emission (see Section \ref{sec:SED_modeling}).
`P' and `NP' correspond to notes related to circular polarization, with `P' denoting sources with detected Stokes V emission above $2.5{\sigma}$ and measured circular polarization fractions ${>}10\%$, and `NP' denoting sources that are not detected in the Stokes V plane but have Stokes V upper limits that imply a circular polarization fraction of ${<}10\%$ (see Section \ref{sec:pol}).
Finally, a `V' is used to denote sources with variable ${6.4}$ GHz emission, based on comparisons with prior flux measurements from \cite{Kounkel14} (see Section \ref{sec:variable}). 
}}
\tablenotetext{\dagger }{Indicates a candidate cluster member (see Section \ref{sec:fitting:detections}).}
\end{deluxetable*}


\begin{figure*}[htp!]
\epsscale{1.0}
\hspace{-0.3in}
    \plotone{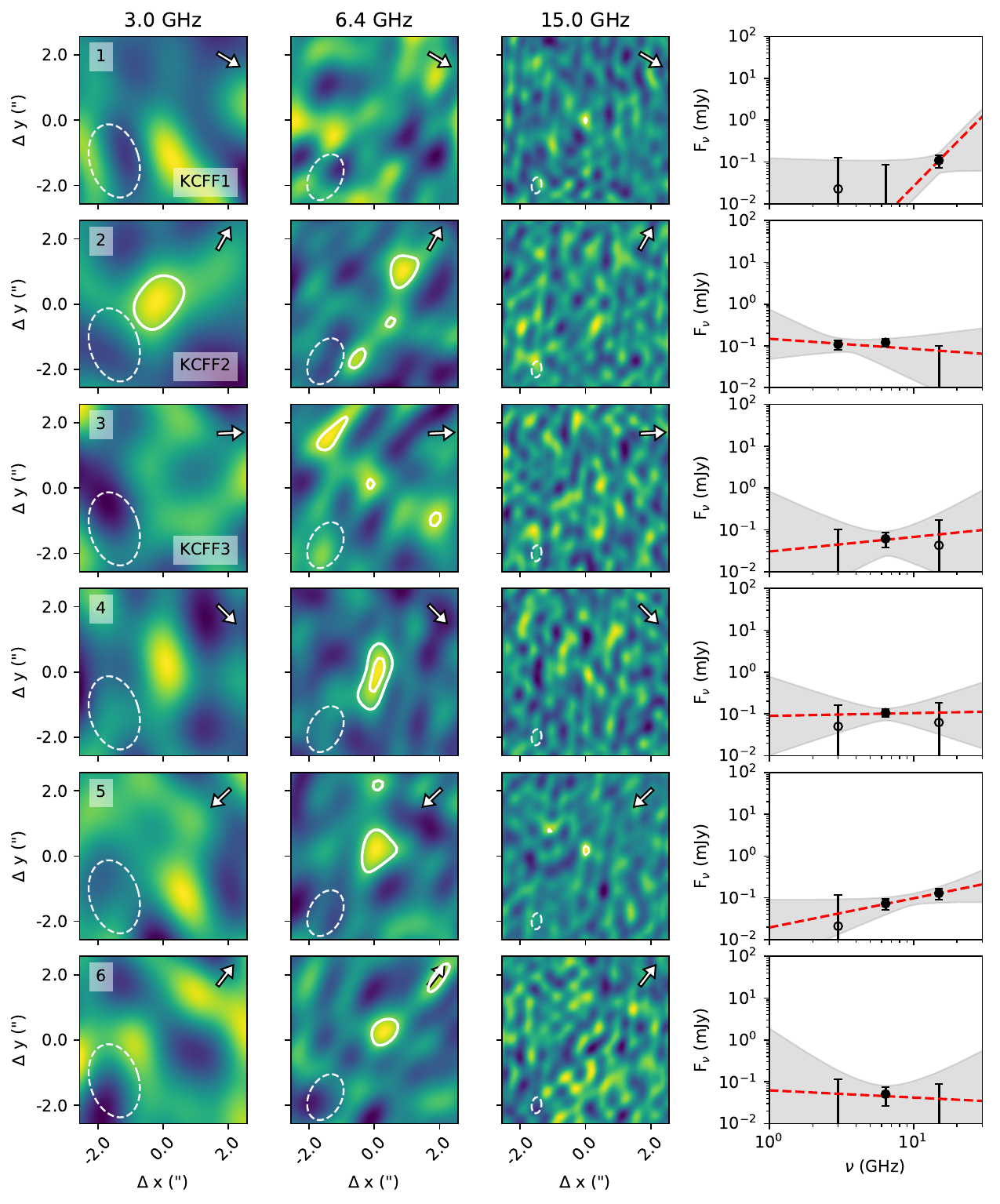}
\caption{\footnotesize 3.0 GHz (S-band) continuum images (left column), 6.4 GHz (C-band) continuum images (center left column), 15.0 GHz (Ku-band) continuum images (center right column), and radio spectral energy distributions (SEDs, right column) of {NGC 1977 sources} 
detected in our VLA maps. 
Each row shows the continuum images and SED of a single source, with our source ID (see Table \ref{tab:VLA_source_properties}) provided in the top-left corner of each 3.0 GHz image panel. For detections with optically identified proplyds, we include the proplyd name from \cite{Kim16} in the bottom right corner of the 3.0 GHz image panel. 
{The white arrow in each image panel points to the direction of the nearest (in projected separation) B- or A-type star in NGC 1977.}
{The {solid} white contours in the  3.0 and 6.4 GHz image panels show 2.5$\sigma$ to 10.5$\sigma$ emission in increments of 2$\sigma$. In the 15.0 GHz panels, the solid white contours show 3.5$\sigma$ to 11.5$\sigma$ emission in increments of 2$\sigma$.}
{The dashed white contour in each panel shows the synthesized beam at each wavelength}. 
In the SED panels, filled black circles show the fluxes measured with Gaussian fitting when a source is detected above our noise threshold, while open black circles indicate fluxes measured in an aperture around the source position when the source not detected above our noise threshold. 
If a detection has prior radio-continuum flux measurements from \cite{Kounkel14}, we show them in the SED panel as blue squares.
{The red dashed lines show the best-fit single power law or piecewise free-free emission power law that is derived for each source (see Section \ref{sec:SED_modeling}). The shaded gray regions show the 1$\sigma$ confidence intervals of each fit. For sources that are modeled with a piecewise power law, we include red circles in the SED panels that show the the 5.25 and 7.5 GHz fluxes that are obtained when imaging the two 6.4 GHz basebands separately (see Section \ref{sec:data}).}
\label{fig:detections_1}}
\end{figure*}


\begin{figure*}[htp!]
\epsscale{1.0}
\hspace{-0.3in}
    \plotone{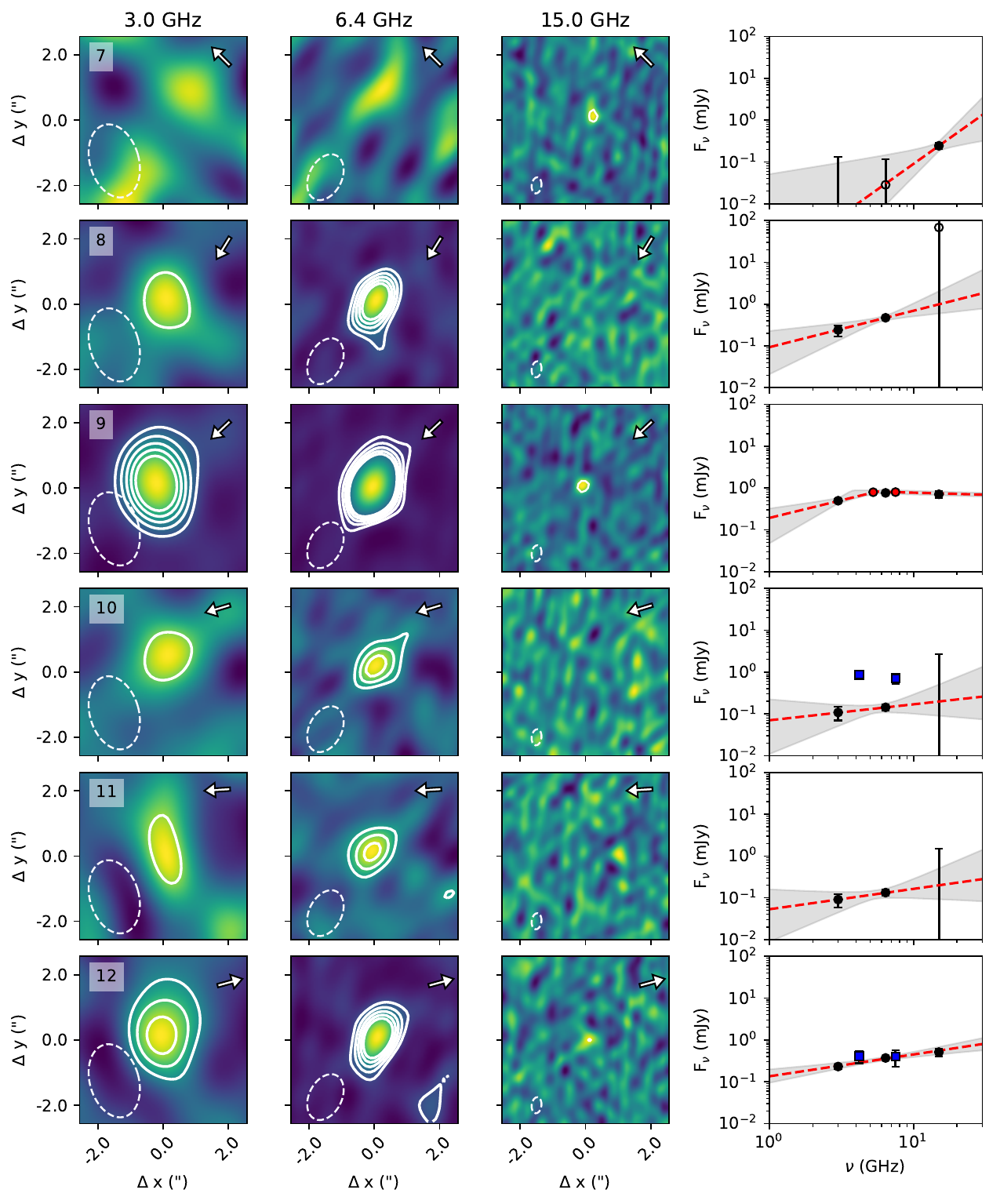}
\caption{\footnotesize {Continuation of Figure \ref{fig:detections_1}.}
\label{fig:detections_2}}
\end{figure*}


\begin{figure*}[htp!]
\epsscale{1.0}
\hspace{-0.3in}
    \plotone{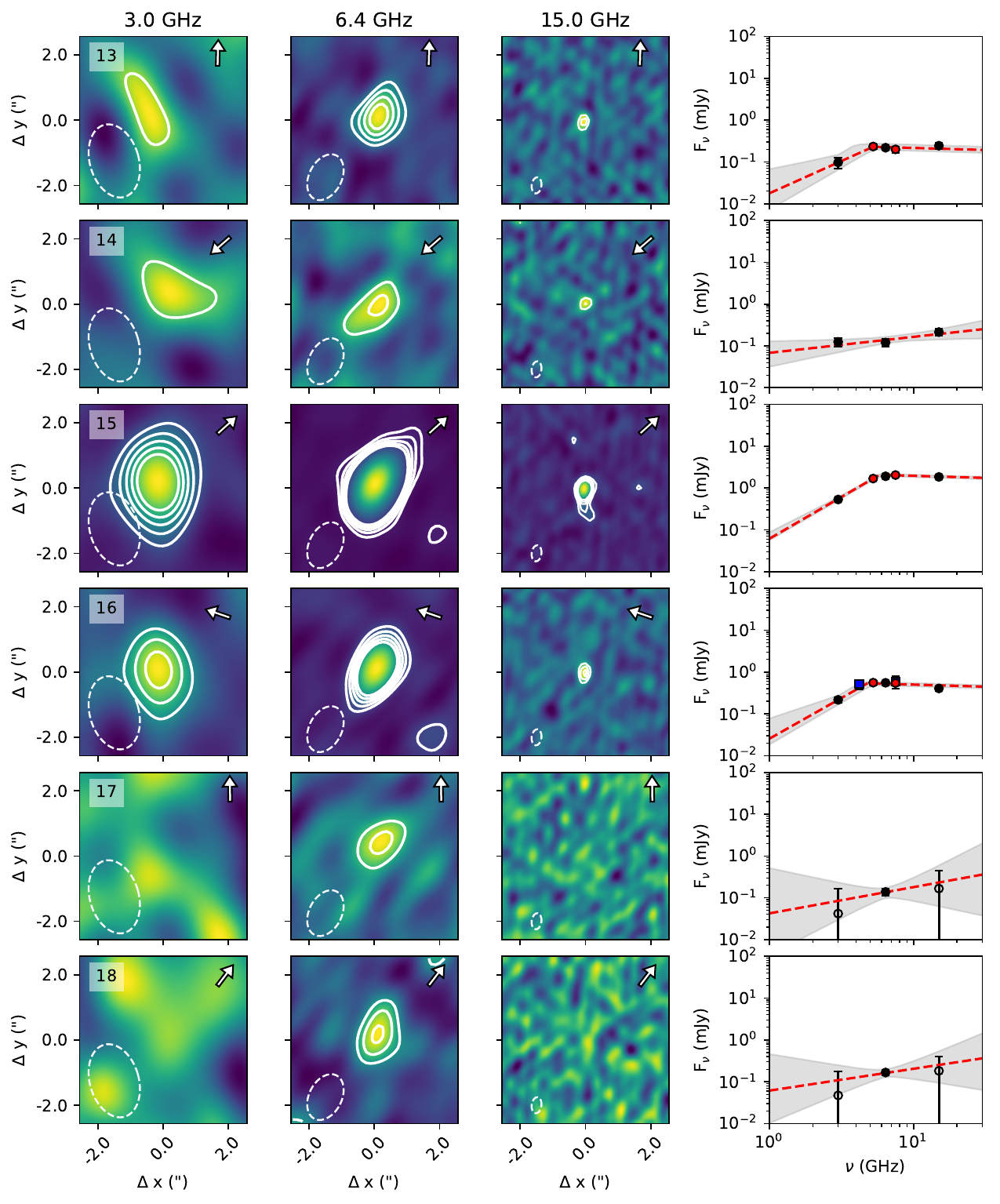}
\caption{\footnotesize {Continuation of Figure \ref{fig:detections_1}.}
\label{fig:detections_3}}
\end{figure*}

\begin{figure*}[htp!]
\epsscale{1.0}
\hspace{-0.3in}
    \plotone{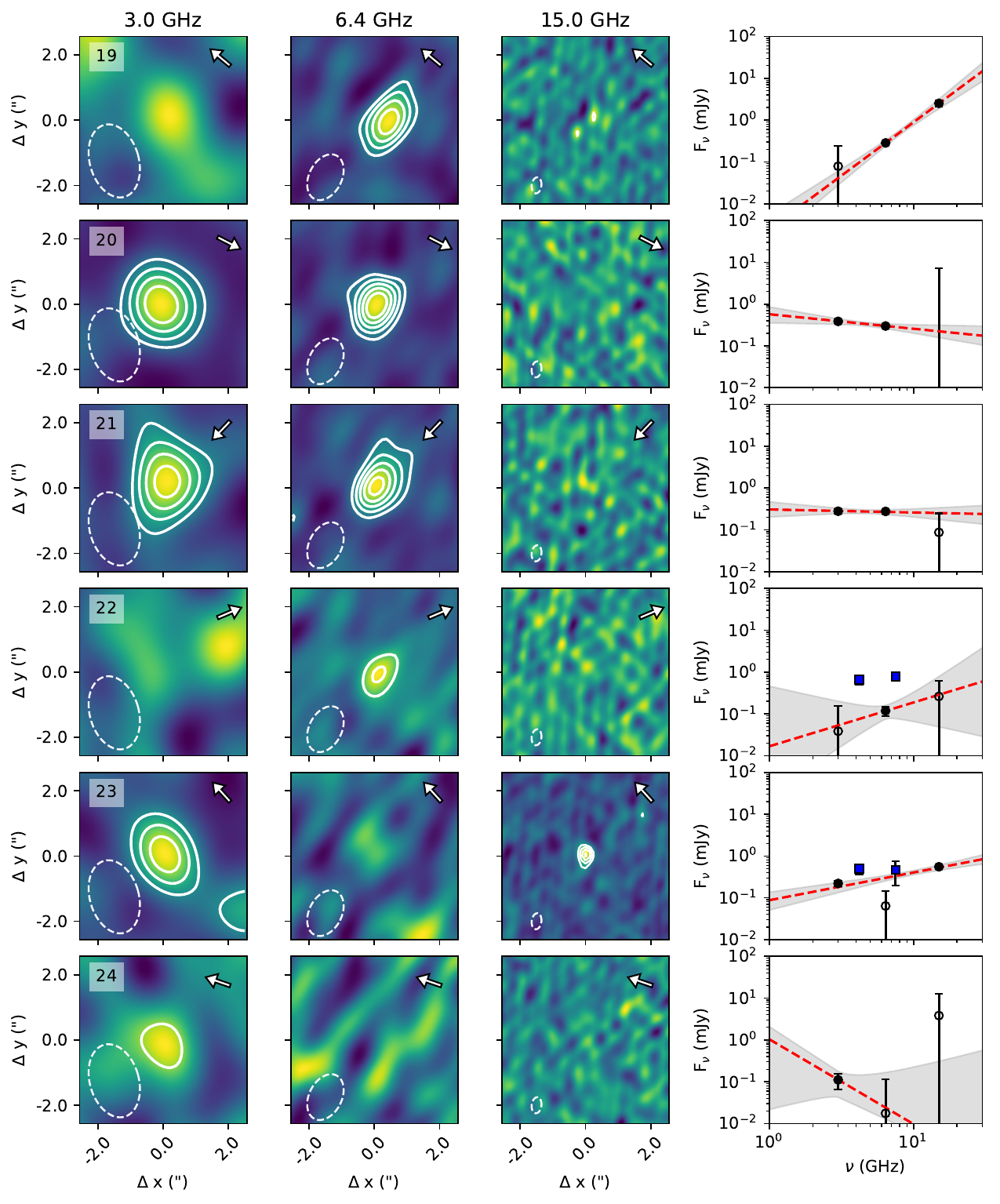}
\caption{\footnotesize {Continuation of Figure \ref{fig:detections_1}.}
\label{fig:detections_4}}
\end{figure*}

\begin{figure*}[htp!]
\epsscale{1.0}
\hspace{-0.3in}
    \plotone{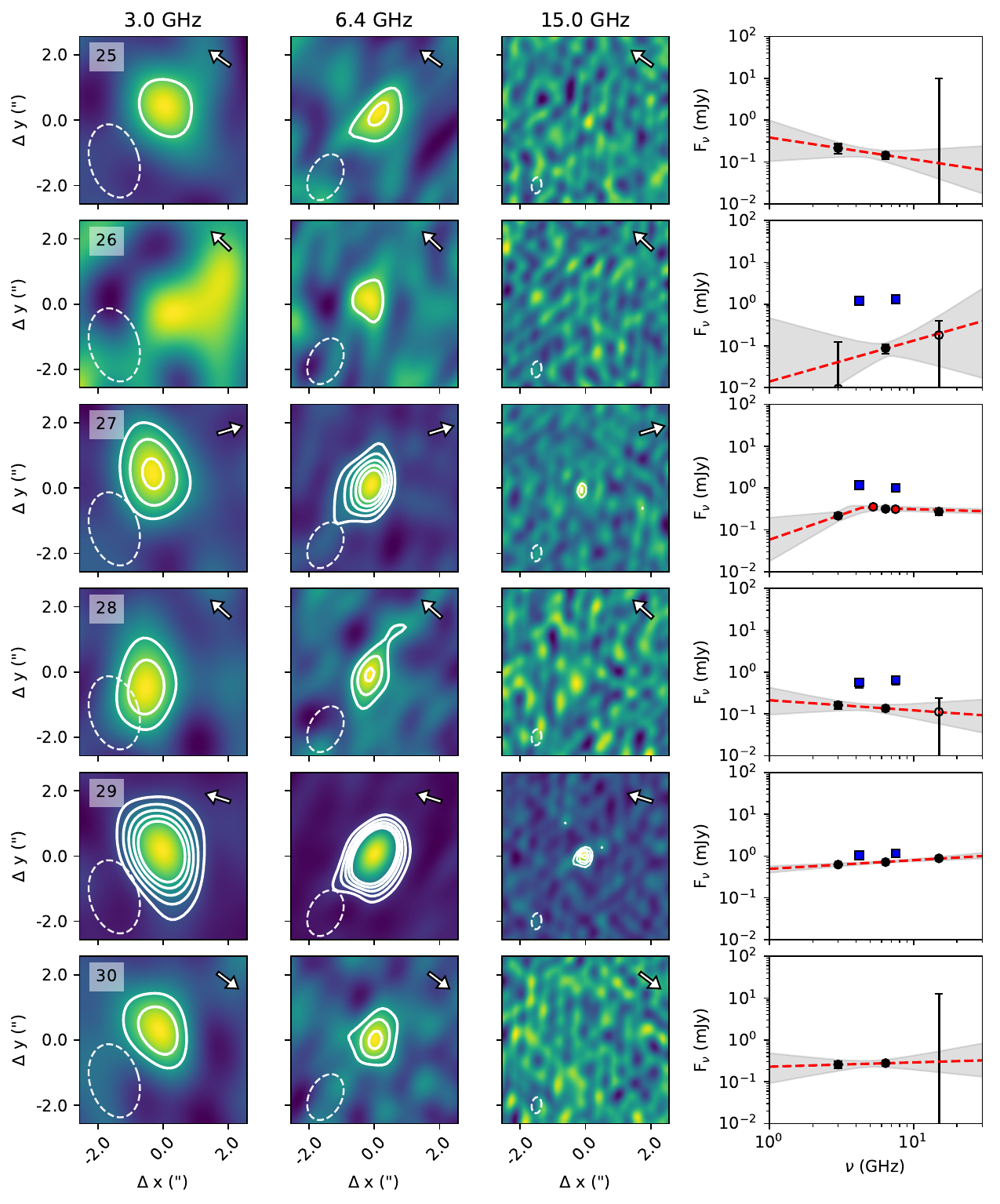}
\caption{\footnotesize {Continuation of Figure \ref{fig:detections_1}.}
\label{fig:detections_5}}
\end{figure*}

\begin{figure*}[htp!]
\epsscale{1.0}
\hspace{-0.3in}
    \plotone{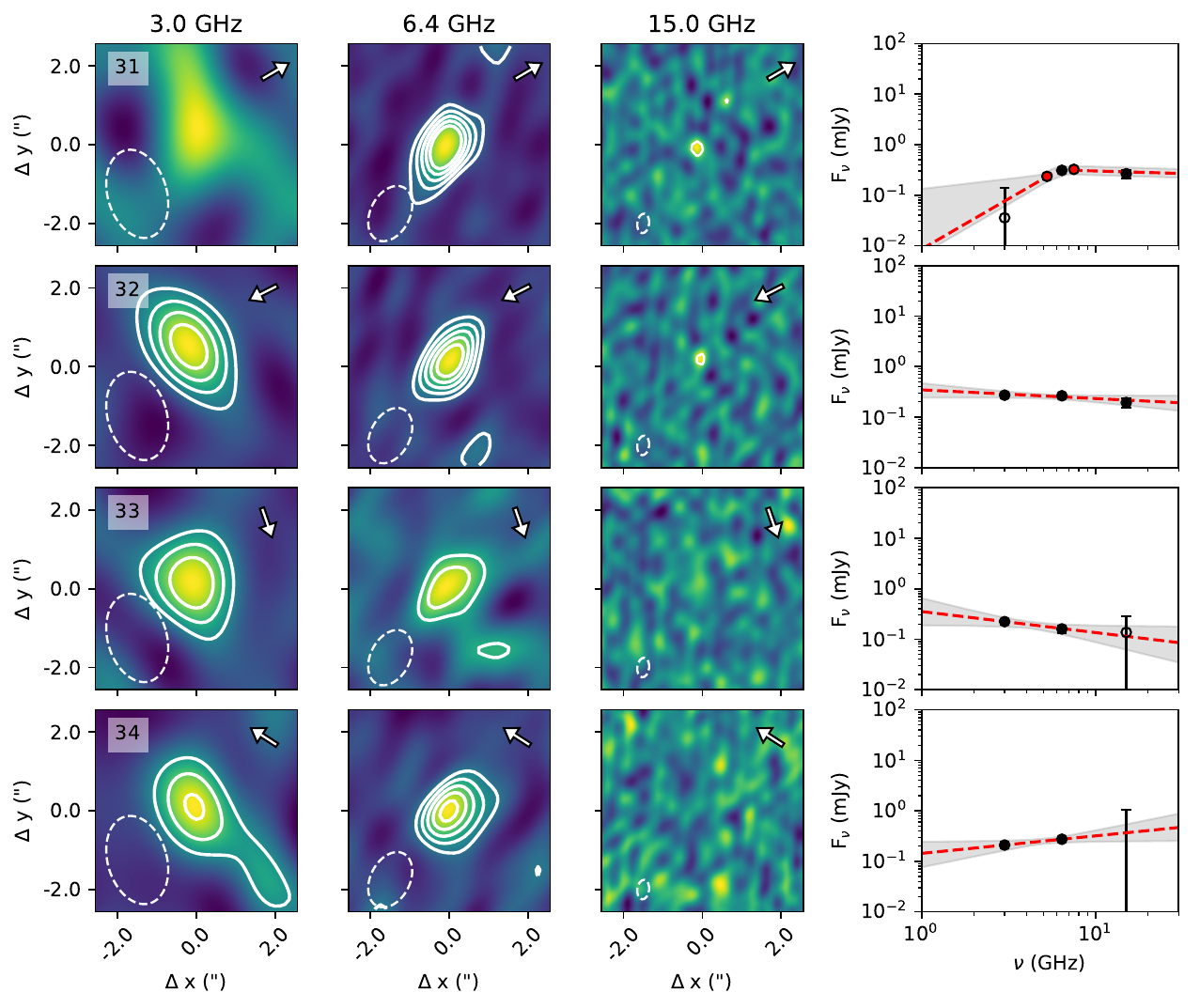}
\caption{\footnotesize {Continuation of Figure \ref{fig:detections_1}.}
\label{fig:detections_6}}
\end{figure*}

%
%
%
%

\clearpage

We use the {\it Spitzer} point source catalog to determine whether the 46 uncataloged radio detections are contaminant background objects or candidate cluster members of NGC 1977.  
{37} of them have {\it Spitzer} counterparts with photometry that are consistent with background objects (see Figure \ref{fig:CMD}), and are thus likely to not be associated with NGC 1977. We classify these {37} sources as probable background objects, and provide further discussion on their observed properties in Appendix \ref{appendix:NGC_A}. The remaining {9} uncataloged detections either have GAIA parallaxes that imply a distance of 400 pc, or {\it Spitzer} point source counterparts that have partial photometric coverage but are consistent with YSOs under two- or three-band selection criteria (see Section \ref{sec:Sample}). 
We consider these {9} sources to be newly-detected cluster members of NGC 1977, and we include them with our samples of radio-detected proplyds and YSOs, for a total of {34} radio-detected NGC 1977 sources. 

\begin{figure*}[ht!]
\epsscale{1.1}
\hspace{-0.3in}
    \plotone{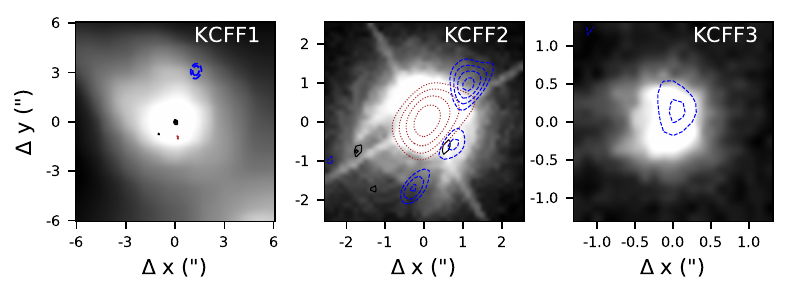}
\caption{  
Spitzer $8$ $\mu$m image (left) and HST/ACS F658N images (center and right) of known proplyds detected in our VLA maps. Contours show VLA radio-continuum emission measured towards the positions of the proplyds, with dotted brown contours showing 3.0 GHz emission, dashed blue contours showing 6.4 GHz emission, and solid black contours showing 15.0 GHz emission. 
All contours start start at $2{\sigma}$ and increase in increments of $0.5{\sigma}$. 
The HST and Spitzer images are registered with respect to the 2MASS coordinate system. 
{
The (0, 0) coordinates for KCFF1 are: 5:35:24.10 --4:50:09.60.
The (0, 0) coordinates for KCFF2 are: 5:35:25.51  --4:51:20.65.
Finally, the (0, 0) coordinates for KCFF3 are 5:35:28.82 --4:50:22.60.}
\label{fig:proplyds_VLA} 
}
\end{figure*}

We measure the fluxes of all detected sources by fitting a 2D Gaussian to the observed radio emission using the CASA task {\tt imfit}. If a source is not 
detected in all radio-continuum bands,  
then we use aperture photometery to obtain an unbiased signal estimate towards the source position in each non-detected band. We also include an additional $5\%$ error on all flux measurements to account for the uncertainty in the absolute flux scale. Table \ref{tab:VLA_source_properties} lists the measured fluxes at 3.0 GHz, 6.4 GHz, and 15.0 GHz for all {34} radio-detected sources that we identify as confirmed or candidate NGC 1977 cluster members, along with the source IDs, coordinates, and individual catalog associations of each source. For the remainder of this paper, we use the source IDs listed in Table \ref{tab:VLA_source_properties} when describing individual {NGC 1977} sources, unless otherwise noted.

{In Figures \ref{fig:detections_1} - \ref{fig:detections_6}, 
we show 3.0 GHz subimages, 6.4 GHz subimages, 15.0 GHz subimages, and radio spectral energy distributions (SEDs) for the {34} VLA-detected NGC 1977 cluster members. 
Each row shows the results for an individual detected source. The sub-images have dimensions of ${\sim}2'' \times 2''$, and the SEDs are produced using the continuum flux measurements provided in Table \ref{tab:VLA_source_properties}. However, if a detection is a known radio source associated with the \cite{Kounkel14} catalog, we include the cm-wavelength flux measurements from \cite{Kounkel14} in the SED panels.  }

\subsubsection{\it Proplyd Detections}\label{sec:proplyds_VLA}

We detect cm-wavelength emission towards the positions of proplyds KCFF1, KCFF2, and KCFF3 in our VLA maps. In Table \ref{tab:VLA_source_properties}, these proplyds {correspond} to source IDs 1, 2, and 3, respectively. All three proplyds are located within $\sim 1\rlap{.}'4$ of 42 Ori (see Figure \ref{fig:NGC1977_region}), and they were initially discovered by \cite{Kim16}, who identified KCFF1 in archival {\it Spitzer}/{\it IRAC} images, and KCFF2 and KCFF3 in archival HST/ACS 658N images.

In Figure \ref{fig:proplyds_VLA}, we show {\it Spitzer} 8$\mu$m observations of KCFF1 and HST 658N images of KCFF2 and KCFF3. We also show  contours of the measured radio-continuum emission measured towards these sources in order to compare the morphologies of the detected radio emission with the morphologies seen in the {\it Spitzer} and HST images.  
The {\it Spitzer} observations of KCFF1 are obtained from the {Spitzer} Heritage Archive\footnote{
\dataset[https://sha.ipac.caltech.edu/applications/Spitzer/SHA/]{https://sha.ipac.caltech.edu/applications/Spitzer/SHA/}} {\citep{SpitzerHeritage22}}, 
and the HST/ACS images of KFCC2 and KCFF3 are obtained from the MAST archive\footnote{\dataset[https://mast.stsci.edu/portal/Mashup/Clients/Mast/Portal.html]{https://mast.stsci.edu/portal/Mashup/Clients/Mast/Portal.html}}. 
We also register the {\it Spitzer} and HST images to the 2MASS coordinate system using the {\tt astrometry.net} software package \citep{Lang10}.
{After registration to the 2MASS coordinate system, the astrometric uncertainties between our radio-wavelength images and the the HST and/or {\it Spitzer} images are  $\lesssim 0.2''$ \citep[e.g.,][]{Eisner05}.}

The multi-wavelength images shown in Figure \ref{fig:proplyds_VLA} reveal that our VLA observations are {tracing the bright ``heads’’ of each proplyd}. Towards the position of KCFF3, the detected 6.4 GHz emission is marginally resolved along the direction of the proplyd head, revealing a morphology at cm-wavelengths that is nearly identical to the morphology seen in the HST image. Towards KCFF1, we see a similar pattern with the detected 15.0 GHz emission, although the Spitzer observations of KCFF1 have a much coarser angular resolution than the 15.0 GHz observations, so the detected cm emission appears very compact in Figure \ref{fig:proplyds_VLA}. 
{For KCFF2, the detected 3.0 GHz emission is concentrated towards the proplyd head, but some of the detected 6.4 GHz emission is spatially offset from the proplyd head.
The  6.4 GHz emission measured towards KCFF2 may therefore be tracing both an ionized jet and a proplyd head, similar to what is observed towards a subset of ONC proplyds \citep[e.g.,][]{Bally98, Bally00, Ricci08}.}

\subsubsection{Structured detections}\label{sec:resolved_new}

{ 
11 of the 34 radio-detected NGC 1977 sources are marginally resolved in our $3.0$, $6.4$, and/or $15.0$ GHz maps: 
sources 2, 4, 10, 15, 19, 21, 23, 27, 28, 29, and 34. 
These sources are labeled as resolved detections in Table \ref{tab:VLA_source_properties}. Resolved detections are identified through a combination of single and multiple Gaussian fitting. If a VLA-detected source has centrally concentrated emission that is well-described by a single Gaussian, then we consider it to be resolved if the fitted Gaussian size is larger than the synthesized beam, following the criteria outlined in \cite{Otter21}. If a source has extended emission that is not well-encapsulated by a single Gaussian, then we perform single and multiple Gaussian fits to the detected emission and consider the source to be resolved if none of the single-Gaussian fits are within the derived confidence intervals. 

We also find that sources 6 and 17 have unresolved emission that is spatially offset from the central stellar coordinates by ${\sim}0\rlap{.}''3 - 0\rlap{.}''5$. These two sources are all detected in the $6.4$ GHz continuum band only, and they are labeled as spatially offset  detections in Table \ref{tab:VLA_source_properties}.

Some of the marginally resolved detections have elongated morphologies that are oriented towards the direction of a nearby B- or A-type star and suggestive of ionized proplyd structures, including sources 15, 21, and 34. For the spatially offset detections, the offset radio emission is closer to a B- or A-type star than the central star, consistent with the expected orientations of proplyds. Despite these suggestive proplyd morphologies, it is worth noting that at our current resolution ({${\sim}0\rlap{.}''5 - 1\rlap{.}''0$}), the observed morphologies of the resolved and offset detections can also be explained by jets with disk orientations that are perpendicular to the direction of a nearby B- or A-type star  \citep{Rodriguez94, Carrasco-Gonzalez12, Tychoniec18, Tobin20}. Hence, from a spatial examination of our VLA maps alone, it is challenging to determine whether the marginally resolved or spatially offset detections are winds or jets.

}

\subsection{Radio Spectral Energy Distributions}\label{sec:SED_modeling}

Ionized gas from photoevaporating disks, stellar winds, and outflows can emit {strong} free-free emission at centimeter wavelengths \citep[e.g.,][]{Wendker73, Garay87, Felli93a, Plambeck95, Anglada98, Reipurth99}.
For a spherically symmetric wind or collimated jet, the free-free emission spectrum 
is expected to follow a power-law dependence with the piecewise form:
\begin{equation}\label{eq:freefree}
F_{\nu}=
    \begin{cases}
        F_{\nu_{\textrm{turn}}} \Big(\frac{\nu}{\nu_{\textrm{turn}}}\Big)^{-0.1} & \nu \geq \nu_{\textrm{turn}} \\[0.125in]
        F_{\nu_{\textrm{turn}}} \Big(\frac{\nu}{\nu_{\textrm{turn}}}\Big)^{\alpha} & \nu < \nu_{\textrm{turn}}
    \end{cases}
\end{equation}
\citep[e.g.,][]{Panagia75, Wright75, Reynolds86}. 
Here, $\nu_{\textrm{turn}}$ is the turnover frequency above which the ionized gas is optically thin, $F_{\nu_{\textrm{turn}}}$ is the flux density at $\nu_{\textrm{turn}}$, and $\alpha$ is the spectral index of the optically thick component of the ionized gas. Depending on the density, geometry, ionization structure, and velocity of the  outflow, $\alpha$ is expected to be a positive number between 0 to 2 \cite[for examples, see][]{Reynolds86}.

{
The turnover frequency is sensitive to the size and density at the inner boundary where the ionized material is launched, with denser and smaller inner boundaries resulting in higher turnover frequencies. Ionized jets from low- and high-mass protostars are thought to have turnover frequencies above 50 GHz \citep[][and references therein]{Anglada18}, 
since large samples of jets are observed to be optically thick at radio wavelengths \citep[e.g.,][]{Anglada98, Tychoniec18b}. 
Ionized disk winds have more extended spatial origins than jets, so the free-free emission spectrum of a disk wind usually turns over and becomes optically thin at lower frequencies \citep[e.g.,][]{Eisner08, Pascucci12, Owen13, Pascucci14}. In the ONC, the majority of free-free emitting proplyds have turnover frequencies below ${\sim}10$ GHz \citep[e.g.,][]{Sheehan16}.
}

{
Here we measure the spectral indices of all VLA-detected NGC 1977 sources to discriminate between optically thin free-free emission from a wind, optically thick free-free emission from a wind or jet, or emission from other mechanisms.  
We initially measure the spectral indices by fitting a power law to the radio SEDs. The power law model has two free parameters: the spectral index and the reference flux at a particular frequency.  
We generate a large suite of power law models over a broad range of spectral indices and reference fluxes, and we fit these models to the measured 3.0, 6.4, and 15.0 GHz radio fluxes via a $\chi^2$ minimization procedure. Since a few of the detections show signs of radio variability {(see Section \ref{sec:variable})}, we opt to use our radio flux measurements alone in the fitting procedure.
}

\begin{figure}[t!]
\epsscale{1.2}
\hspace{-0.15in}
    \plotone{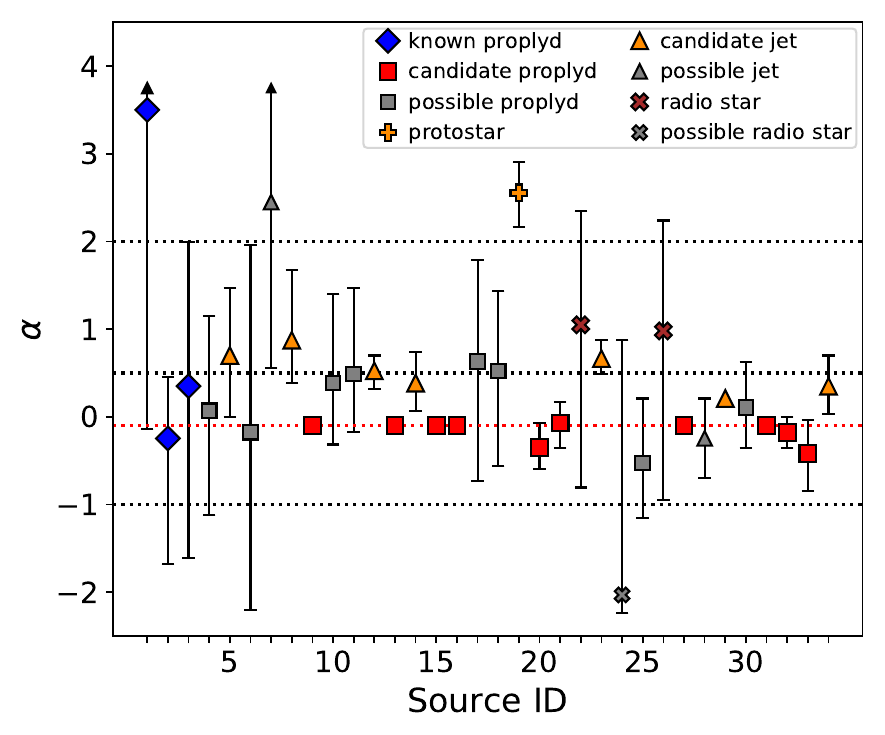}
\caption{Radio spectral indices derived for each VLA-detected NGC 1977 cluster member. For piecewise-power-law-modeled sources, we plot the inferred spectral index above the best-fit turnover frequency, i.e., $\alpha = -0.1$.  All sources are color- and symbol-coded source following their radio classifications in Table \ref{tab:VLA_source_properties}.
The red dotted line depicts $\alpha = -0.1$, the expected spectral index of optically thin free-free emission. The dotted black lines show $\alpha = -1$, $\alpha = -0.5$, and $\alpha = 2$.  \label{fig:alpha} 
}
\end{figure}

\begin{deluxetable}{llrrrch}\tablenum{4}
\tablecaption{Radio spectral energy distribution modeling results \tablenotemark{}}\label{tab:SED_modeling}
\tablehead{ 
    \colhead{ID} & 
    \colhead{Model} & 
    \colhead{  $F_{\nu{\textrm{, ref}}}$   }  &  
    \colhead{$\alpha$ }  & 
    \colhead{$\nu_{\textrm{turn}}$ }  &
    \colhead{Notes}        & 
    \nocolhead{$\chi^2_{min}$   }\\ 
    \colhead{}        &  
    \colhead{}        &  
    \colhead{(mJy)}   & 
    \colhead{}  & 
     \colhead{(GHz)}        &
     \colhead{}        & 
    \nocolhead{} }
\colnumbers
\startdata 
          1 &     Single Power Law &   $1.222^{0.628}_{1.161}$ &               ${>}-0.14$ &                       &                       &             0.0532 \\
          2 &     Single Power Law &   $0.064^{0.199}_{0.062}$ &    $-0.24^{0.70}_{1.44}$ &                       &                       &             0.9888 \\
          3 &     Single Power Law &   $0.100^{0.777}_{0.097}$ &     $0.35^{1.64}_{1.96}$ &                       &                       &             0.2828 \\
          4 &     Single Power Law &   $0.113^{0.450}_{0.098}$ &     $0.07^{1.08}_{1.19}$ &                       &                       &             0.3314 \\
          5 &     Single Power Law &   $0.211^{0.247}_{0.133}$ &     $0.70^{0.77}_{0.70}$ &                       &                    PS &             0.0528 \\
          6 &     Single Power Law &   $0.034^{0.513}_{0.033}$ &    $-0.17^{2.14}_{2.03}$ &                       &                       &             0.5835 \\
          7 &     Single Power Law &   $1.328^{2.120}_{1.013}$ &                ${>}0.56$ &                       &                       &             0.0007 \\
          8 &     Single Power Law &   $1.825^{4.688}_{1.051}$ &     $0.88^{0.81}_{0.49}$ &                       &                    PS &             0.6233 \\
9 &  Piecewise Power Law &   $0.824^{0.080}_{0.073}$ &     $0.85^{1.15}_{0.35}$ &  $5.52^{1.31}_{1.89}$ &                    FS &             0.1016 \\
         10 &     Single Power Law &   $0.260^{1.068}_{0.186}$ &     $0.39^{1.02}_{0.70}$ &                       &                       &             0.0702 \\
         11 &     Single Power Law &   $0.282^{1.121}_{0.200}$ &     $0.49^{0.98}_{0.66}$ &                       &                       &             0.0522 \\
12 &     Single Power Law &   $0.807^{0.302}_{0.244}$ &     $0.53^{0.17}_{0.21}$ &                       &                    PS &             1.8993 \\
13 &  Piecewise Power Law &   $0.231^{0.042}_{0.032}$ &     $1.55^{0.45}_{0.85}$ &  $5.23^{1.60}_{1.46}$ &                    FS &             1.3317 \\
         14 &     Single Power Law &   $0.253^{0.152}_{0.103}$ &     $0.39^{0.35}_{0.31}$ &                       &                    PS &             1.4193 \\
15 &  Piecewise Power Law &   $2.073^{0.118}_{0.148}$ &     $2.00^{0.00}_{0.25}$ &  $5.81^{0.44}_{0.15}$ &                    FS &             0.4253 \\
16 &  Piecewise Power Law &   $0.539^{0.041}_{0.056}$ &     $1.95^{0.05}_{0.80}$ &  $4.79^{0.87}_{0.58}$ &                    FS &             2.8067 \\
         17 &     Single Power Law &   $0.362^{1.649}_{0.324}$ &     $0.63^{1.15}_{1.37}$ &                       &                       &             0.1794 \\
         18 &     Single Power Law &   $0.367^{1.076}_{0.304}$ &     $0.53^{0.91}_{1.08}$ &                       &                       &             0.3627 \\
         19 &     Single Power Law &  $14.729^{8.518}_{6.374}$ &     $2.56^{0.35}_{0.38}$ &                       &                    SS &             0.0528 \\
         20 &     Single Power Law &   $0.174^{0.124}_{0.070}$ &    $-0.35^{0.28}_{0.25}$ &                       &                    FS &             0.4353 \\
         21 &     Single Power Law &   $0.242^{0.140}_{0.105}$ &    $-0.07^{0.25}_{0.28}$ &                       &                    FS &             1.1606 \\
         22 &     Single Power Law &   $0.595^{3.203}_{0.567}$ &     $1.05^{1.30}_{1.86}$ &                       &                       &             0.0225 \\
         23 &     Single Power Law &   $0.841^{0.223}_{0.203}$ &     $0.66^{0.21}_{0.17}$ &                       &                    PS &             9.5074 \\
         24 &     Single Power Law &   $0.001^{0.562}_{0.001}$ &    $-2.03^{2.90}_{0.21}$ &                       &                       &             0.1991 \\
         25 &     Single Power Law &   $0.064^{0.175}_{0.047}$ &    $-0.52^{0.74}_{0.63}$ &                       &                       &             0.0266 \\
         26 &     Single Power Law &   $0.393^{1.948}_{0.376}$ &     $0.98^{1.26}_{1.93}$ &                       &                       &             0.0847 \\
27 &  Piecewise Power Law &   $0.340^{0.040}_{0.047}$ &     $1.20^{0.80}_{0.90}$ &  $4.35^{2.77}_{1.02}$ &                    FS &             0.7525 \\
         28 &     Single Power Law &   $0.092^{0.128}_{0.057}$ &    $-0.24^{0.46}_{0.46}$ &                       &                    FS &             0.0008 \\
         29 &     Single Power Law &   $1.007^{0.182}_{0.142}$ &     $0.21^{0.10}_{0.07}$ &                       &                    PS &             0.0945 \\
         30 &     Single Power Law &   $0.328^{0.501}_{0.197}$ &     $0.11^{0.53}_{0.46}$ &                       &                       &             0.0011 \\
31 &  Piecewise Power Law &   $0.315^{0.057}_{0.053}$ &     $2.00^{0.00}_{1.60}$ &  $6.10^{3.79}_{0.73}$ &                    FS &             0.5262 \\
         32 &     Single Power Law &   $0.192^{0.075}_{0.058}$ &    $-0.17^{0.18}_{0.17}$ &                       &                    FS &             0.7474 \\
         33 &     Single Power Law &   $0.085^{0.094}_{0.050}$ &    $-0.42^{0.39}_{0.42}$ &                       &                    FS &             0.0322 \\
         34 &     Single Power Law &   $0.471^{0.406}_{0.218}$ &     $0.35^{0.35}_{0.32}$ &                       &                    PS &             0.1594 \\
\enddata
\tablenotetext{ }{{\bf Notes.} 
Column (1): source IDs.  
Column (2): 
model type that is used to fit the source radio spectral energy distribution. 
The single power law model has two free parameters: the reference flux and the spectral index. The piecewise power law is constructed from Equation \ref{eq:freefree} and has three free parameters: the turnover flux, the turnover frequency, and the spectral index below the turnover frequency. Column (3): best-fit reference fluxes (single-power-law-modeled sources) or turnover fluxes (piecewise-power-law-modeled sources), along with their $1{\sigma}$ uncertainties. For sources that are fit with a single power law, the reference flux is evaluated at 30 GHz. Column (4): best-fit spectral indices and $1{\sigma}$ uncertainties. For piecewise-power-law-modeled sources, these spectral indices are valid below the turnover frequency. Above the turnover frequency, the best-fit spectral index is $-0.1$.
Column (5): best-fit turnover frequencies and $1{\sigma}$ uncertainties for all sources that are modeled with a piecewise power law. 
Column (6): Notes on individual sources. A `PS' denotes a positive spectrum source with a best-fit spectral index between 0 and 2. An `FS' denotes a flat spectrum source that is either a piecewise-power-law-modeled source, or a single-power-law modeled source with a best-fit spectral index that is consistent with $-0.1$ and between  $-1$ and $0.5$. An `SS' denotes a steep spectrum source with a best-fit spectral index ${>}2$.}
\end{deluxetable}

{ 
Most of our sources are fit quite well by a single power law model, but a subset shows evidence of having a free-free turnover within the frequency range covered by our observations (e.g., sources 9, 15, 16). We construct a piecewise free-free emission model via Equation \ref{eq:freefree} and fit this model to the SEDs of all sources whose single power-law fits yield a reduced $\chi^2$ $>2$. If the piecewise power law fit yields an improved reduced $\chi^2$, then we use the piecewise power law to characterize the radio spectrum. Otherwise, we use the single power law model. The piecewise power law model has three free parameters: the turnover frequency, the turnover flux, and the spectral index below the turnover frequency. 
To ensure that our piecewise fits have one degree of freedom, we utilize the 5.25 and 7.5 GHz maps that are generated by imaging the two C-band basebands separately (see Section \ref{sec:data}). We construct SEDs from flux measurements at 3.0, 5.25, 7.5, and 15.0 GHz, and fit the piecewise power law models to these SEDs. 
}

{
Table \ref{tab:SED_modeling} indicates which sources are characterized with a single or piecewise power law. For the sources that are characterized with a single power law, we show the best-fit spectral indices and reference fluxes in Table \ref{tab:SED_modeling}. For the sources that are characterized with a piecewise power law, we instead show the best-fit turnover frequencies, best-fit turnover fluxes, and best-fit spectral indices below the turnover frequency. Figure \ref{fig:alpha} shows the best-fit spectral indices and their uncertanties for all single-power-law-modeled sources. For the piecewise-power-law-modeled sources, we plot $\alpha = -0.1$, i.e., the best-fit spectral index above the turnover frequency. Finally, in the SED panels in Figures \ref{fig:detections_1} - \ref{fig:detections_6}, 
we show the best-fit SEDs and 1$\sigma$ confidence intervals that are derived from the single or piecewise power-law fits. 
}

{
In general, the VLA-detected NGC 1977 sources have radio-SEDs that are 
consistent with free-free emission. 
Six sources\textemdash sources 9, 13, 15, 16, 27, and 31\textemdash  prefer the piecewise power-law fit over the single power law fit and have best-fit turnover frequencies between ${\sim}$1 and 10 GHz. These sources are fit very well by the piecewise power-law model, suggesting that they are emitting free-free emission from a wind rather than a jet.
Six of the single-power-law-modeled sources have relatively flat spectral indices that fall below the typical values measured towards jets \citep[${\sim}0.5$,][]{Anglada18}, but overlap with the $-0.1$ value expected for optically thin free-free emission.  
These sources may be emitting optically thin free-free emission from a wind, although it is possible that their flat spectral indices are produced by a combination of optically thick free-free emission and non-thermal emission with a steep negative spectral index (see Section \ref{sec:pol}). 
Seven single-power-law-modeled sources have positive best-fit spectral indices that are above  $-0.1$ and consistent with 
the typical values measured towards optically thick jets. 
Finally, one source, source 19, has a best-fit spectral index ${>}2$, which cannot be explained by optically thick free-free emission, but can be explained by dust emission \citep[e.g.,][]{Hildebrand83}. 
}

{
In Tables \ref{tab:VLA_source_properties} and \ref{tab:SED_modeling}, we label sources as flat spectrum (FS), positive spectrum (PS), or steep spectrum (SS) depending on the preferred model fit and the range of allowed spectral indices. 
The flat spectrum label is applied to all piecewise-power-law-modeled sources, and to all single-power-law-modeled sources with best-fit spectral indices that are consistent with optically thin free-free emission and between $-1.0$ and $0.5$. Above $0.5$, we expect values consistent with optically thick jets. Below $-1.0$, we expect values indicative of  non-thermal emission dominating the continuum.
The positive spectrum label is applied to all single-power-law-modeled sources with positive best-fit spectral indices between $0$ and $2$. Finally, the steep spectrum label is applied to source 19, the one single-power-law-modeled source with a best-fit spectral index ${>}2$.
We do not label any sources whose range of allowed spectral indices overlaps with the above categories, since it is less clear whether they are emitting optically thin free-free emission, optically thick free-free emission, or emission from another mechanism.
}

\subsection{\it Circularly Polarized Emission}\label{sec:pol}

{ 

Radio observations of ionized winds or outflows can in some cases be contaminated by non-thermal gyrosynchrotron emission produced from stellar magnetospheric activity \citep[][]{Dulk85}. Gyrosynchrotron emission typically exhibits a steep negative spectral index at centimeter wavelengths, although for some electron energy distributions, the spectral index can be flatter and more similar to optically thin free-free emission \citep[e.g.,][]{Gudel02}. In this scenario, one of the main ways to discriminate between optically thin free-free emission and flat-sloped gyrosynchrotron emission is through measurements of circular polarization, computed as the ratio of the Stokes V (circularly polarized) and Stokes I (total intensity) flux densities \cite[e.g.,][]{Feigelson98, Zapata04b}. This is because optically thin gyrosynchrotron emission has moderate-to-high levels of circular polarization at radio wavelengths \citep[${>}10\%$; ][]{Dulk85}, whereas optically thin free-free emission does not.

Here we use measurements of circular polarization to examine whether the radio-SEDs of any of our detected NGC 1977 sources are contaminated with gyrosynchrotron emission. For each cluster member that we detect in the Stokes I plane, we search for circularly polarized emission in the Stokes V plane towards the same position as the detected Stokes I emission. If Stokes V emission is detected above $2.5{\sigma}$ (see Section \ref{sec:fitting:detections}), then we calculate the Stokes V flux density using the same aperture that was employed in the Stokes I plane, and compute the circular polarization fraction as the ratio of the measured Stokes V and Stokes I fluxes. 
If no Stokes V emission is detected, then 
we calculate an upper limit on the circular polarization fraction using the measured Stokes I flux and the $2.5{\sigma}$ upper limit on the Stokes V flux. 

In total, circularly polarized emission was detected towards only 4 of the 34 Stokes-I-detected NGC 1977 cluster members: sources 2, 22, 28, and 29.  For sources 2, 22, and 28, the Stokes V emission is detected in the 6.4 GHz band only. For source 29, the detected Stokes V emission is in the 3.0 GHz band only. In all cases, the detecections are at marginal statistical significance ($2.5{\sigma}-3.5{\sigma}$). 
Nevertheless, we find that the circular polarization fractions implied by the detected Stokes V emission are ${>}10\%$ by more than $1{\sigma}$ and, thus, consistent with the polarization fractions expected from gyrosynchrotron emission. 
The detected radio emission towards sources 2, 22, 28, and 29, may therefore be tracing gyrosynchrotron emission in addition to, or instead of, free-free emission. 
In Table \ref{tab:VLA_source_properties}, we labels sources 2, 22, 28, and 29 as likely gyrosynchrotron emitters based on their measured circular polarization fractions. 

Because source 2 (proplyd KCFF2) has 6.4 GHz emission that is spatially separated from both central star and detected 3.0 GHz emission, 
it is possible that the 3.0 GHz emission is tracing free-free emission from the proplyd head (see Section \ref{sec:proplyds_VLA}), while some of the detected 6.4 GHz emission is tracing polarized emission from another component of the YSO, such as a nonthermal jet. 
Non-thermal jets are rarer than thermal jets, but a handful have been detected towards low-mass YSOs \citep[e.g.,][]{Tychoniec18}. These jets typically have a double-lobed morphology with peaked emission that is offset from the central protostar, similar to what is observed at 6.4 GHz towards source 2 (see Figure \ref{fig:proplyds_VLA}). 

For sources 9, 15, and 16, no Stokes V emission is detected, but the upper limits imply circular polarization fractions of ${<}10\%$. These limits are below the minimum circular polarization fraction expected for radio-wavelength gyrosynchrotron emission from an individual star, so the radio emission observed towards sources 9, 15, and 16 is unlikely to be contaminated with gyrosynchrotron emission. These 3 sources are bright in the Stokes I plane, detected in all three continuum bands, and have SEDs that are well reproduced by a piecewise power-law model (see Figures \ref{fig:detections_1} - \ref{fig:detections_6}). 
In Table \ref{tab:VLA_source_properties}, we label sources 9, 15, and 16 as sources whose SEDs are not contaminated with gyrosynchrotron emission.

For the majority of NGC 1977 cluster members with non-detected Stokes V emission, the upper limits on the circular polarization fractions vary from ${\lesssim}20\%$ to ${\lesssim}80\%$. 
These limits are not stringent enough to determine whether the radio-wavelength SEDs are contaminated with gyrosynchrotron emission. 
Deeper observations are needed to explicitly rule out the possibility of   gyrosynchrotron contamination for these sources. 

Finally, we note that the Stokes V detection rate is higher among the background radio sources that are detected in the Stokes I plane of our VLA maps (see Appendix \ref{appendix:NGC_A}). 
Extragalactic objects are known to emit strong levels of gyrosynchtrotron emission \citep[e.g.,][]{Condon92}, and so we expect some background sources in our maps to be emitting circularly polarized radio emission. 

}

\subsection{\it Variable Sources}\label{sec:variable}

{
Here we search for evidence of variability among the subset of the NGC 1977 cluster members that are detected in both our VLA maps and in the previous VLA maps of \cite{Kounkel14}. For each of these sources, we compare our 6.4 GHz flux measurement with the average of the 4.5 and 7.5 GHz flux measurements from \cite{Kounkel14}, and then identify variable sources as the ones for which the 6.4 GHz flux measurement and average \cite{Kounkel14} flux measurement differ by more than $3\sigma$. Taking the average of the \cite{Kounkel14} flux measurements allows us to obtain an approximate 6.4 GHz flux estimate at this earlier epoch. 
To ensure that none of the sources with intrinsically steep spectral indices are misidentified as variable as a result of comparing flux measurements at slightly different frequencies,
we include a minimum variability percentage of $20\%$ in our criteria for variable source identification. We define the variability percentage as the ratio of the standard deviation to the mean of our 6.4 GHz flux measurement and the average \cite{Kounkel14} flux measurement.

With these criteria, we identify 6 cluster members as variable: sources 10, 22, 26, 27, 28, and 29. 
Of these 6 sources, 4 are spatially resolved (sources 10, 27, 28, and 29), and 3 are 
also detected in the Stokes V plane (sources 22, 28, and 29). In Table \ref{tab:VLA_source_properties}, we label sources 10, 22, 26, 27, 28, and 29 as variable. 

Sources 22 and 26 show a particularly strong degree of radio variability. For these sources, our measured 6.4 GHz fluxes are more than $5$ times lower than the previously reported fluxes from \cite{Kounkel14}. These levels of variability are stronger than the typical values measured towards thermal  jets or winds \citep[e.g.,][and references therein]{Anglada18}, but they are consistent with the values expected from stellar gyrosynchrotron emission, which is known to be highly variably on short timescales \citep[e.g.,][]{Feigelson85}. 

}

\subsection{\it Classification of NGC 1977 Radio Sources}\label{sec:classification}

{
Following the discussion in Sections \ref{sec:fitting:detections} - \ref{sec:variable}, we classify the radio-detected NGC cluster members into different categories based on the observed characteristics of the detected radio emission. In Table \ref{tab:VLA_source_properties}, we include a column that indicates all source classifications.

We first classify sources 1, 2, 3 (proplyds KCFF1, KCFF2, and KCFF3) as ``Known Proplyds,'' since the detected radio emission of these sources is spatially coincident with the {HST- or Spitzer-identified proplyd heads} and spectrally consistent with optically thin free-free emission. For source 2, however, we note that the measured 6.4 GHz emission may be tracing a nonthermal jet in addition to the proplyd head. 

We then classify additional radio-detected NGC 1977 cluster members as ``Candidate Proplyds,'' ``Candidate Jets,'' or ``Radio Stars'' based on their radio spectral indices, circular polarization, and/or radio variability. In principle, morphology would be the most effective characteristic for distinguishing between free-free  emission from a wind, free-free emission from a jet, or gyrosynchrotron emission from a compact region of magnetic reconnection near the central star. 
However, most of our detections are unresolved, 
and the resolved detections exhibit radio morphologies that, at our current resolution, are consistent with the morphologies of both winds and jets.
We therefore exclude morphology as a criterion when classifying radio detections without HST- or Spitzer-identified proplyds.  We note, however, that some of the sources that we classify as ``Candidate Proplyds'' or ``Candidate Jets'' are marginally resolved, whereas none of the sources that we classify as ``Radio Stars'' are resolved (see Table \ref{tab:VLA_source_properties}).

We classify all flat spectrum (FS) sources with non-detected Stokes V emission as candidate proplyds, since the low ($<10$ GHz) turnover frequencies and/or flat spectral indices of these sources are more likely to be explained by free-free emission from a wind than from a jet. As shown in Table \ref{tab:VLA_source_properties}, this selection criteria yields 10 candidate proplyds: sources 9, 13, 15, 16, 20, 21, 27, 31, 32, and 33. Three of these candidate proplyds have Stokes V upper limits that explicitly rule out potential contamination from gyrosynchrotron emission (sources 9, 15, 16). For the remaining candidate proplyds, deeper imaging, Stokes V observations, and/or time monitoring is needed to firmly rule out contributions from gyrosynchrotron emission. 

We classify all positive spectrum (PS) sources (sources 5, 8, 12, 14, 23, 29, 34) as candidate jets, since these sources have positive radio spectral indices that are consistent with the free-free emission being optically thick, as expected for jets. For most jet candidates, the spectral indices are comparable to the typical values measured towards jets. For source 29, the spectral index is shallower than typical jet values, but the emission is circularly polarized (see Table  \ref{tab:VLA_source_properties}), suggesting that gyrosynchrotron emission is contributing to the continuum and causing the overall spectral index to appear shallow. It is important to note that some of these jet candidates may be disk winds with free-free turnovers located just above the frequency range covered with our VLA observations (see Section \ref{sec:discussion_internal_external}). This can be tested with future observations at higher radio frequencies.   

We classify sources 22 and 26 as radio stars since they exhibit a high level of variability that cannot be explained by free-free emission from a wind or jet but can be explained by gyrosynchrotron emission from stellar magnetic reconnection. Both of these sources are unresolved and detected in the 6.4 GHz band only, and source 22 is circularly polarized, further pointing to stellar gyrosynchrotron emission as the sole emission mechanism.

\begin{figure*}[ht!]
\epsscale{1.0}
\hspace{-0.3in}
\plotone{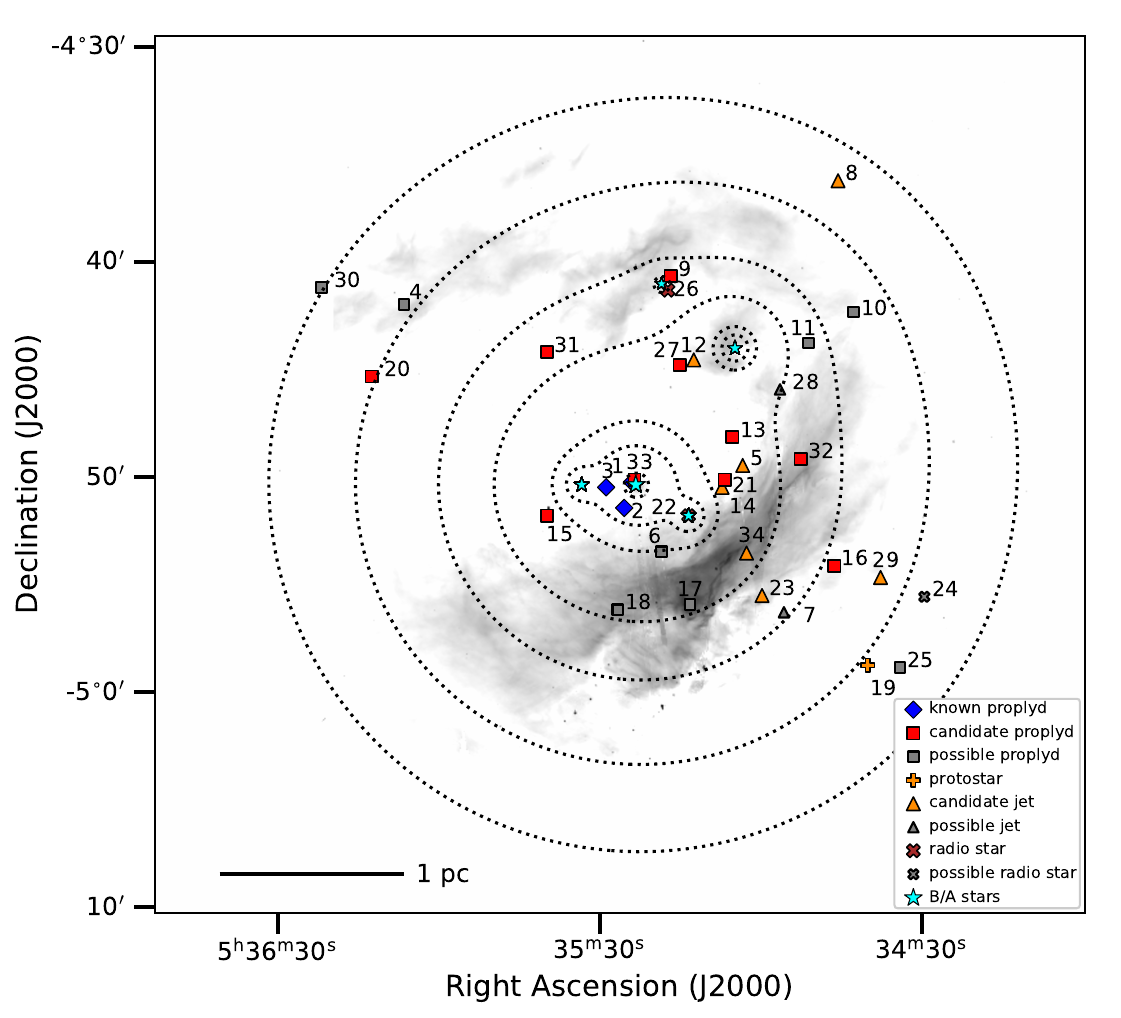}
\caption{Spatial distribution of VLA-detected NGC 1977 cluster members. 
The layout of this plot is similar to that of Figure \ref{fig:NGC1977_region}, except here we include dotted contours that show the local FUV radiation field strength at $15$ $G_0$, $25$ $G_0$, $50$ $G_0$, $100$ $G_0$, $500$ $G_0$, $1000$ $G_0$, and $10,000$ $G_0$, where $G_0 = 1.6 \times 10^{-3}$ erg cm$^{-2}$ s${-1}$ \citep{Habing68}.
{Sources are color- and symbol-coded based on their radio classifications in Table \ref{tab:VLA_source_properties}. The source IDs are listed near the position of each source symbol. \label{fig:detections_spatial} }
}
\end{figure*}

We classify source 19 as a protostar. The steep spectral index of source 19 suggests that its radio continuum is dominated entirely by thermal dust emission. Assuming optically thin dust with a constant opacity and constant dust temperature of 20 K  \citep[][]{Beckwith90}, the measured 15.0 GHz continuum flux of source 19 would imply an unreasonably high disk mass. We note, however, that  the measured flux is compatible with hotter, optically thick dust surrounding less evolved class 0/I protostars  \citep[e.g.,][]{Sheehan22}.

The remaining 11 radio-detected cluster members are labeled as a `Possible Proplyd,' `Possible Jet,' or `Possible Radio Star', depending on the uncertainties on their derived spectral indices and on whether or not they are detected in the Stokes V plane. If a source is undetected in the Stokes V plane and has a best-fit spectral index that overlaps with our criteria for a flat or positive spectrum (see Section \ref{sec:SED_modeling}), we refer to it as a possible proplyd, since its spectrum is consistent with, albeit broadly, free-free emission. Alternatively, if a source has a best-fit spectral index that overlaps with our criteria for a positive or steep spectrum, we refer to it as a possible jet. The only exceptions to these two classification schemes are sources 24 and 28. We classify source 24 as a possible radio star, because its best-fit spectral index trends toward steep negative values suggestive of non-thermal emission (see Figure \ref{fig:alpha}). And finally, we classify source 28 as a possible jet rather than a possible proplyd, since it is detected in the Stokes V plane and, thus, likely has a steeper-than-measured free-free spectral index. For all of these 11 sources, we emphasize that while we are somewhat able to narrow down their nature, additional data is needed to determine they are winds, jets, or gyrosychrotron-dominated radio sources.
}

\section{\bf{Discussion}}\label{sec:discussion}


\subsection{Disk Photoevaporation in an Intermediately-irradiated Star-forming Region}\label{sec:nature}

{The detection of several-dozen NGC 1977 cluster members at radio wavelengths allows for one of the first systematic investigations of YSOs in an intermediately-irradiated star-forming environment.}
In  Figure \ref{fig:detections_spatial}, we show the spatial distribution of the 34 radio-detected NGC 1977 sources, compared with the spatial distribution of B- or A- type stars in NGC 1977, namely, 42 Ori, HD 37058, HD 294264, HD 369658, and HC 294262. 
We also include contours that show the local external FUV radiation field strength in NGC 1977. Here, the external FUV field is calculated by estimating the FUV-continuum luminosity of each B- or A- type star using the stellar evolutionary models of \cite{DiazMiller98}, and then computing a localized FUV flux at all positions in our VLA maps. The spectral types for 42 Ori ($\sim$B1V), HD 37058 ($\sim$B3V), HD 294264 ($\sim$B3V), HD 369658 ($\sim$B3V), and HC 294262 ($\sim$A0) are taken from the literature \citep[][and references therein]{Peterson08}, and they imply FUV-continuum luminosities of ${\sim}5 \times 10^{47}$ s$^{-1}$ for 42 Ori , ${\sim}2 \times 10^{46}$ s$^{-1}$ for HD 37058, HD 294264, and HD 369658, and ${\sim}3 \times 10^{44}$ s$^{-1}$ for HD 294262. Altogether, the B- or A- type stars in NGC 1977 stars produce an external FUV field that ranges from ${\sim}10^4$ $G_0$ towards the center of NGC 1977, to ${\sim}10$ $G_0$ in the outskirts of NGC 1977.

{The} radio-detected NGC 1977 sources are distributed throughout the cluster and, thus, exposed to a  range of intermediate FUV fields (between ${\sim}10$ and $10^4$ $G_0$). The known proplyds that we detect are all located in the cluster center where the FUV field is ${\sim}1,000 - 10,000$ $G_0$, {as this was the region of NGC 1977 that was previously observed with HST (see Section \ref{sec:Sample})}. {The newly-detected candidate proplyds tend to be located in regions where the  FUV field is ${\sim}50 - 1,000$ $G_0$, although we find one candidate proplyd in the inner ${\sim}10,000$ $G_0$ region of NGC 1977, and another in the outskirts where the FUV field is ${\sim}20$ $G_0$. 
Half of jet candidates are positioned near one of more candidate proplyds, while the rest are located in the outskirts along with source 19, the candidate young protostar that we have identified. 
Finally, the two sources that we classify as radio stars are in very close proximity to a B- or A-type star. This, along with the close proximity of some jet candidates and proplyd candidates, demonstrates how both gyrosynchrotron-dominated radio sources and free-free-emitting jets can be located in the same regions of a cluster where expect to find populations of photoevaporating disks. 
}

{Numerical simulations widely predict that intermediate {FUV} fields are sufficient enough to launch photoevaporative winds 
that dominate over internally-driven mechanisms of disk dispersal and drive disk mass loss at rates 
$\gtrsim 10^{-8} - 10^{-6}$ M$_{\odot}$ yr$^{-1}$ } \citep[e.g.,][]{Adams04, Clarke07, Facchini16, Haworth18, Haworth18b, Haworth19, Coleman22}. Recent ALMA surveys in Orion have also found indirect evidence of external photoevaporation in intermediate-UV environments, revealing that disks close to B- and A-type stars are on average ${\sim}2-4$ times less massive than the disks further away from B- and A-type stars \citep[][]{Terwisga19, vanTerwisga23}. 
{Despite these findings, identifying direct signatures of external photoevaporation has proven challenging outside of the $>10^4$ $G_0$ regions of the ONC and other O-star-hosting clusters, where many disks are exposed not only to intense FUV irradiation, but also to intense EUV irradiation that can externally ionize FUV-driven photoevaporative winds to point where large samples of ionized winds become bright and easy to see with the VLA \citep[e.g.,][]{Churchwell87, Garay87, Felli93a, Felli93b, Forbrich16, Sheehan16}, HST \citep[e.g.,][]{Odell94, Bally98, Bally00, Ricci08, Smith10, Fang12, Haworth21b}, {VLT-MUSE \citep[e.g.,][]{Haworth23, Aru24}}, JWST \citep{Berne22, Habart23, McCaughrean23}, and ALMA {\citep{Ballering23}}.}

{If the newly-detected candidate proplyds in our sample are indeed YSOs with externally-evaporating disks, as suggested by their free-free emission spectra and, in several cases, their lack of circular polarization and/or radio variability, this would demonstrate the use of broadband radio photometry for identifying externally-evaporating disks in intermediatly irradiated regions of star formation.  Higher-resolution imaging of the NGC 1977 region\textemdash with the VLA, ALMA, HST, VLT-MUSE, and/or JWST\textemdash is needed to spatially resolve free-free emission structures, confirm the disk-wind nature of our photometrically-identified candidate proplyds, 
and explicitly rule out the possibility of 
gyrosynchrotron contamination for the majority of Stokes-V-nondetected candidate proplyds.  Deeper imaging can also improve uncertainties in the flux measurements and hence the inferred radio spectral indices of all VLA-detected sources in our maps, including possible proplyds, possible jets, and background extragalactic objects. 
}

\subsection{Mass-loss rates}\label{sec:mdot}

\begin{figure*}[ht!]
\epsscale{1.0}
    \hspace{-0.7in}
\plotone{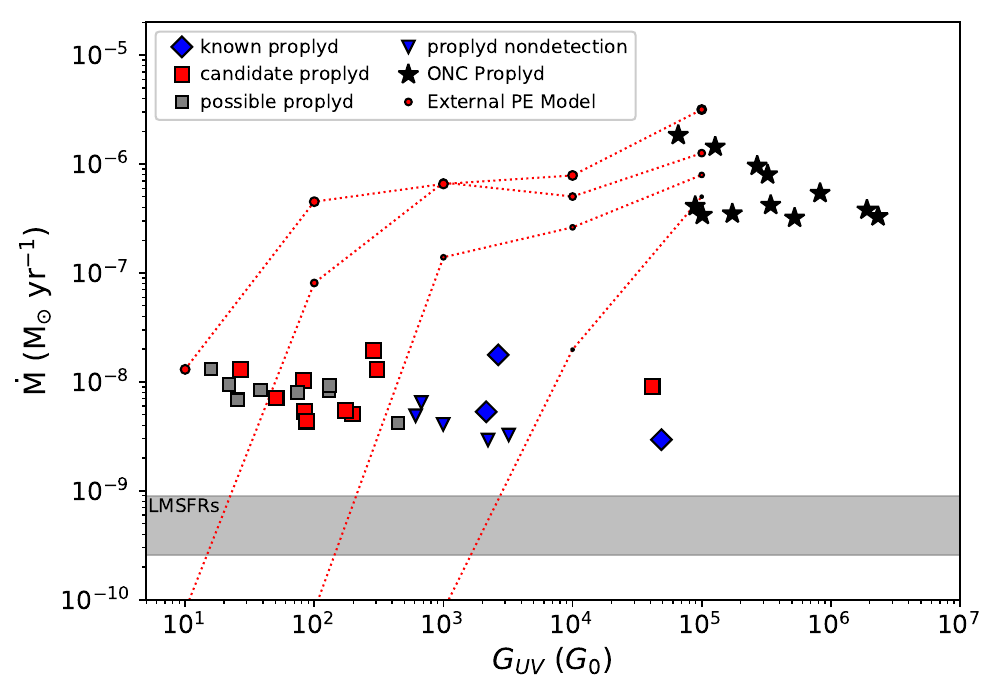}
\caption{Photoevaporative mass loss rates ($\dot{M}$) derived from Equation \ref{eq:Mdot} 
{for all known, candidate, and possible NGC 1977 proplyds detected in our VLA maps,}
plotted as a function of the local FUV radiation field strength.
{We also include points showing the mass-loss rates inferred from the $4{\sigma}$  6.4 GHz upper flux limits that are measured towards the VLA--non-detected known proplyds from 
\cite{Bally12} and \cite{Kim16}.}
We also show the mass loss rates derived by {\cite{Ballering23}} for a subset of ALMA-detected proplyds in the ONC, where the FUV radiation field strength towards the ONC proplyds is calculated using a FUV photon luminositiy of $7.0 \times 10^{48}$ s$^{-1}$ for $\theta^1$ Ori C \citep[][]{Johnstone98, Storzer99, Odell17}, which we assume as the sole, dominant source of external FUV radiation in our calculations. 
The shaded grey region indicates the $1{\sigma}$ region around the median {internally-driven wind mass-loss rates} derived by \cite{Fang18} towards a sample of disks in low-mass star-forming regions (LMSFRs).
The dotted red lines show predicted mass loss rates (as a function of the FUV field strength) from the {\tt FRIED} grid of externally evaporating disk models \citep{Haworth18b}, with the different lines corresponding to models with different disk radii. The model disk size is directly proportional to the marker size of the plotted red circles, and we show the predicted mass loss rates for disk radii of 20 AU (smallest circles), 40 AU, 75 AU, and 150 AU (largest circles). All plotted models share a common initial disk mass of $\sim 5$ Jupiter masses, and a common central stellar mass of $\sim 0.5$ M$_{\odot}$. \label{fig:Mdot_plot}   }
\end{figure*}

Measuring the mass-loss rates of protoplanetary disks provides crucial tests for understanding the impact of the stellar cluster environment on disk properties and evolution. Here we measure the mass-loss rates of our VLA-detected proplyds and candidate proplyds {by computing the steady-state mass flow through a sphere at the location where the ionized wind is launched, via $\dot{M} = 4 \pi r^2 \rho v$. We assume a fully ionized flow, such that the gas density can be expressed in terms of the electron density, $n_e$, as $\rho = \mu m_H n_e$, where $\mu = 1.35$ is the mean molecular weight of the ionized gas and $m_H$ is the mass of a hydrogen atom. For the outflow velocity, we adopt $v = 10$ km s$^{-1}$ \citep[typical for ionized photoevaporative winds; e.g.,][]{Johnstone98}. }

Since most of our VLA detections are unresolved or marginally resolved, we are unable to spatially isolate the wind launching radius with our VLA observations. Sources 2 and 3, however, have proplyd ionization fronts that are well-resolved with {HST} (see Figure \ref{fig:proplyds_VLA}). 
Since these ionization fronts mark the approximate location where the photovaporative winds become fully ionized, 
we calculate $\dot{M}$ for sources 2 and 3 at the location of the HST-identified ionization fronts, using the ionization front radii derived by \cite{Kim16}.
For the remaining sources, we calculate $\dot{M}$ assuming an upper limit to the wind launching radius. We approximate the launching radius as the half-width-at-half maximum of the synthesized beam major axis, {using the beam size from the highest frequency observations at which the detected free-free emission is consistent with being optically thin.} As shown in Table \ref{tab:VLA_obs}, this typically results in $r \sim 0\rlap{.}''5 - 1''$ (${\sim}200-500$ AU), i.e., values that are similar to the ionization front radii of proplyds in the ONC {\citep[e.g.,][]{Ballering23}}.

{For an optically thin ionized wind, the electron density can be derived from the emission measure, $EM \approx n_e^2 L$, where $L$ is line-of-sight path length through the emitting region. We calculate the emission measure using} Equation A.1b of \cite{Mezger67}:
\begin{equation}
    \Big(\frac{EM}{\textrm{pc cm}^{-6}}\Big) = \Big(\frac{\tau_{ff}}{3.28 \times 10^{-7}}\Big) \Big(\frac{T_e}{10^{4} \textrm{ K}}\Big)
    \Big(\frac{\nu}{\textrm{GHz}}\Big),
\end{equation}
where $\tau_{ff}$ denotes the line-of-sight optical depth of the ionized gas, and $T_e$ denotes the electron gas temperature  \citep{Mezger67}. {Again,} we compute $\tau_{ff}$ at the highest frequency at which the detected free-free emission is optically thin via $\tau_{ff} = I_\nu / B_{\nu}(T_e)$, where $I_\nu$ is the measured peak surface brightness of the free-free emission, and $B_{\nu}(T_e)$ is the Planck function. For an electron temperature of $10^{4}$ K, we obtain emission measures in the range $\sim 10^3 - 10^5$ pc cm$^{-6}$ for our VLA-detected sources.

{Previous studies of the ONC proplyds have computed the line-of-sight path length assuming that the ionized winds have a hemispherical geometry or a thin spherical shell geometry
\cite[e.g.,][]{Bally98, Ballering23}. 
We compute the line-of-sight path length through our NGC 1977 proplyds and candidate proplyds using the same thin shell geometry assumed by \cite{Ballering23},  in which case the path length is approximated as $L = 2\sqrt{2r / n_e \sigma}$, where $\sigma = 6.3 \times 10^{-18}$ {cm$^{2}$} is the ionization cross section. This choice allows for a direct comparison of the photoevaporative mass-loss rates derived from measurements of free-free emission in NGC 1977 vs. the ONC. 
If any our NGC 1977 sources have more spatially extended distributions of ionized gas, 
as expected for an isotropic or internal ionization source (see Section \ref{sec:ni_sources}), 
we would expect the inferred electron densities and, thus, mass loss rates to be, at most, an order of magnitude lower as a result of the increased path length.
For example, if were to assume a uniform sphere of ionized gas (i.e., $L \approx 2 r$), the electron densities would be ${\sim} 3-4$ times lower than the values inferred under the hemispherical shell geometry. 
}

{With our choice of path length geometry, the mass-loss rate can be expressed in terms of the flow velocity, launching radius, and emission measure (i.e., free-free emission surface brightness) as
\begin{equation}\label{eq:Mdot} 
    \dot{M} = 2 \pi \mu m_H v \sigma^{1/3} EM^{2/3} r^{5/3}.
\end{equation}
This equation is valid over frequencies where the ionized winds are optically thin. 
In the optically thick regime, the mass-loss rates can also be derived from the mass-flow equation provided that the free-free fluxes, turnover frequencies, spectral indices, source distances, and source geometries are constrained \citep[e.g.,][]{Panagia75, Reynolds86}.
Assuming a spherical wind geometry, we find that for the NGC 1977 sources that are fitted with a piecewise power law model and have evidence for a free-free turnover, the mass-loss rates derived in the optically thick regime are lower than, but within a order of magnitude of, the values derived in the optically thin regime via Equation \ref{eq:Mdot}.}

Figure \ref{fig:Mdot_plot} shows the mass-loss rates obtained from Equation \ref{eq:Mdot} for all {VLA-detected proplyds and candidate proplyds in NGC 1977 as a function of the local FUV radiation field strength. We also include points showing 
the inferred mass-loss rates for possible NGC 1977 proplyds in our maps, and for all 
non-detected proplyds assuming an upper flux limit of $4{\sigma}$ and an ionization front radius from the literature. Finally, we include points showing the 
mass-loss rates derived by {\cite{Ballering23}} for the subset of ALMA-detected ONC proplyds. }
Typically, we derive mass-loss rates of $\sim10^{-9} - 10^{-8}$ M$_{\odot}$ yr$^{-1}$ for the NGC 1977 sources, although in most cases, the derived mass-loss rates {may be} upper limits due to the values we have assumed for the {wind launching radii and/or line-of-sight path-lengths}. 
Overall, the mass-loss rates of our NGC 1977 sources 
are significantly lower than the mass loss rates of the ALMA-detected ONC proplyds, by up to two orders of magnitude. 
{If any of our newly-identified candidate proplyds have smaller wind launching radii than the values assumed in our calculations\textemdash as expected for an internally-driven photoevaporative wind (see Section \ref{sec:discussion_internal_external})\textemdash 
then the derived mass-loss rates would be even lower. }

The lower mass-loss rates derived towards our NGC 1977 sources suggest that disks in NGC 1977 may be less prone to the ``proplyd lifetime problem'' than disks in the ONC. The lifetimes of disks in clustered star-forming regions can be estimated by dividing the total mass of a photoevaporating disk by the derived photoevaporative mass-loss rate, and in the ONC, these calculations typically result in lifetimes of ${\sim}1-100$ kyr, which are orders of magnitude lower than the ${\sim}1$ Myr age of the ONC 
{\citep[for recent mass loss estimates, see][]{Ballering23}.}
The mismatch in the derived lifetimes vs. stellar ages suggest that disk-proplyd systems in the ONC should have dispersed well before the present time in which they are being observed, unless they are more massive than typically assumed {\citep[e.g.,][]{Clarke07}}, and/or have more recently begun to photoevaporate due to extinction effects, dynamical effects, or younger-than-assumed stellar ages \citep[e.g.,][]{Scally01, Winter19, Qiao22}. If we assume that the VLA-detected NGC 1977 sources have similar disk masses as the ONC proplyds, then the derived lifetimes would be ${\sim}0.1-10$ Myr, which are longer than the derived lifetimes of ONC proplyds, and consistent with the ${\sim}1$ Myr age of NGC 1977 \citep[e.g.,][]{DaRio16}.

\subsection{Evidence for spatially-extended externally-evaporating disks in NGC 1977 }\label{sec:discussion_internal_external}

Although we find lower photoevaporative mass-loss rates in NGC 1977 than in the ONC (see Figure \ref{fig:Mdot_plot}), the derived mass-loss rates are consistent with the values predicted by models of external photoevaporation in intermediate-UV environments.
In Figure \ref{fig:Mdot_plot}, we include red dashed lines that show the predicted mass-loss rates of externally evaporating disks with different disk sizes. The predicted mass-loss rates are taken from the {\tt FRIED} grid of photoevaporating disk models \citep{Haworth18b}, and we plot the mass-loss rates of models with an initial disk mass of ${\sim}5$ Jupiter masses, a stellar mass of $0.5$ M$_{\odot}$, and a disk radius of either 20, 40, 75, or 150 AU. 

For the {candidate proplyds} located at ${\sim}20-500$ $G_0$, the derived mass-loss rates match up well with the predicted values of extended photoevaporating disk models with radii between $\sim 50$ and 150 AU. For the two VLA-detected proplyds at ${\sim}1000$ $G_0$ (Sources 1 and 2, i.e., KCFF2 and KCFF3), we find that smaller evaporating disk models, with radii $\sim 20-40$ AU, are needed to produce an agreement between the inferred and predicted mass-loss rates. Finally, the {candidate proplyd and known proplyd} at ${\sim}10,000$ $G_0$ (Source 33 and Source 3, i.e., KCFF 1) appear to have mass-loss rates that are lower than the values predicted by external photevaporation models. This discrepancy can be reconciled if the two sources are exposed to weaker FUV fields than what is implied by their projected separations from nearby B and A stars, due to projection or extinction effects \citep[e.g.,][]{Winter19, Parker21b, Qiao22}. It can also be reconciled if the two sources have higher stellar masses than the $0.5$ M$_{\odot}$ model values plotted in Figure \ref{fig:Mdot_plot}, which seems plausible given the positions of cluster members on our near-infrared color-magnitude diagram (see Figure \ref{fig:CMD}).

The model comparisons shown in Figure \ref{fig:Mdot_plot} suggest that NGC 1977 may contain a population of extended disks with systematically larger disk sizes than the compact disks typically found in the ONC \citep[e.g.,][]{Eisner18, Boyden20, Otter21}. In the harsh ${>}10^5$ $G_0$ regions of the ONC, the external FUV field is strong enough such that even compact ($R = 20-30$ AU) disks are expected to undergo intense external photoevaporation and truncate down to smaller disk sizes (e.g., $R \lesssim 5$ AU) before no longer being subject to significant mass loss \citep[][see Figure \ref{fig:Mdot_plot}]{Haworth18b}. In the more intermediately-irradiated environment of NGC 1977, however, only extended disks can experience significant mass loss (i.e., $\dot{M} > 10^{-10}$ M$_{\odot}$ yr$^{-1}$) from external photoevaporation, with ${\gtrsim}50$ AU radii required at $100$ $G_0$, and ${\gtrsim}100$ AU radii required at $10$ $G_0$. {Our VLA-detected candidate proplyds at ${\sim}10 - 1,000$ $G_0$ may therefore host
spatially extended photoevaporating disks with sizes that are similar to the values commonly found in lower-mass star-forming regions \citep[e.g.,][]{Ansdell18, Sanchis21}. The VLA-detected {proplyds and candidate proplyds} at ${\sim}1000 - 10,000$ $G_0$, however, appear likely to host compact disks with comparable sizes as the ONC proplyds (see Figure \ref{fig:Mdot_plot}). }

While FUV-driven winds due to external photoevaporation are expected to dominate the mass evolution of the outer regions of protoplanetary disks in intermediate-UV environments, we can also examine whether 
some of the observed free-free fluxes in NGC 1977 are produced by 
strong, internally-generated disk winds. The inner regions of protoplanetary disks are thought to launch magnetohydrodynamic (MHD) and photoevaporative winds that can also deplete disks of planet-forming material, where these winds are driven by internal magneto-centrifugal processes and ionizing photons from the central star, respectively \citep[for a recent review of inner disk winds, see][]{Pascucci22}. 
Surveys of disk forbidden line emission typically derive mass-loss rates of around $\sim 10^{-9}$ M$_{\odot}$ yr$^{-1}$ for internally-generated disk winds \citep[e.g.,][]{Natta14, Fang18}, which are somewhat lower than the inferred mass loss rates of our NGC 1977 sources (see Figure \ref{fig:Mdot_plot}). 
Some class II objects, however, are found to have higher wind mass-loss rates of ${\gtrsim}10^{-8}$ M$_{\odot}$ yr$^{-1}$, where these sources typically have high accretion rates \citep[e.g.,][]{Fang18}, and are theorized to be X-ray active \citep[e.g.,][]{Owen12, Picogna19, Ercolano21}.

{If our VLA-detected sources in NGC 1977 are biased towards high accretion rates or high X-ray luminosities, then we might expect some of the derived mass-loss rates to be consistent with a strong internal disk wind. 
However, this scenario seems unlikely to apply to the candidate proplyds with evidence for a free-free turnover (see Table \ref{tab:SED_modeling}). The free-free emission spectrum of a vigorous X--ray-driven disk wind (i.e., $\dot{M} > 10^{-9}$ M$_{\odot}$ yr$^{-1}$) is predicted to have a relatively high turnover frequency (${>}20$ GHz) as a result of intense X--ray-driven heating and ionization of bound material close (${<}1$ AU) to the central star \citep[e.g.,][]{Owen13}. 
Six of our photometrically-identified candidate proplyds have turnover frequencies between 1 and 10 GHz (see Table \ref{tab:SED_modeling}), suggesting that they are not irradiated by enough X-ray photons to launch a strong inner disk wind that is optically thick over the full radio spectrum.   
}

{
{Spatially resolved imaging of the central disks and surrounding winds,} 
combined with constraints on the free-free turnover frequency, 
could help identify which cluster members in NGC 1977  
are launching internally- versus externally-driven disk winds. 
If any of the VLA-detected sources in our sample have compact disk sizes that are smaller than the values consistent with external photoevaporation models, then these would be more likely to be launching internally, rather than externally, driven disk winds.  
}

\subsection{Proplyd Illumination Sources in NGC 1977}\label{sec:ni_sources}

\begin{figure*}[ht!]
\epsscale{1.1}
\plotone{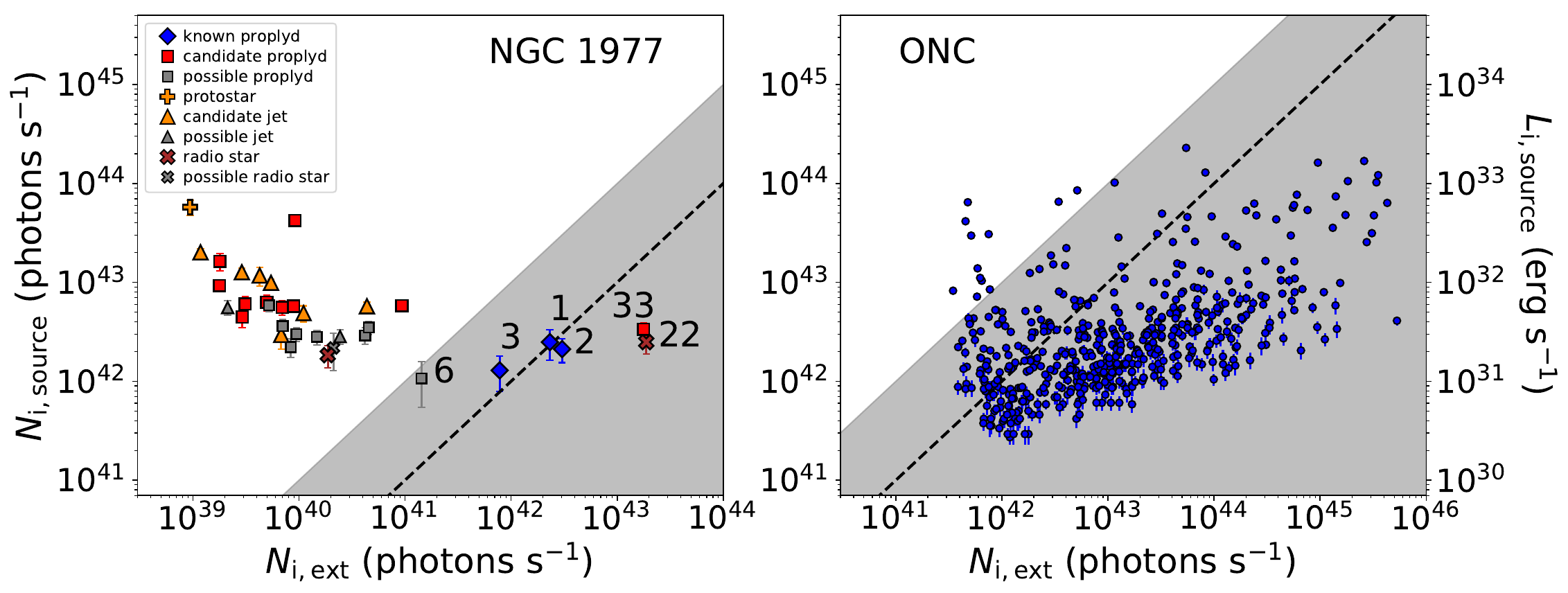}
\caption{
Ionizing photon {rates} ($N_{\mathrm{i, source}}$) derived from Equation \ref{eq:Ni} for all 
for all NGC 1977 {cluster members} detected in our VLA maps 
(left panel), 
and all compact radio sources in the central $20' \times 20'$ region of the ONC (right panel). The ionizing photon {rates} are plotted as a function of the  {local rate} of {ionizing EUV photons} ($N_{\mathrm{i, ext}}$) provided by massive stars in each region. 
The {ionizing photon rates} are shown in both units of photons s$^{-1}$ and in units of erg s$^{-1}$, {where the conversion factor from photon s$^{-1}$ to erg s$^{-1}$ is ${\sim}2.2 \times 10^{-11}$}. The dashed black line indicates where $N_{\mathrm{i, source}} = N_{\mathrm{i, ext}}$, while the shaded gray region depicts where $N_{\mathrm{i, source}} \leq 10 \times N_{\mathrm{i, ext}}$. NGC 1977 sources within the shaded gray region are indicated with their source IDs listed in Table \ref{tab:VLA_source_properties}. To derive ionizing photon {rates} for the ONC sources, we use the 6.0 GHz continuum flux measurements from \cite{VargasGonzalez21}.
\label{fig:Ni_plot}   
}
\end{figure*}

Here we examine which ionization sources in NGC 1977 are responsible for ionizing and, thus, illuminating our radio-detected proplyds and candidate proplyds. 
Measurements of optically thin free-free emission relate directly to an {ionizing photon flux} via the ``Volume Emission Measure,'' defined as: 
\begin{equation}\label{eq:VEM}
    \frac{VEM}{\mathrm{cm}^{-3}} = 2.9 \times 10^{50} \bigg(\frac{S_{\nu}}{\mathrm{mJy}}\bigg)  \bigg(\frac{\nu}{\mathrm{GHz}}\bigg)^{0.1}  \bigg(\frac{T_e}{10^4 \mathrm{ K}}\bigg)^{0.35} \bigg(\frac{d}{\mathrm{pc}}\bigg)^{2},
\end{equation}
where $S_{\nu}$ is the measured flux of the free-free emission, $\nu$ is the frequency at which the free-free emission flux is measured, $T_e$ is the electron temperature of the ionized gas, and $d$ is the source distance \citep{Mezger67}. {The rate of ionizing photons required to ionize the gas}
is $N_i = \alpha VEM$, where $\alpha$ is the recombination coefficient. Assuming $\alpha = 2.6 \times 10^{-13} (T_e / 10^4 \mathrm{K})^{-0.8}$ \citep[e.g.,][]{Churchwell87, Garay87}, $T_e = 10^4$ K, and $d = 400$ pc, $N_i$ can be written in terms of the free-free flux as:
\begin{equation}\label{eq:Ni}
    \frac{N_i}{\mathrm{s}^{-1}} = {1.2 \times 10^{43}}\bigg(\frac{S_{\nu}}{\mathrm{mJy}}\bigg)  \bigg(\frac{\nu}{\mathrm{GHz}}\bigg)^{0.1}.
\end{equation}
{We use this equation to determine the ionizing photon rates that are required to produce the observed free-free fluxes of known proplyds and new candidate proplyds in NGC 1977. 
We also use this equation to compute the 
ionizing photon rates implied by the measured fluxes of other radio-detected cluster members,
including jet candidates, possible proplyds or jets that are potentially emitting free-free emission, and gyrosynchrotron- or dust-dominated radio sources.  
For the sources where we clearly identify the optically thick and thin components of the free-free emission spectrum, 
we compute the ionizing photon rate at the highest frequency in which the free-free emission is detected and consistent with being optically thin. 
For all other sources, we use the measured radio fluxes at the highest detected frequency to compute the ionizing photon rate.}

{Figure \ref{fig:Ni_plot} shows the rate of ionizing photons derived from Equation \ref{eq:Ni} as a function of 
the local rate of incident ionizing photons from  known B- and A- type stars in NGC 1977, i.e., 42 Ori, HD 37058, HD 294264, HD 369658, and HD 294262. For each B- or A-type star, we compute an ionizing photon rate incident on an NGC 1977 radio source using the {EUV-continuum} luminosities from the evolutionary models of \cite{DiazMiller98} and the same source sizes used to derive the photoevaporative mass-loss rates (see Section \ref{sec:mdot}). We then sum the individual external ionizing photon rates provided by each B or A star to obtain a total external ionizing photon rate. We assume no intracluster extinction in these calculations, and use an {EUV-continuum} luminosity of $10^{46}$ s$^{-1}$ for 42 Ori, an {EUV-continuum} luminosity of  $10^{43}$ s$^{-1}$ for HD 37058, HD 294264, and HD 369658, and an {EUV-continuum} luminosity of $10^{37}$ s$^{-1}$ for HD 294262. 
}

We also include a panel in Figure \ref{fig:Ni_plot} that shows the number of ionizing photons derived for compact radio sources in the ONC, in order to compare the number ionizing photons derived for sources in NGC 1977 vs. the ONC. Radio continuum flux measurements for the ONC sources are taken from \cite{VargasGonzalez21}, who compiled $6.0$ GHz continuum flux measurements for a sample of ${\sim}500$ sources within the central $20' \times 20'$ region of the ONC. To calculate the number of {EUV} photons provided by massive stars in the ONC, we assume that $\theta^1$ Ori C is the sole, dominant producer of external {EUV} radiation, and we adopt an  {EUV-continuum}  luminosity of $7 \times 10^{48}$ erg s$^{-1}$ \citep{Johnstone98, Storzer99, Odell17}.

Figure \ref{fig:Ni_plot} suggests that B- and A-type stars in NGC 1977 do not produce enough EUV radiation to externally ionize and externally illuminate most of the radio-detected NGC 1977 sources in our VLA maps. In order for a free-free emitting source to be illuminated by external ionizing radiation, the number of ionizing photons derived from Equation \ref{eq:Ni} must be less than or equal to the localized number of ionizing photons provided by massive stars, as found towards proplyds and YSOs in the core of the ONC {\cite[e.g.,][see also Figure \ref{fig:Ni_plot}]{Churchwell87, Garay87, Zapata04a, Zapata04b, Forbrich16, Sheehan16, VargasGonzalez21, Ballering23}}. 
{Of the 34 radio-detected NGC 1977 sources plotted in Figure \ref{fig:Ni_plot}, only 4 satisfy this requirement: sources 1, 2, 22, and 33. If the EUV-continuum luminosities of NGC 1977’s B and A stars were ${\sim}2-10$ times more luminous than assumed, then 2 additional sources would satisfy this requirement: sources 3 and 6. As shown in Figure \ref{fig:detections_spatial}, sources 1, 2, 3,  6, 22, and 33 are all located in the core of NGC 1977 near 42 Ori, with sources 1, 2, and 3 being known proplyds, source 33 being a newly-identified candidate proplyd, source 6 being a possible proplyd, and source 22 being a radio star that is likely emitting gyrosynchrotron emission rather than free-free emission. }

{The free-free emission that we measure towards sources 1, 2, and 33 can  therefore be produced by gas that is externally ionized by {EUV} radiation from 42 Ori, with additional contributions from HD 37058, HD 294264, HD 369658, and HD 294262. 
If we have underestimated the EUV luminosities of 42 Ori, HD 37058, HD 294264, HD 369658, and/or HD 294262, then we might also expect the radio emission from source 3 and, if emitting free-free emission, source 6 to be produced by gas that is externally ionized by neighboring B- or A-type stars.
For all other free-free emitting NGC 1977 radio sources in our sample, including 
any possible proplyds that are potentially emitting free-free emission, 
other ionization sources are needed to produce the ${\gtrsim}0.1$ mJy radio fluxes that we are measuring.}

{Unless a photoevaporating disk is located in the core of NGC 1977, we might expect its wind to be illuminated by internal ionizing radiation from the central star \citep[e.g.,][]{Pascucci12, Owen13, Pascucci14}, regardless of whether the wind is internally or externally driven.}
In Figure \ref{fig:Ni_plot}, we include an additional y-axis that shows the ionizing luminosities implied by the {ionizing photon rates} derived from Equation \ref{eq:Ni}. These luminosities are consistent with the expected {ionizing} luminosities of pre-main-sequence stars \citep[e.g.,][]{Alexander05, Getman05, Herczeg07, Shkolnik14, Hartmann16}, although they tend to be on the more luminous end of expected values. The detection of radio emission towards a relatively small number of {UV-luminous stars}, however, matches up well with the overall {low detection rates that we observe towards known YSOs in 
NGC 1977 (see Figure \ref{fig:detection_profile})}. 

{
Given the potential prevalence of internally illuminated disk winds in NGC 1977, it is worth pointing out that externally driven but internally illuminated disk winds can have similar morphologies as externally driven and externally illuminated winds. The d203-506 disk in the Orion Bar, for example, is irradiated by a ${\sim}10^4$ $G_0$ FUV field but shielded from the external EUV field \citep[e.g.,][]{Champion17}, and recent infrared- and submillimeter-wavelength imaging of this disk has revealed that the photoevaporative wind surrounding d203-506 has the same cometary shape as the externally illuminated ONC proplyds \citep[e.g.,][]{Haworth23, Berne24}. This new imaging demonstrates that anisotropic FUV radiation fields play a dominant role in sculpting cometary wind morphologies when an external EUV radiation field is not present.  Proplyd-like structures in NGC 1977 may therefore be expected outside of the more intensely EUV-irradiated cluster center, although in these regions the externally FUV-driven winds may have more uniform distributions of ionized gas, rather than distributions that are concentrated in a thin shell near an ionization front (see Section \ref{sec:mdot}).
}

{ 
A deeper, higher-resolution VLA survey in NGC 1977 is needed to establish a larger sample of free-free emitting YSOs, rule out the possibility of gyrosychrotron contamination, and test whether most free-free-emitting sources in NGC 1977 are illuminated by internally or externally sourced radiation. If deeper observations were to reveal diffuse ionized wind structures whose free-free fluxes correlate with stellar UV or X-ray luminosity, this would point to internal ionization as the main illuminating source for ionized winds in NGC 1977, in which case some of the FUV-driven photoevaporative winds in NGC 1977 may only be partially ionized \citep[e.g., ][]{Haworth19}. We note, however, that in order to enable investigations with stellar properties, additional spectroscopic data at other wavelengths are needed, as the majority of YSOs in NGC 1977 do not have published accretion rates or X-ray luminosities. If such measurements were obtained, and if deeper imaging were to instead reveal that most externally FUV-driven disk winds in NGC 1977 are surrounded by thin shells of ionized gas with brighter free-free fluxes than expected from internal illumination, this would point to other ionization sources as the dominant illumination mechanism and suggest that we are underestimating the amount of external ionization in NGC 1977.

Shock ionization has been proposed as an explanation for some of the  bright ionization fronts observed towards intermediately irradiated disks in the outskirts of the ONC, where the EUV field is low and the FUV field is ${\sim}1000$ $G_0$ \citep[e.g.,][]{Odell97b, Smith05}, and towards a subset of the more intensely irradiated ONC disks near the cluster center \citep[e.g.,][]{Bally98}. Candidate proplyds in NGC 1977 may be externally illuminated by a similar mechanism if they are located near dense interstellar gas or high-mass stellar outflows. 
On the other hand, Figure \ref{fig:Ni_plot} suggests that at least one known proplyd in NGC 1977 has a higher free-free flux than what is implied by the expected EUV luminosities of neighboring B and A stars.
A similar inconsistency has been found by \cite{Bally12} towards the NGC 1977 proplyd Parengo 2042, but with HST-based measurements of electron density. 
The EUV radiation field in NGC 1977 is therefore likely to be stronger than currently predicted, although it remains unclear whether this is due to an incorrect spectral type for one of the region's massive stars, or to a systematic underprediction of the EUV luminosities of young B-type stars,
as has been observed towards the handful of evolved B stars where direct measurements of the ionizing UV flux are possible \cite[e.g.,][]{Cassinelli95}. 
}

\section{\bf{Conclusions}}\label{sec:conclusions}

We presented deep, multi-wavelength VLA observations covering the central $\sim 30' \times 45'$ region of NGC 1977 at 3.0 GHz, 6.4 GHz, and 15.0 GHz. We searched for compact radio sources in these maps using blind and catalog-driven approaches, and identified 56 sources at 3.0 GHz, 65 sources at 6.4 GHz, and 26 sources at 15.0 GHz, for a total 71 unique sources. 
{Of these 71 VLA-detected sources, 3 are associated with HST- and/or {\it Spitzer}-identified proplyds in NGC 1977, 22 are associated with other YSOs in NGC 1977, {9} are newly-identified candidate members of NGC 1977, and {37} are  background extragalactic objects that are not associated with NGC 1977.} {Thus, we detected a total of {34} confirmed or candidate NGC 1977 cluster members at centimeter wavelengths.}

For each cluster member that we detected in our maps, we measured its flux density at 3.0, 6.4, and 15.0 GHz, {computed the radio spectral index via single or piecewise power-law fits, and searched for signatures of circular polarization and/or radio variability. 
We then used the observed spectral, polarization, and temporal characteristics of the detected radio emission to 
identify candidate proplyds, candidate jets, and gyrosynchrotron-dominated radio sources.}

{
We identify 10 new candidate proplyds in NGC 1977, effectively doubling the total number of candidate proplyds found in the region. Our newly-identified candidate proplyds have radio spectral energy distributions that are well-characterized by power-law free-free emission models with low turnover frequencies and/or flat spectral indices, as expected for ionized, externally evaporating disk winds. 9 of them are located in regions of NGC 1977 where the external FUV radiation field is ${\sim}10-1000$ $G_0$, while one is located in the same ${>}1000$ $G_0$ region as the proplyds previously identified with HST and/or Spitzer imaging. High angular resolution observations are needed to confirm the disk-wind nature of our photometrically identified candidate proplyds. If confirmed, this would demonstrate the ability of broadband radio photometry to identify bright, externally evaporating disks in intermediately irradiated regions of star formation. 
}

{We have also used measurements of centimeter-wavelength free-free emission to calculate the mass-loss rates of known  proplyds and newly-identified candidate proplyds in NGC 1977.} The derived mass loss rates are $1-2$ orders of magnitude lower than the values obtained for proplyds in the central region of the ONC, suggesting that disks in NGC 1977 have longer lifetimes than disk-proplyd systems in the ONC. 
{If the photoevaporative winds are internally driven, which we argued is possible for a subset of the VLA-detected cluster members, then the mass loss rates derived here would be upper limits. Our current VLA-derived mass-loss rates in NGC 1977} 
are consistent with the values predicted for spatially extended externally-evaporating disks that are irradiated by intermediate FUV fields. 
The intermediately-irradiated NGC 1977 cluster {may therefore} 
host a population of extended photoevaporating disks that are larger than the disks typically found in the ONC. 
{Spatially resolved observations of the central disks and surrounding winds, however, are needed to test this hypothesis} and to help identify any compact disks in NGC 1977 that are more likely to be launching internally, rather than externally, driven winds.

{ 
Finally, we computed the ionizing photon rates required to produce the measured free-free emission of our VLA-detected proplyds and candidate proplyds. We find that most of the candidate proplyds in NGC 1977, including one HST-identified proplyd, 
have free-free fluxes that are higher than the values implied by photoionizing radiation from neighboring B- and A-type stars. This marks a difference from photoevaporating disks in the core of the ONC, the majority of which are externally ionized by EUV radiation from the O star $\theta^1$ Ori C.
Photoevaporative disk winds in NGC 1977 may be illuminated by ionizing radiation from their central stars or by other sources of external ionization, unless they are located very close to known B- and A-type stars in the region. The external EUV radiation field in NGC 1977 may also be larger than currently assumed, as suggested by this study and by previous studies in the region, although the exact degree of underestimation remains uncertain. 
}

Our findings suggest that intermediately-irradiated {regions of star formation} host a less hostile environment for disk evolution and planet formation than the intensely-irradiated, proplyd hosting regions of the ONC and other O-star-hosting clusters. 
Disks in the ONC are severely truncated and photoevaporated by strong external UV fields, whereas disks in NGC 1977 appear likely to remain intact and long-lived despite ongoing external photoevaporation. 
The VLA-detected NGC 1977 sources presented in this study represent an ideal set of targets for future follow-up studies of disk demographics in intermediate-UV environments. Such environments represent the typical conditions of star and planet formation in the Galaxy.

\

{\it \noindent  Acknowledgements:} We are grateful to P. Sheehan, who provided useful ideas for the initial development of this work. {We also thank NRAO staff, especially Amy Kimball and Drew Medlin, 
for assistance with VLA scheduling blocks and Stokes V imaging.}
We also acknowledge the use of NRAO computing facilities for the reduction and imaging of the VLA data. 
{Finally, we thank the anonymous referee, whose feedback improved the clarity and organization of this manuscript.}
This work was supported by NSF AAG grant 1811290, and by NRAO Student Observing Support Award SOPS21A-001. R. Boyden also acknowledges support from the University of Arizona's Marshall Foundation Dissertation Fellowship, the Virginia Initiative on Cosmic Origins (VICO), and  NSF grant no. AST-2206437. The National Radio Astronomy Observatory is a facility of the National Science Foundation operated under cooperative agreement by Associated Universities, Inc. 
{Some of the data presented in this paper were obtained from the Mikulski Archive for Space Telescopes (MAST) at the Space Telescope Science Institute. The specific observations analyzed can be accessed via \dataset[10.17909/bagj-yf51]{https://doi.org/10.17909/bagj-yf51}.}
This material is based upon work  supported by the National Aeronautics and Space Administration under Agreement No. 80NSSC21K0593 for the program ``Alien Earths.''
The results reported herein benefitted from collaborations and/or information exchange within NASA's Nexus for Exoplanet System Science (NExSS) research coordination network sponsored by NASA's Science Mission Directorate. 

{\it Facility: NSF's Karl G. Jansky Very Large Array (VLA)}

{\it Software:} {\tt Astropy} \citep{astropy13, Astropy18, Astropy22}, {\tt CASA} \citep{CASA22}, {\tt matplotlib} \citep{Hunter07}

\bibliography{new.ms}

\appendix

\twocolumngrid

\section{Background Source Detections}\label{appendix:NGC_A}

{ 
Here we discuss the radio flux measurements, continuum images, and SEDs for the 37 VLA detections that we classify as probable background objects. {In Table \ref{tab:VLA_source_properties_BKG}, we list the source IDs, coordinates, radio flux measurements, and radio spectral indices for each background source. The radio spectral indices are measured by fitting a single power law to the measured 3.0, 6.4, and 15.0 GHz fluxes. Table \ref{tab:VLA_source_properties_BKG}}
also indicates whether a background source is a known radio source from the \cite{Kounkel14} catalog, {as well as whether the detected emission is resolved, circularly polarized, and/or variable.}

In general, the sample of candidate background sources that we have identified from our infrared photometric selection criteria (see Section \ref{sec:Sample}) is consistent with the expected  distribution of background radio sources in our VLA maps. Following \cite{Rodriguez89}, the number of background radio sources per square degree with a flux greater than or equal to $S_{\nu}$ can be estimated via the equation
\begin{equation}\label{eq:background_sources}
    \langle N \rangle = 40  \bigg(\frac{S_{\nu}} {\textrm{mJy}}\bigg)^{-0.75} \textrm{deg}^{-2}.
\end{equation}
The top panel of Figure \ref{fig:background_sources_plot} compares the flux distributions of our sample of probable background sources with the distribution obtained for our full sample of 46 unclassified radio-detected sources (i.e., detections not associated with a proplyd or YSO catalog), and the distribution predicted from Equation \ref{eq:background_sources} for a field size of $\sim 25' \times 25'$. 
{The} bottom panel of Figure  \ref{fig:detection_profile} shows the spatial distribution of background sources compared with both the spatial distribution of the unclassified radio sources {and the predicted spatial distribution for detected background radio sources in our VLA maps.} 
These figures reveal that both the flux and spatial distributions of the candidate background sources match up well with the predicted distributions from Equation \ref{eq:background_sources}, particularly towards the brighter end of the flux distribution and the outskirts of our maps, where all detections are consistent with being background objects. They also demonstrate that a subset of the faint (${<1}$ mJy) unclassified detections within ${\sim}10'$ of 42 Ori are likely newly-identified cluster members of NGC 1977, rather than additional background sources.

\begin{figure}[ht!]
\epsscale{1.1}
\hspace{-0.15in}
    \plotone{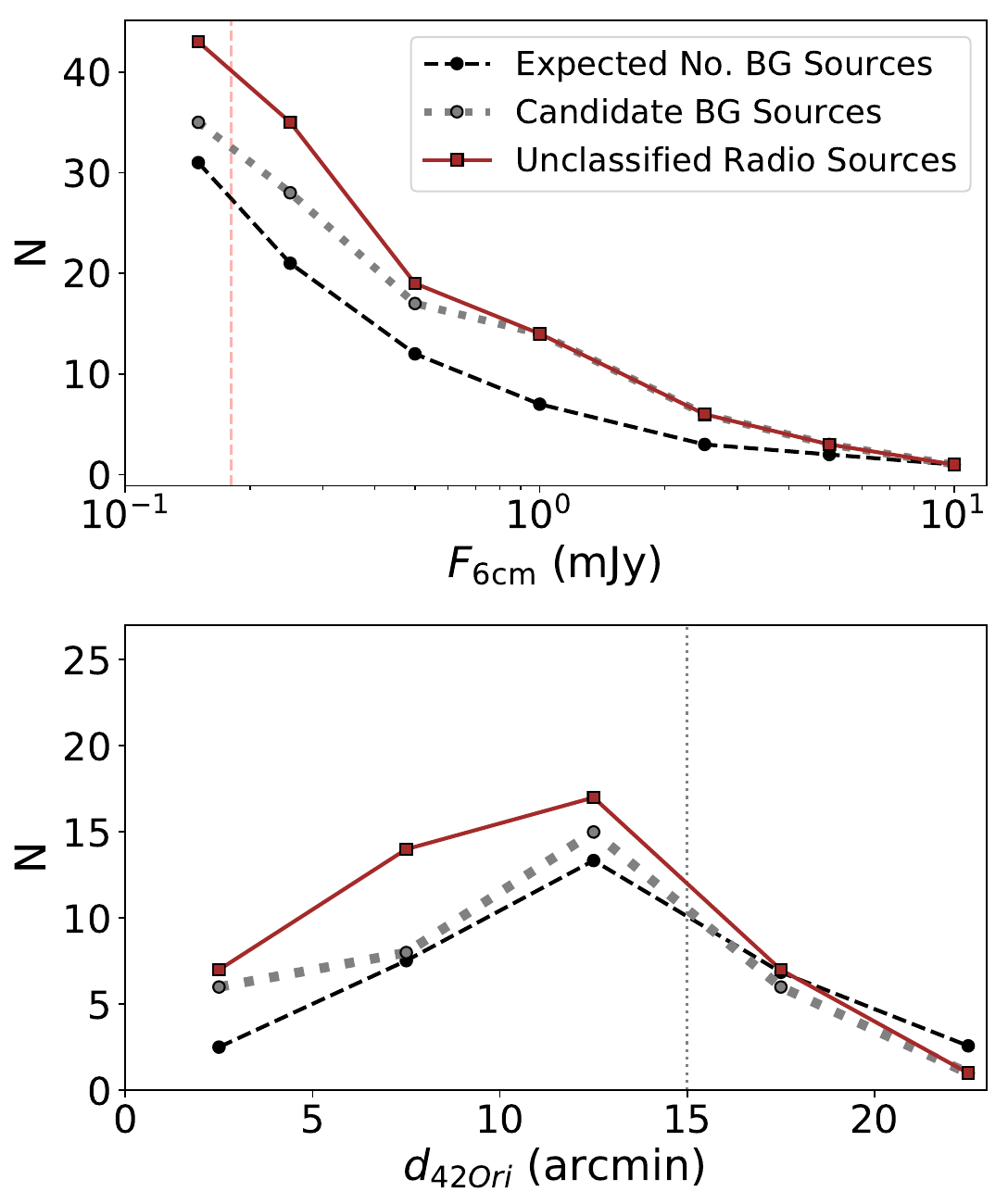}
\caption{Top: Cumulative number of {unclassified radio sources} (brown) and candidate background sources (gray) detected above a certain flux level, compared with the expected number of background sources within the central $25’$ region of NGC 1977 (black). {Unclassified radio sources consist of the 46 detections that are not associated with the proplyd or YSO catalogs assembled in Section \ref{sec:Sample}}. 
The red dashed line depicts the {typical 6$\sigma$ rms level in our VLA maps.} {Bottom: Radially-averaged spatial distribution of the unclassified radio sources and candidate background sources, compared with  {the predicted number of background radio sources in our VLA maps with fluxes greater than the local 6$\sigma$ rms level}.}
\label{fig:background_sources_plot}  
}
\end{figure}

In Figures \ref{fig:BG_detections_1} - \ref{fig:BG_detections_6},  
we show the 3.0 GHz subimages, 6.4 GHz subimages, 15.0 GHz subimages, and radio SEDs for each background object using a similar layout as in Figure \ref{fig:detections_1}. 
In the SED panels of  Figures \ref{fig:BG_detections_1} - \ref{fig:BG_detections_6}, 
we also include the best-fit {power law} and 1$\sigma$ confidence intervals derived for each background source (see Section \ref{sec:SED_modeling}).

{Many of the candidate background objects have steep negative radio spectral indices that are much lower than the values expected for optically thin or optically thick free-free emission  
(e.g., sources B9, B14, B22, and B26).}
These background objects are likely to be emitting strong levels of non-thermal emission, such as gyrosynchrotron emission \citep[e.g.,][]{Feigelson99}. 
This is expected for extragalactic radio sources, as the radio SEDs of galaxies are known to be dominated by gyrosynchrotron radiation under a range of physical conditions \citep[][and references therein]{Condon92}. 

We also find that a subset of background sources shown in Figures \ref{fig:BG_detections_1} - \ref{fig:BG_detections_6} 
has flat spectra that are consistent with optically thin free-free emission 
(e.g., sources B6, B16, and B23). 
Star-forming galaxies can, at times, exhibit flat radio spectra that spectrally resemble optically thin free-free emission \citep[see review in][]{Condon92}, so it seems reasonable for some of the background objects detected in our maps to be fit well by flat-sloped power laws even if they are extragalactic. }

\begin{deluxetable*}{lllcrrrrrrr}\tablenum{5}
\tabletypesize{\small}
\tablewidth{0pt}
\tablecaption{Properties of background radio sources detected in our VLA maps \tablenotemark{ }}\label{tab:VLA_source_properties_BKG}
\tablehead{ 
    \colhead{ID} & 
    \colhead{R.A.}  &
    \colhead{Decl.}  & 
    \colhead{Catalog}  & 
    \colhead{$F_{\nu, \ \textrm{\footnotesize 3.0 GHz}}$} &  
    \colhead{$F_{\nu, \ \textrm{\footnotesize 6.4 GHz}}$}  & 
    \colhead{$F_{\nu, \ \textrm{\footnotesize 15.0 GHz}}$}  &
    \colhead{Sp. Index} &
    \colhead{Res.?} &
    \colhead{Pol.?} &
    \colhead{Var.?} \\ 
    \colhead{}        &  
    \colhead{(J2000)} & 
    \colhead{(J2000)}    & 
    \colhead{}    & 
    \colhead{ (mJy)}   &    
    \colhead{(mJy)}  & 
    \colhead{(mJy)}  & \colhead{} & \colhead{} &  \colhead{} & \colhead{} }
\colnumbers
\startdata 
B1  &  05:36:08.64 &  --4:34:45.40 &          \nodata &   $5.231 \pm 0.379$ &   $6.015 \pm 0.431$ &                \nodata &   $0.18^{0.14}_{0.10}$ &                        &                  Y &                        \\
B2  &  05:36:23.36 &  --4:37:04.83 &          \nodata &  $11.272 \pm 0.801$ &   $4.061 \pm 0.304$ &                \nodata &  $-1.36^{0.14}_{0.10}$ &                      Y &                  Y &                        \\
B3  &  05:35:21.84 &  --4:37:59.30 &              K14 &   $1.992 \pm 0.148$ &   $1.269 \pm 0.093$ &      $3.031 \pm 5.080$ &  $-0.59^{0.14}_{0.14}$ &                      Y &                    &                        \\
B4  &  05:35:50.28 &  --4:39:32.60 &              K14 &   $0.917 \pm 0.075$ &   $0.665 \pm 0.052$ &     $2.463 \pm 10.005$ &  $-0.42^{0.14}_{0.17}$ &                        &                    &                        \\
B5  &  05:36:03.67 &  --4:40:30.80 &              K14 &   $0.602 \pm 0.055$ &   $1.414 \pm 0.103$ &      $2.481 \pm 0.185$ &   $0.84^{0.07}_{0.04}$ &                        &                    &                        \\
B6  &  05:35:30.52 &  --4:40:35.99 &          \nodata &   $0.169 \pm 0.036$ &   $0.177 \pm 0.025$ &      $0.196 \pm 0.038$ &   $0.11^{0.25}_{0.28}$ &                        &                    &                        \\
B7  &  05:34:28.70 &  --4:40:38.61 &          \nodata &   $0.328 \pm 0.069$ &   $0.308 \pm 0.041$ &                \nodata &  $-0.07^{0.49}_{0.46}$ &                        &                    &                        \\
B8  &  05:36:25.10 &  --4:41:51.57 &          \nodata &   $1.554 \pm 0.122$ &   $1.378 \pm 0.106$ &   $46.165 \pm 112.296$ &  $-0.17^{0.18}_{0.14}$ &                      Y &                    &                        \\
B9  &  05:35:50.63 &  --4:42:02.00 &              K14 &   $2.793 \pm 0.200$ &   $1.774 \pm 0.127$ &      $0.673 \pm 0.068$ &  $-0.84^{0.07}_{0.07}$ &                      Y &                    &                        \\
B10 &  05:36:09.39 &  --4:42:53.65 &          \nodata &   $0.252 \pm 0.038$ &   $0.172 \pm 0.026$ &     $-0.037 \pm 0.221$ &  $-0.52^{0.35}_{0.38}$ &                        &                    &                        \\
B11 &  05:35:35.97 &  --4:43:41.24 &          \nodata &   $0.218 \pm 0.034$ &   $0.164 \pm 0.025$ &      $0.012 \pm 0.137$ &  $-0.42^{0.35}_{0.38}$ &                        &                    &                        \\
B12 &  05:34:30.54 &  --4:44:29.40 &              K14 &   $0.604 \pm 0.062$ &   $0.765 \pm 0.059$ &   $63.094 \pm 184.289$ &   $0.32^{0.21}_{0.21}$ &                        &                    &                      Y \\
B13 &  05:34:31.50 &  --4:44:42.87 &          \nodata &   $0.088 \pm 0.150$ &   $0.181 \pm 0.027$ &     $0.713 \pm 78.576$ &             ${>}-0.77$ &                        &                    &                        \\
B14 &  05:36:23.14 &  --4:44:59.60 &              K14 &   $2.100 \pm 0.155$ &   $1.018 \pm 0.079$ &                \nodata &  $-0.94^{0.14}_{0.14}$ &                      Y &                    &                      Y \\
B15 &  05:34:36.52 &  --4:45:18.40 &          \nodata &   $0.041 \pm 0.130$ &   $0.239 \pm 0.028$ &      $4.019 \pm 9.329$ &              ${>}0.04$ &                        &                    &                        \\
B16 &  05:34:50.92 &  --4:47:04.45 &          \nodata &   $0.162 \pm 0.031$ &   $0.155 \pm 0.023$ &      $0.220 \pm 0.203$ &  $-0.03^{0.46}_{0.42}$ &                        &                    &                        \\
B17 &  05:36:11.72 &  --4:47:07.00 &              K14 &   $0.868 \pm 0.072$ &   $1.506 \pm 0.109$ &      $0.839 \pm 2.495$ &   $0.70^{0.17}_{0.14}$ &                        &                    &                        \\
B18 &  05:35:17.81 &  --4:48:03.49 &          \nodata &   $0.403 \pm 0.041$ &   $0.303 \pm 0.032$ &      $0.156 \pm 0.242$ &  $-0.38^{0.25}_{0.25}$ &                      Y &                    &                        \\
B19 &  05:35:07.24 &  --4:48:53.20 &              K14 &   $0.297 \pm 0.036$ &   $0.411 \pm 0.036$ &      $0.241 \pm 0.046$ &  $-0.03^{0.10}_{0.14}$ &                        &                    &                        \\
B20 &  05:35:14.70 &  --4:49:05.49 &          \nodata &   $0.132 \pm 0.030$ &   $0.160 \pm 0.025$ &      $0.118 \pm 0.145$ &   $0.18^{0.46}_{0.46}$ &                        &                  Y &                        \\
B21 &  05:35:47.43 &  --4:49:52.60 &              K14 &   $1.935 \pm 0.142$ &   $1.608 \pm 0.116$ &      $0.462 \pm 0.049$ &  $-0.80^{0.04}_{0.07}$ &                        &                    &                        \\
B22 &  05:35:47.56 &  --4:49:53.30 &              K14 &   $1.943 \pm 0.143$ &   $1.491 \pm 0.108$ &      $0.472 \pm 0.049$ &  $-0.80^{0.04}_{0.07}$ &                        &                  Y &                        \\
B23 &  05:35:31.03 &  --4:50:38.40 &              K14 &   $0.362 \pm 0.040$ &   $0.469 \pm 0.040$ &      $0.301 \pm 0.042$ &  $-0.07^{0.10}_{0.10}$ &                        &                  Y &                        \\
B24 &  05:35:58.42 &  --4:51:23.19 &          \nodata &   $0.272 \pm 0.040$ &   $0.272 \pm 0.032$ &     $-0.079 \pm 1.320$ &   $0.00^{0.35}_{0.31}$ &                        &                    &                        \\
B25 &  05:35:13.90 &  --4:51:59.30 &              K14 &   $2.290 \pm 0.165$ &   $0.974 \pm 0.076$ &      $0.218 \pm 0.063$ &  $-1.26^{0.10}_{0.10}$ &                      Y &                  Y &                        \\
B26 &  05:35:13.73 &  --4:52:01.90 &              K14 &  $12.769 \pm 0.904$ &   $8.419 \pm 0.596$ &      $2.011 \pm 0.154$ &  $-1.15^{0.07}_{0.04}$ &                      Y &                  Y &                        \\
B27 &  05:35:57.91 &  --4:52:32.44 &          \nodata &   $0.325 \pm 0.041$ &   $0.330 \pm 0.038$ &      $0.457 \pm 1.435$ &   $0.04^{0.28}_{0.31}$ &                        &                    &                        \\
B28 &  05:34:24.63 &  --4:53:03.84 &          \nodata &   $2.693 \pm 0.196$ &   $2.688 \pm 0.193$ &     $-1.957 \pm 7.527$ &   $0.00^{0.14}_{0.14}$ &                      Y &                  Y &                        \\
B29 &  05:34:19.75 &  --4:53:22.31 &          \nodata &   $0.425 \pm 0.062$ &   $0.446 \pm 0.047$ &    $18.413 \pm 88.576$ &   $0.07^{0.31}_{0.31}$ &                      Y &                    &                        \\
B30 &  05:36:14.03 &  --4:53:29.38 &          \nodata &   $0.397 \pm 0.047$ &   $0.658 \pm 0.057$ &     $2.285 \pm 13.865$ &   $0.66^{0.25}_{0.24}$ &                        &                    &                        \\
B31 &  05:36:13.78 &  --4:54:40.13 &          \nodata &   $0.273 \pm 0.044$ &   $0.328 \pm 0.045$ &                \nodata &   $0.25^{0.39}_{0.39}$ &                        &                    &                        \\
B32\tablenotemark{a} &  05:35:58.88 &  --4:55:37.70 &              K14 &  $25.922 \pm 1.833$ &  $61.653 \pm 4.361$ &     $63.940 \pm 4.544$ &   $0.56^{0.04}_{0.04}$ &                        &                  Y &                        \\
B33 &  05:36:11.91 &  --4:56:34.14 &          \nodata &   $0.535 \pm 0.054$ &   $0.468 \pm 0.057$ &                \nodata &  $-0.17^{0.25}_{0.28}$ &                        &                    &                        \\
B34 &  05:35:00.69 &  --4:57:53.50 &              K14 &   $4.521 \pm 0.322$ &   $2.955 \pm 0.211$ &      $0.862 \pm 0.111$ &  $-0.91^{0.07}_{0.07}$ &                      Y &                  Y &                        \\
B35 &  05:36:04.43 &  --4:59:08.40 &              K14 &   $0.315 \pm 0.050$ &   $0.435 \pm 0.057$ &  $-56.370 \pm 111.789$ &   $0.42^{0.39}_{0.35}$ &                        &                  Y &                      Y \\
B36 &  05:35:37.21 &  --5:00:55.20 &              K14 &   $0.097 \pm 0.198$ &   $0.372 \pm 0.053$ &                \nodata &              ${>}-0.1$ &                        &                    &                        \\
B37 &  05:35:55.90 &  --5:04:25.20 &              K14 &   $0.406 \pm 0.087$ &             \nodata &                \nodata &                \nodata &                        &                    &                        \\                  
\enddata
\tablenotetext{ }{{\bf Notes.} 
Column (1): source IDs used in this article. Columns (2) and (3): phase center coordinates. 
Column (3): catalog association of VLA-detected sources. Here we use the same notation as in Table \ref{tab:VLA_source_properties}.  
Columns (5), (6), and (7): measured fluxes at 3.0 GHz, 6.4 GHz, and 15.0 GHz.
{Column (8): Best-fit spectral index derived from fitting a power law to the measued radio fluxes. }
{Column (9): Source morphology tracker. A `Y' is used to denote sources that are spatially resolved in one or more bands. Empty entries indicate detections that are unresolved in our VLA maps. }
{Column (10): Circular polarization indicator. A `Y' denotes a source with Stokes V emission detected above a $2.5{\sigma}$ level. }
{Column (11): Variable source indicator. A `Y' is used to denote sources with variable ${\sim}{6.4}$ GHz emission, based on comparisons with prior flux measurements from \cite{Kounkel14} (see Section \ref{sec:variable}).}}
\tablenotemark{a}{Indicates the bright radio source J053558.88-045537.7}
\end{deluxetable*}

\begin{figure*}[ht!]
\epsscale{1.1}
\hspace{-0.3in}
    \plotone{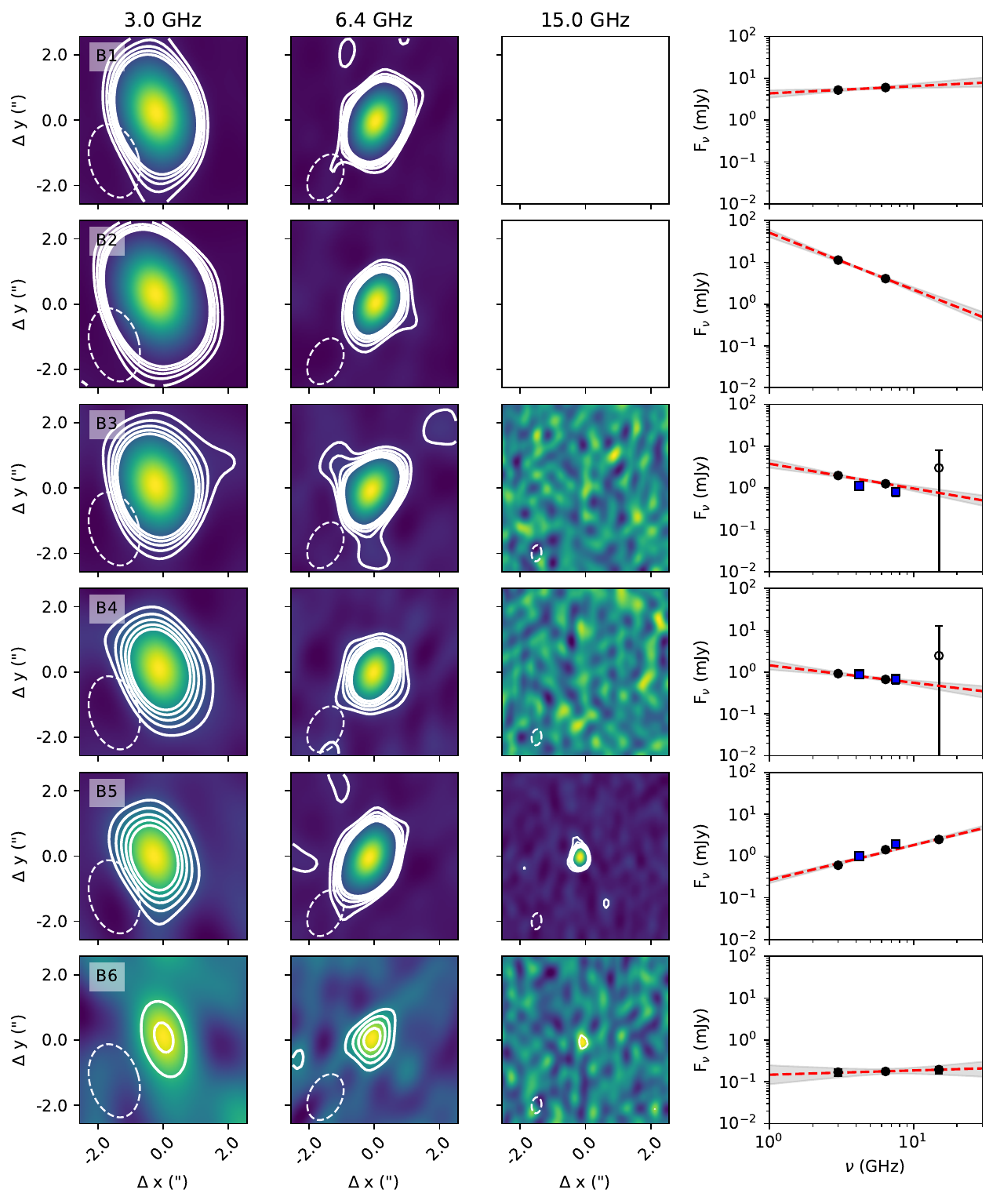}
\caption{Same as Figure \ref{fig:detections_1}, 
 but for background objects detected in our VLA maps. { For sources that lie outside of the field of views of our 6.4 GHz and/or 15.0 GHz observations (see Figure \ref{fig:NGC1977_region}), we use whitespace to fill in the corresponding image subpanels.}
\label{fig:BG_detections_1}}
\end{figure*}

\begin{figure*}[ht!]
\epsscale{1.1}
\hspace{-0.3in}
    \plotone{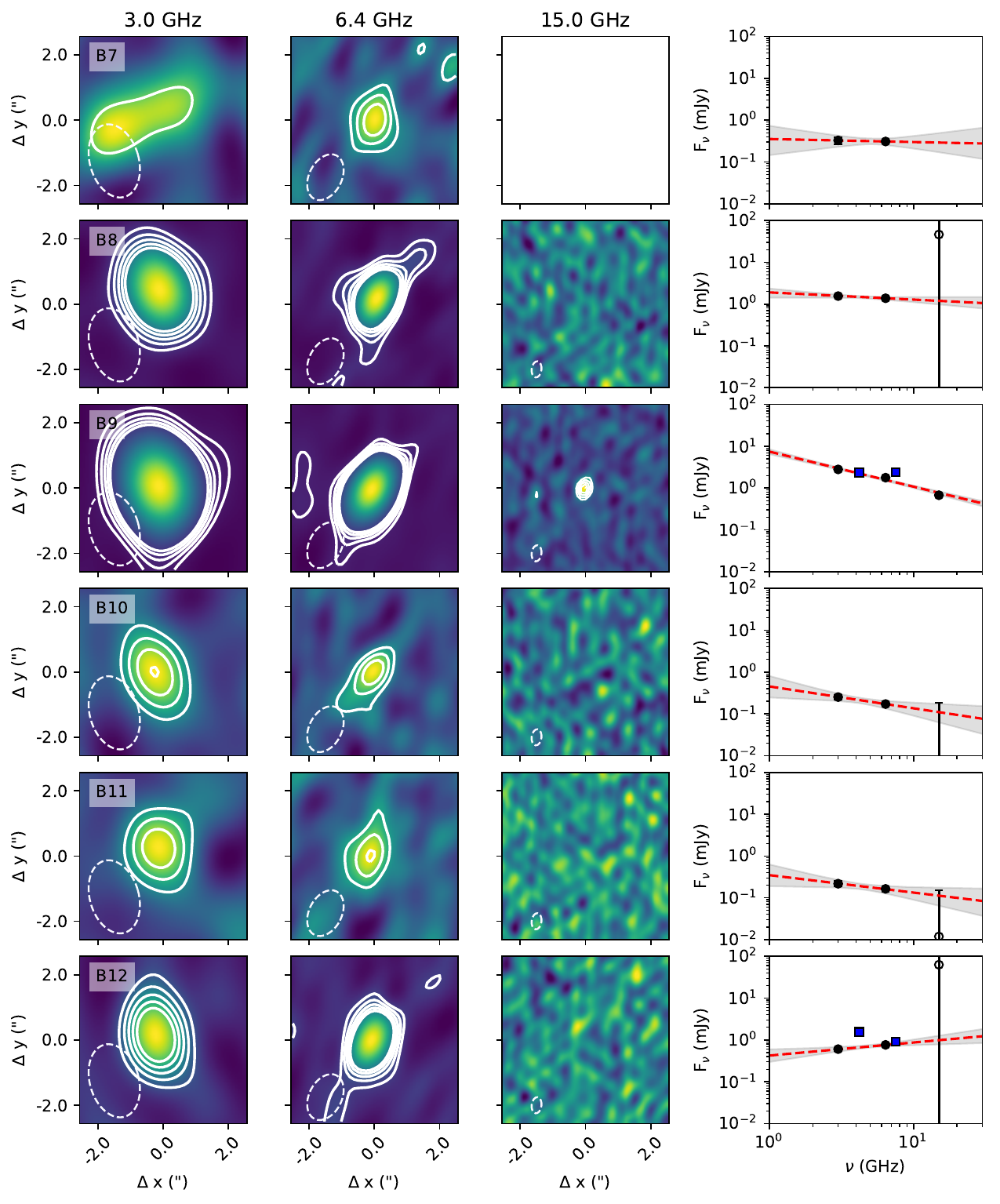}
\caption{{  Continuation of Figure \ref{fig:BG_detections_1}.}
\label{fig:BG_detections_2}}
\end{figure*}

\begin{figure*}[ht!]
\epsscale{1.1}
\hspace{-0.3in}
    \plotone{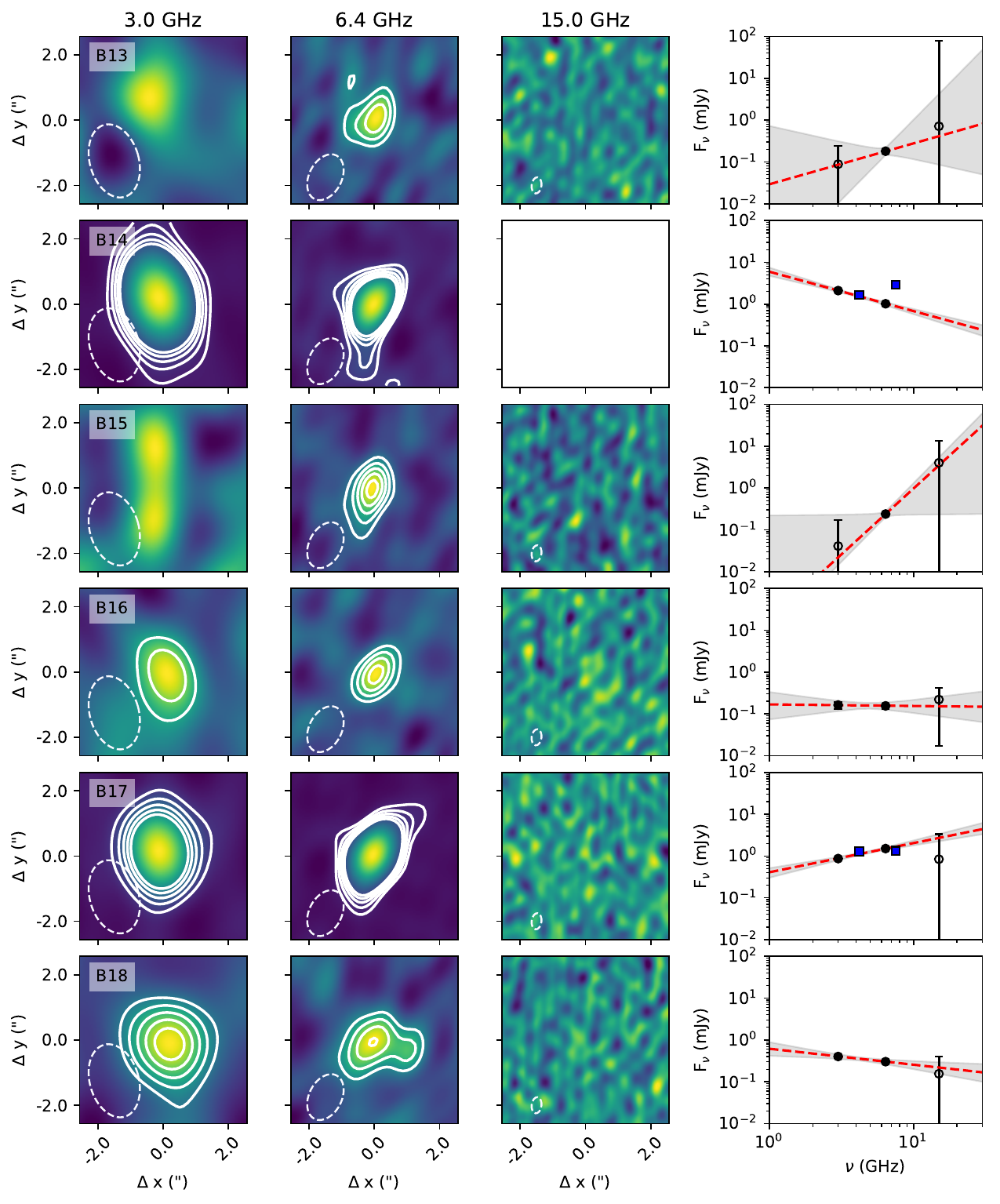}
\caption{{  Continuation of Figure \ref{fig:BG_detections_1}.}
\label{fig:BG_detections_3}}
\end{figure*}

\begin{figure*}[ht!]
\epsscale{1.1}
\hspace{-0.3in}
    \plotone{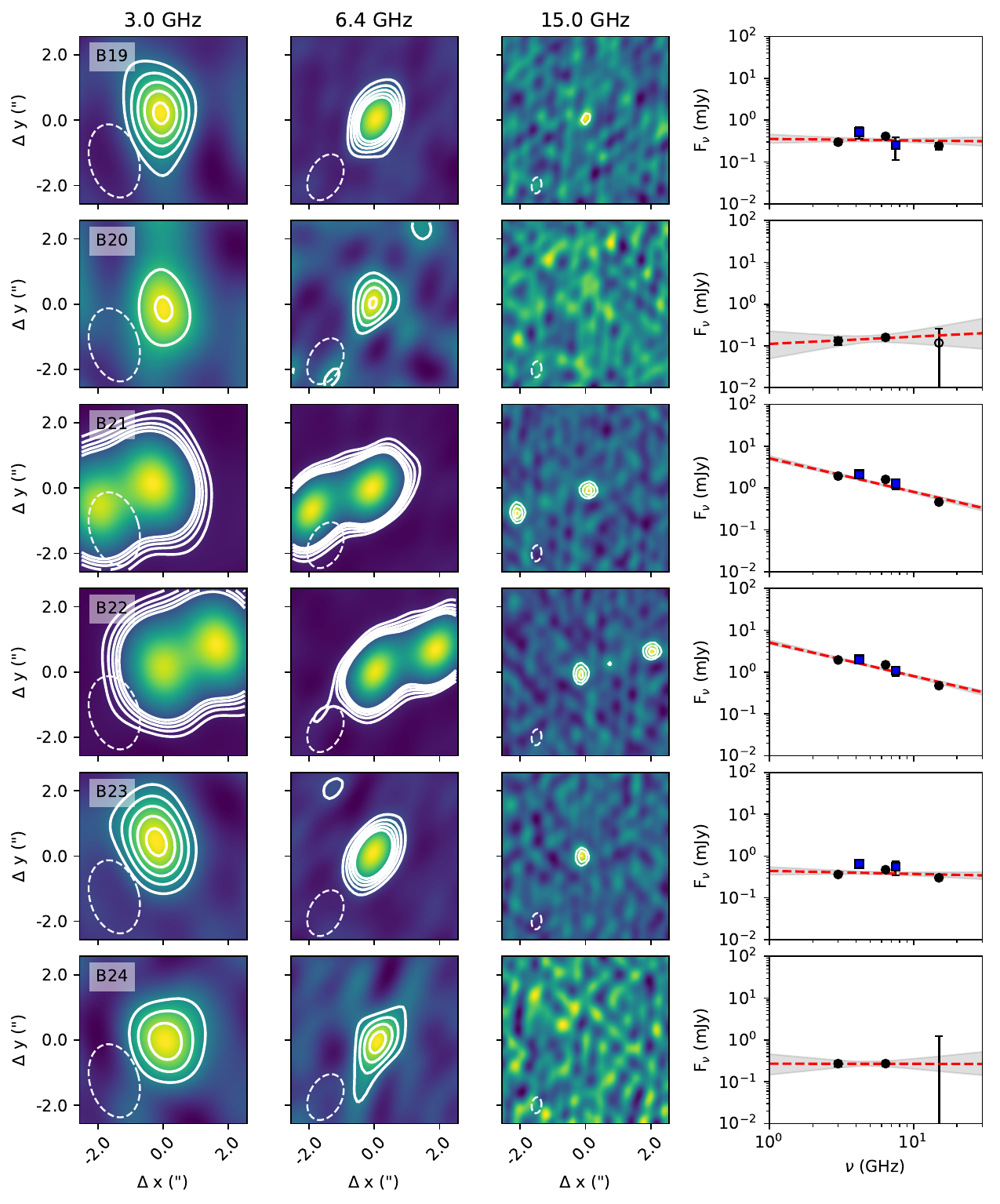}
\caption{{  Continuation of Figure \ref{fig:BG_detections_1}.}
\label{fig:BG_detections_4}}
\end{figure*}

\begin{figure*}[ht!]
\epsscale{1.1}
\hspace{-0.3in}
    \plotone{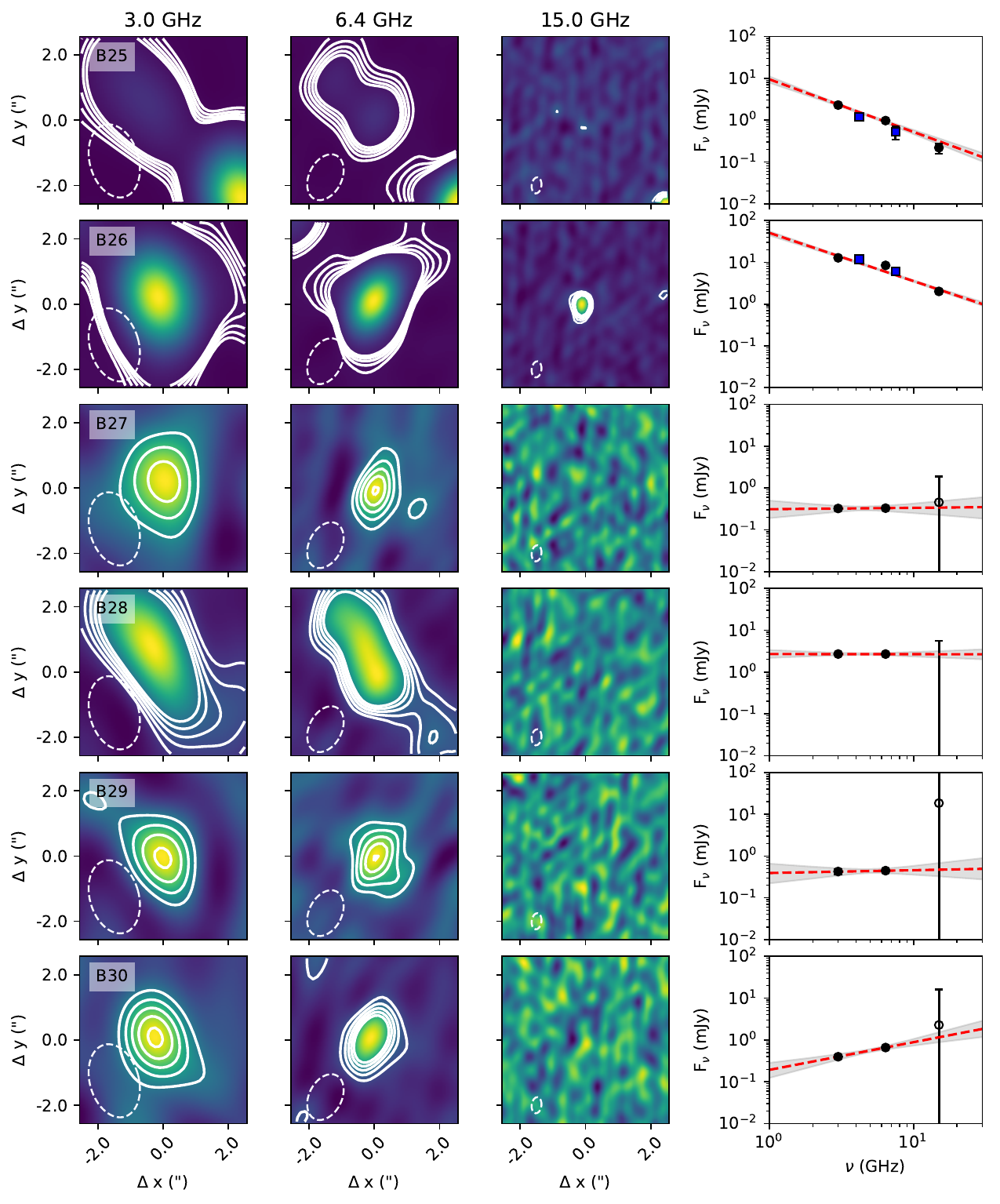}
\caption{{  Continuation of Figure \ref{fig:BG_detections_1}.}
\label{fig:BG_detections_5}}
\end{figure*}

\begin{figure*}[ht!]
\epsscale{1.1}
\hspace{-0.3in}
    \plotone{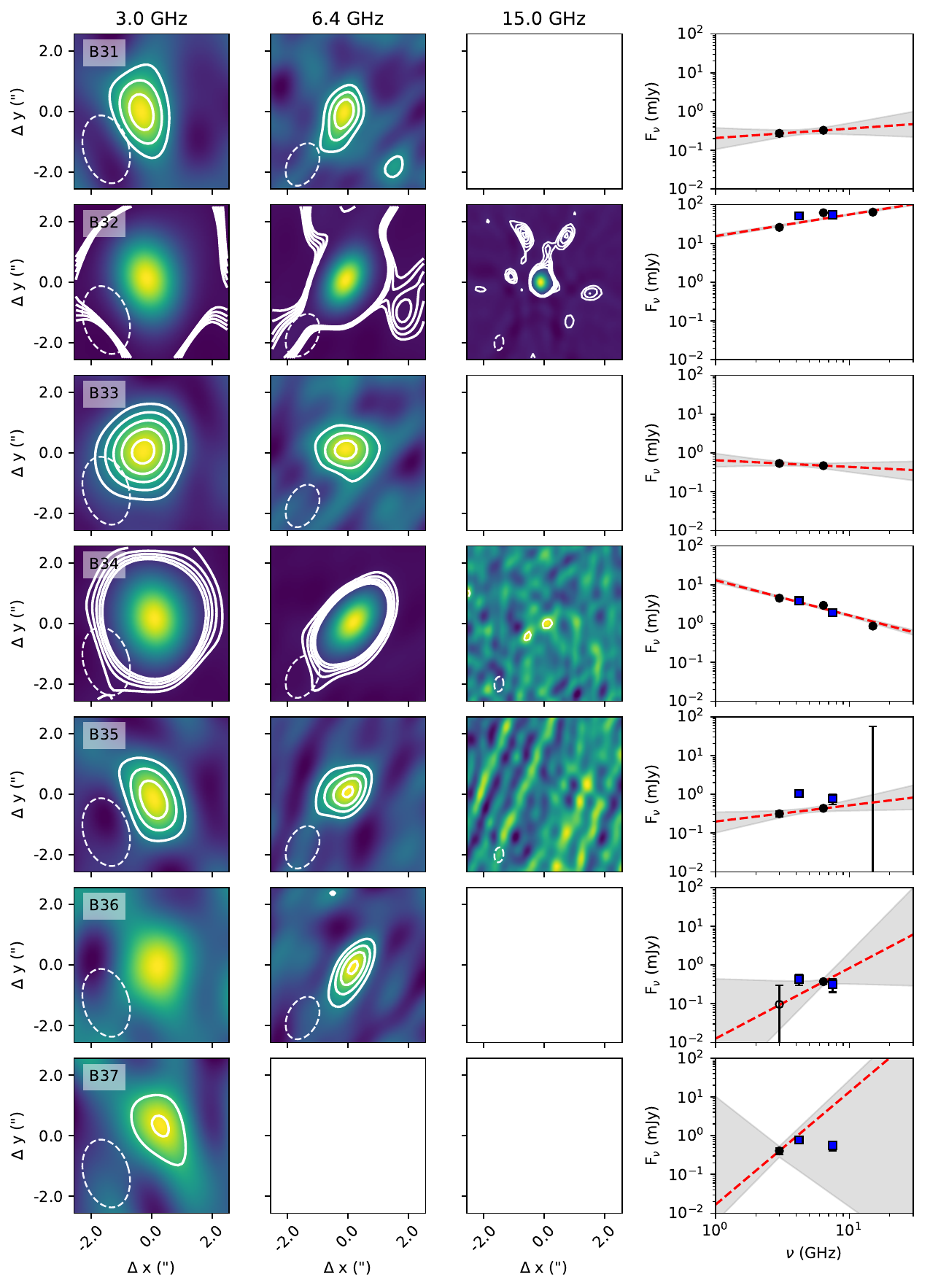}
\caption{{  Continuation of Figure \ref{fig:BG_detections_1}.}
\label{fig:BG_detections_6}}
\end{figure*}

%
%
%
%
%
\end{document}